\documentclass[prd,twocolumn,floatfix,preprintnumbers,showpacs,nofootinbib,superscriptaddress]{revtex4}

\usepackage{amsmath,amsthm,latexsym,amssymb,amsfonts,epsfig, psfrag}
\usepackage{dsfont}
\def\be{\begin{equation}}
\def\ee{\end{equation}}
\def\ba{\begin{eqnarray}}
\def\ea{\end{eqnarray}}
\def\q{{\quad}}
\def\nn{\nonumber}

\newcommand{\cc}{\mathcal C}

\newcommand{\cg}{\mathcal G}

\newcommand{\cp}{\mathcal P}
\newcommand{\cq}{\mathcal Q}

\newcommand{\fg}{\mathfrak{g}}

\begin{document}

\preprint{IGC--10/11--3, ITP-UU-10/42,
SPIN-10/35}

\title{Effective approach to the problem of time: \\ general features and examples}

\author{Martin Bojowald}
\email[ ]{bojowald@gravity.psu.edu}
\affiliation{Institute for Gravitation and the Cosmos, The
Pennsylvania State
University,\\
104 Davey Lab, University Park, PA 16802, USA}
\author{Philipp A.~H\"ohn}
\email[]{p.a.hohn@uu.nl}
\affiliation{Institute for Theoretical Physics,
 Universiteit Utrecht,\\
Leuvenlaan 4, NL-3584 CE Utrecht, The Netherlands}
\affiliation{Institute for Gravitation and the Cosmos, The
Pennsylvania State
University,\\
104 Davey Lab, University Park, PA 16802, USA}
\author{Artur Tsobanjan}
\email[]{axt236@psu.edu}
\affiliation{Institute for Gravitation and the Cosmos, The
Pennsylvania State
University,\\
104 Davey Lab, University Park, PA 16802, USA}



\setcounter{footnote}{0}

\begin{abstract}
 The effective approach to quantum dynamics allows a reformulation of
 the Dirac quantization procedure for constrained systems in terms of
 an infinite-dimensional constrained system of classical type. For
 semiclassical approximations, the quantum constrained system can be
 truncated to finite size and solved by the reduced phase space or
 gauge-fixing methods. In particular, the classical feasibility of
 local internal times is directly generalized to quantum systems,
 overcoming the main difficulties associated with the general problem
 of time in the semiclassical realm. The key features of local
 internal times and the procedure of patching global solutions using
 overlapping intervals of local internal times are described and
 illustrated by two quantum mechanical examples. Relational evolution in a given choice of
 internal time is most conveniently described and interpreted in a
 corresponding choice of gauge at the effective level and changing the
 internal clock is, therefore, essentially achieved by a gauge
 transformation. This article complements the conceptual discussion in
 \cite{EffTime1}.
\end{abstract}

\maketitle

\section{Introduction}

One of the most pressing issues in the development of a consistent
theory of quantum gravity is the problem of time
\cite{Kuc1,Ish,anderson,Rovbook}. As a generally covariant theory, its
dynamics is fully constrained, without a true Hamiltonian generating
evolution with respect to a distinguished or absolute time.
Within the classical treatment, using the conventional spacetime
(manifold) picture, this does not immediately pose a serious problem
since there are different notions of time available in general relativity.
The physical notion of time as experienced by a specific observer is
supplied in an invariant and unambiguous manner by the proper time
along that observer's worldline. The second notion appears in the
context of the canonical initial-value formulation, often constructed
by introducing a foliation of spacetime by spatial
hypersurfaces. However, the time coordinate that labels these
hypersurfaces, in contrast to proper time, has no invariant physical
meaning. It
is simply the gauge
parameter for orbits of the Hamiltonian constraint and, classically,
these orbits lie entirely within the constraint surface. Evolution
along the orbits may be interpreted with respect to this time
coordinate which provides an ordering to physical relations. When
quantizing the theory via the Dirac procedure, however, physical
states are to be annihilated by the quantum constraints and are,
therefore, gauge invariant by construction. The gauge flow, along with
the gauge parameters of the constraints, is absent in the physical
Hilbert space. In the presence of a Hamiltonian constraint this means
that physical states are timeless. Furthermore, physical observables
should be gauge invariant and must thus be constant along classical
dynamical trajectories and commute with the constraints in the quantum
theory.\footnote{The viewpoint that physically observable quantities
  in parametrized systems should commute with all constraints,
  including the Hamiltonian constraint, has been challenged by
  Kucha\v{r} (and, more recently, by Barbour and Foster \cite{Bar}).
  For instance, in \cite{Kuc2} he argues for a difference between
  conventional gauge systems and parametrized systems, leading to the
  proposal that states along the orbit of the Hamiltonian constraint
  should not be identified since this would stand in contradiction to
  our every-day experience of the flow of time. He advocates that,
  instead, in general relativity physically observable quantities
  should only commute with the diffeomorphism constraints, but not
  necessarily with the Hamiltonian constraint. Nevertheless, in this
  article we take the conventional standpoint of requiring that
  physically observable quantities should commute with all constraints
  and, consequently, that in this sense no distinction ought to be
  made between the Hamiltonian and the other constraints.} It appears
as if ``nothing moves'', or, as if ``dynamics is frozen''.

Change and dynamics, however, can be untangled from this static
world by taking the underlying principles of general relativity
seriously, according to which physics is purely relational.
Evolution is not measured with respect to an absolute external
parameter but time can be chosen among the internal degrees of
freedom. Evolution is then interpreted relative to such an
internal clock, where internal time is more general than and not necessarily directly
linked to the proper time of any observer. While proper time is
practical for describing dynamics {\it in} a gravitational field
since it depends on the worldlines of observers and has meaning only
after solving the Einstein equations, in quantum gravity one is rather
interested in the dynamics {\it of} the gravitational field, for which
internal time is useful. This concept has led to the so-called {\it
evolving constants of motion} \cite{Rovbook,Rovmod}, which are
relational Dirac observables measuring physical correlations between
the internal clock and other degrees of freedom. Significant progress in this
direction and generalizations of such relational observables have
been undertaken in \cite{Bianca1,Bianca2,Haj1}, and some criticism
concerning their capability of solving the problem of time has been
raised in \cite{Kuc1,Ish,Kuc2,Hartle}. In the sequel, we will adopt
the relational viewpoint and employ internal clocks as
measures of a relational time. (Some interesting real-world aspects
also relevant to internal clocks have been discussed, for instance, in
\cite{gampul}.) As regards evolution, the choice and corresponding
notion of time are inherently connected to the choice of the
internal clock variable.

Apart from this conceptual issue, the problem of time usually comes
with a whole plethora of technical problems
\cite{Kuc1,Ish,anderson}, of which the ones touched upon in this
article may be summarized as follows:
\begin{itemize}
\item {\it The multiple-choice problem}. Which internal time should one choose as a clock? There is no natural choice of an internal clock variable and different internal times may provide different quantum theories \cite{Kuc1,Ish,Hajlec}. Furthermore,  one must impose restrictions on the choice of internal time functions, since some choices lead to inconsistent probabilistic predictions in the quantum theory and time orderings which are not well-defined \cite{Hartle}.
\item {\it The Hilbert space problem}. Which Hilbert space representation is one to choose and how is one to construct a positive-definite physical inner product on the space of solutions to the quantum constraints?
\item {\it The operator-ordering problem}. The usual ordering problems arise upon promoting classical constraints to operator equivalents. The choice of a time variable also plays a role in the ordering problem \cite{Kuc1}.
\item {\it The global time problem}. Similarly to the Gribov problem in non-abelian gauge theories, there may exist global obstructions to singling out good internal clock variables which provide good parametrizations of the gauge orbits in the sense that each classical trajectory intersects every hypersurface of constant clock time once and only once \cite{Kuc1,Ish,Haj1,Haj2,Rovmod}.
\item {\it The problem of observables}. It is very difficult to construct
a sufficient set of explicit observables for gravitational and
parametrized theories and even the existence of a sufficient set has
been questioned \cite{anderson,Kuc2,Haj1}. In fact, no general Dirac
observables are known for general relativity. While classically
significant  progress has been made in this area
\cite{Bianca1,Bianca2,Haj1}, the problem worsens in the quantum
theory due to the previous technical issues since no general scheme
exists for converting such observables --- if found at all --- into
suitable operators.
\end{itemize}

The relational interpretation of evolution is complicated by the
fact that internal clock functions are neither universal nor perfect. A
globally valid choice of internal time is difficult to find and, due
to the {\it global time problem}, may not exist. For specific matter
systems, such as a free massless scalar field or pressurelss dust,
deparameterizations with a matter clock can be performed, but these
models seem rather special. In order to evaluate the dynamics of
quantum gravity and derive potentially observable information from
first principles, the various problems of time must be overcome
without requiring specific adaptations.

The imperfect nature of internal clocks does not constitute a
problem at the classical level, however, since, in principle, we can
always make use of the gauge parameter along the flow of the
Hamiltonian constraint and evolve in this coordinate time with
respect to which the internal clock, say $T(x)$, and the other
variables of interest, say $Q_i(x)$, have a given evolution.
Comparing the values of the internal clock and the $Q_i(x)$ along
the coordinate time then gives a relational evolution. If $T(x)$
fails to be a good global clock, the system will eventually go
backwards in it, the observable correlations $Q_i(T(x))$ will, in
general, be multi-valued and, consequently, the evolution of the
correlations $Q_i(T)$ will be ``patched up'', where on each patch
$T$ will be a good clock. Thus, classically, in principle, we do not
even need to switch clocks if one takes the evolution in some good
time coordinate into account which does not know about non-global
clocks and provides an ordering to the patches. With respect to this
time coordinate we can solve a well-defined initial value problem
(IVP) (as long as a time direction is given). One can even encode
this relational evolution entirely with physical correlations
without referring to any gauge parameter, if one keeps not only the
relational configuration observables but also the relational
momentum observables in mind to determine an orientation in
which to evolve even at a turning point of a non-global clock. If a
time direction is provided, one can also impose relational initial
data to completely specify a classical solution. The classical
solution may then be obtained by choosing a physical Hamiltonian
which moves the surfaces of constant $T$ in phase space.  In the
case of a non-global clock, this reconstruction is complicated by
the fact that a given trajectory may intersect a constant time
hypersurface more than once or not at all. In this case one will
have to choose more than one Hamiltonian but this is merely a
technical difficulty, not a fundamental problem. We will come back
to this point in the main body of this article.

Due to the quantum uncertainties and the lack of a classical gauge
parameter, performing a ``patching" as above will no longer be
possible in the full quantum theory and we are forced to employ purely
relational information which will require the switching of non-global
clocks. If relational time is defined for only a finite range, a
unitary relational state evolution can not be
accomplished and, as we will see, will break down earlier than the
corresponding Hamiltonian evolution in the classical
theory.\footnote{The finite range of a clock and the resulting
apparent non-unitarity are what one could call a ``classical symptom''
and a ``quantum illness'' which prevent an acceptable quantum
dynamical solution in a conventional sense \cite{Klauder}.  The point
is, however, that this non-unitarity in internal time is only the
result of a local dynamical interpretation of an a priori timeless
system which, in itself is not non-unitary.  These considerations are
relevant for quantum gravity, since, from a certain point of view,
there might not exist a fundamental notion of time at the Planck scale
which would allow for a meaningful, conventional unitary evolution
\cite{Rovbook,Rovmod}.} While classical evolution in non-global clocks
is, in principle, unproblematic, non-unitary quantum evolution can
lead to meaningless results long before the end of a local time is
reached and it is not clear how to define relational quantum
observables in this case.

Even though coordinate time may not exist in full quantum gravity at
the Planck scale, one would heuristically expect that on the way to
larger scales --- in a semiclassical regime which ought to provide
the connection to the classical solutions of general relativity ---
one can reconstruct a (certainly non-unique) coordinate time (for a
discussion of this within loop quantum cosmology see
\cite{bojsisk}). Indeed, the notion of a time coordinate and
evolution trajectory should become meaningful for coherent states
whose expectation values follow the classical trajectory at least
for a certain range. In a semi-classical regime, the notion of
coordinate time should, therefore, make sense and we should be able
to follow a similar strategy here as in the classical situation.

For most applications of quantum gravity related to potential
observable effects, semiclassical evolution is sufficient, or, at
least provides a large amount of information. One may then hope that
such a situation makes dealing with the problem of time more
feasible since this problem does not play a handicapping role
classically; at the very least a dedicated analysis of semiclassical
evolution should provide insights which may help in attacking the
problem in full generality.

This article complements the conceptual discussion in~\cite{EffTime1}
with concrete examples and a concrete discussion of the general
features they exhibit. We use the effective approach to quantum
constraints developed in \cite{EffCons,EffConsRel} in the context of
the problem of time; truncation at semiclassical order reintroduces
some notion of classical gauge parameters. It is the aim of the
present article to sidestep a number of technical issues associated to
an explicit Dirac type approach and to specifically cope with the {\it
global time problem}, while the other technical problems alluded to
above will automatically be addressed in the course of the
discussion. It is our goal to make physical predictions based on some
set of (relational) input data, also in non-deparametrizable systems.
We will make use of (local) deparametrizations in order to locally
scan through an a priori timeless physical state, thereby introducing
a notion of quantum evolution.  We propose a practical solution
employing local, rather than global internal times and adopt and
emphasize the viewpoint that the relational interpretation is,
generally, only of local and semiclassical meaning, as was argued
in~\cite{EffTime1}.
For explicit calculations, our methods will lend themselves easily to
gauge-fixing techniques, avoiding complicated derivations of complete
observables.  In analogy to local coordinates on a manifold, we cover
the evolution trajectories by patches of local time and translate
between them in order to evolve through pathologies of local
clocks. The choice of time is best described and interpreted in a
corresponding
choice of
gauge at the effective level and translating between different local
clocks, therefore, requires nothing more than a gauge
transformation. In addition, we find that non-unitarity at the state
level translates into complex internal time. To begin with, we will
focus on simple mechanical toy models which we will treat in the
classical, effective and for comparison, where feasible, in a
Hilbert-space approach.  The first model is deparametrizable, even
though we employ a non-global clock for the relational evolution,
while the second model is a true example of a ``timeless,''
non-deparametrizable system which has previously been discussed by
Rovelli \cite{Rovbook,Rovmod}.

The rest of the article is organized as follows.
Sec.~\ref{sec:effcons} reviews the effective treatment of a quantum
Hamiltonian constraint and summarizes features of the example of the
``relativistic'' harmonic oscillator. In Sec.~\ref{lt} we study the
first of the two models, discussing its classical and quantum behavior
before going through the full effective treatment truncated using the
semiclassical approximation. In this model we opt to use a time
variable which is non-monotonic along every classical trajectory. We
find that a consistent effective treatment of this model requires
assigning a complex expectation value to the kinematical time
operator.  We find an explicit gauge transformation which allows us to
evolve the model of Sec.~\ref{lt} through the turning point of the
non-global clock. A detailed discussion of general features of such
transformations, as well as of the close relationship between the
choice of an internal time variable and suitable gauge fixing follows
in Sec.~\ref{gaugec} and Sec.~\ref{gtmoment}.  The second model is
studied in Sec.~\ref{rovmod}, where the effective treatment is
performed following the footsteps of Sec.~\ref{lt}.  Effective
evolution relative to a local time is compared to the (Hilbert space)
dynamics obtained using a locally deparametrized version of the
constraint, demonstrating good agreement. This model does not possess
a global clock and transformations between local internal times are
necessary for full dynamical evolution. At the effective level these
are once again performed using gauge transformations allowing
``patched-up'' global evolution.  Sec.~\ref{sec:conclusions} contains
several concluding remarks.

\section{Effective constraints}\label{sec:effcons}

All examples in this article are quantum systems with a single
constraint operator $\hat{C}$ playing a role analogous to that of
the Hamiltonian constraint in general relativity. According to the
Dirac quantization procedure, physical states $|\psi\rangle$ satisfy
the condition $\hat{C}|\psi\rangle=0$. When one solves for specific
states represented in a Hilbert space and attempts to equip the
solution space with a physical inner product, spectral properties of
the zero eigenvalue of $\hat{C}$ are important: if zero is in the
discrete part of the spectrum, physical states form a subspace of
the kinematical Hilbert space in which the quantum constraint
equation is formulated; for zero in the continuous part, on the
other hand, a new physical Hilbert space must be constructed for
which some methods exist \cite{PhysHilbert}. These methods in
practical applications, however, have a rather limited range of
applicability, and so finding physical Hilbert spaces remains a
challenge. For our effective procedures, assumptions about the
spectrum of $\hat{C}$ need not be made; effective techniques work
equally well for zero in the discrete as well as the continuous part
of the spectrum of constraint operators.

Effective descriptions for canonical quantum theories
\cite{EffCons,EffConsRel} are based on a description of states not
in terms of wave functions (or density matrices) but by using
expectation values $\langle\hat{q}\rangle$ and
$\langle\hat{p}\rangle$ and moments
\[
 \Delta(q^ap^b):=\langle(\hat{q}-\langle\hat{q}\rangle)^{a}
(\hat{p}-\langle\hat{p}\rangle)^b\rangle_{\rm Weyl}
\]
(ordered totally symmetrically and defined for $a+b\geq 2$).  (For
instance, $\Delta(q^2)=(\Delta q)^2$ is the position fluctuation
with only a slight change of the standard notation.)

The state space is equipped with a Poisson structure defined by
\ba\label{poisson} \{\langle\hat{A}\rangle,\langle\hat{B}\rangle\}=
\frac{\langle[\hat{A},\hat{B}]\rangle}{i\hbar} \ea
for any pair of operators $\hat{A}$ and $\hat{B}$, extended to the
moments using the Leibnitz rule and linearity.  In the case of
dynamics given by a true Hamiltonian, the Schr\"odinger evolution of
states is equivalent to the evolution of expectation values and
moments generated by the quantum Hamiltonian
$H_Q(\langle\hat{q}\rangle,\langle\hat{p}\rangle,\Delta(\cdots))=
\langle\hat{H}\rangle$ through the Poisson bracket defined above.

For physical states parameterized by their expectation values and
moments, the equation $\langle\hat{C}\rangle(\langle\hat{q}\rangle,
\langle\hat{p}\rangle, \Delta(\cdots))=0$ defines a constraint
function on the quantum phase space. In this way, classical
techniques for the reduction of constrained systems can be applied
even in the quantum case, one of the key features exploited in this
article to address the problem of time. The quantum nature of the
problem is manifest in moment-dependent correction terms in the
function $\langle\hat{C}\rangle$ as opposed to the classical
constraint, as well as the infinite dimensionality of the quantum
phase space even for a system with finitely many classical degrees
of freedom. Moreover, since the moments are 
a priori
independent degrees of
freedom, they are restricted by further constraints
\[
 C_{\rm pol}(\langle\hat{q}\rangle,
\langle\hat{p}\rangle, \Delta(\cdots))
 := \langle (\widehat{\rm pol}-\langle\widehat{\rm
   pol}\rangle) \hat{C}\rangle=0
\]
for all polynomials $\widehat{\rm pol}$ in basic operators.%
\footnote{The condition $\langle\hat{C}\rangle=0$ cannot be sufficient
to determine the physical state, since the mean value of $\hat{C}$ may
vanish even if $\hat{C}|\psi\rangle\neq0$.}
This set of functions contains infinitely many first-class constraints
for infinitely many variables;
the quantum constraint functions, therefore, generate gauge
transformations and solving the constraints does not directly lead to
gauge invariance. The latter is only achieved after constructing Dirac
observables on the quantum phase space, which provide the correct
number of physical degrees of freedom. In this aspect, the effective
formalism differs from standard Dirac quantization where the physical
Hilbert space is devoid of gauge flows. This may be understood from
noting that states in the physical Hilbert space only assign
expectation values to Dirac observables, while in the effective
formalism expectation values are a priori assigned to all kinematical
variables, which even at the classical level are not gauge invariant.

For the first-class nature, the ordering of operators in the
products $\widehat{\rm pol}\hat{C}$ is important, which, as shown
explicitly in the form written above, is not ordered symmetrically.
Some of the quantum constraints then take complex values, which does
not cause problems as already shown for deparameterizable systems.
This complex nature of the constrained system is also rooted in the
fact that the effective expectation values are assigned to all
kinematical variables. It is not surprising that only some
kinematical moments satisfy reality conditions after the constraints
are implemented.
Reality will be imposed on the physical expectation
values and moments --- the Dirac observables of the constrained
system --- and contact with the physical Hilbert space is made. We
will provide further examples in this article.

Regarding the construction of Dirac observables for the constrained
system defined here, we note that observables which commute with the
quantum constraints translate into Dirac observables for the
effective system, Poisson-commuting with all the quantum constraint
functions:
\ba\label{effobs} \delta\langle \hat{O}\rangle &=& \{\langle \hat{O}
\rangle,\langle (\widehat{\rm pol}-\langle\widehat{\rm pol}\rangle)
\hat{C}\rangle\} \\ &=&\frac{1}{i\hbar}\left(\langle (\widehat{\rm
pol}-\langle\widehat{\rm pol}\rangle)[\hat{O},\hat{C}]\rangle
+\langle[\hat{O},\widehat{\rm pol}](\hat{C} - \langle \hat{C}
\rangle) \rangle\right), \nn \ea
vanishes weakly if $\hat{O}$ is a Dirac observable. By the same
token, moments computed for Dirac observables are Dirac observables
in the effective approach.

The set of infinitely many constraints for infinitely many variables
is directly tractable by exact means only if the constraints
decouple into finite sets, a situation realized only for constraints
linear in canonical variables. More interesting systems can be dealt
with by approximations which reduce the system to finite size when
subdominant terms are ignored. The prime example for such an
approximation is the semiclassical expansion, in which moments of
high orders are suppressed compared to expectation values and
lower-order moments. Semiclassicality in a very general form is
implemented by the condition $\Delta(q^ap^b)=O(\hbar^{(a+b)/2})$;
considering only finite orders in $\hbar$ thus allows one to
restrict the infinite set of constraints to a finite one, and
physical moments up to the order considered can be found more
easily. When the system of all quantum constraints is reduced to
finite size, we call the resulting constraints ``effective,''
motivated by the fact that an analogous reduction in
quantum-mechanical systems (combined with an adiabatic
approximation) reproduces equations of motion that follow from the
low-energy effective action
\cite{EffAc}.

Despite the fact that the moments can be varied independently at the effective level, they must, in general, satisfy an infinite tower of inequalities in order to represent a true quantum state.
Namely, in ordinary quantum mechanics, the values assigned by a state to the
various quantum moments are subject to inequalities that follow
directly from the Schwarz inequality of the Hilbert space. In
particular, for any two observables represented by Hermitian
operators $\hat{A}$\ and $\hat{B}$, we have
\begin{widetext}
\begin{eqnarray*}
\left\langle ( \hat{A} - \langle \hat{A} \rangle )^2
\right\rangle \left\langle ( \hat{B} - \langle \hat{B} \rangle
)^2 \right\rangle \geq \frac{1}{4} \left| \left\langle
-i [ \hat{A}, \hat{B} ] \right\rangle \right|^2
 + \frac{1}{4} \left| \left\langle \big[ ( \hat{A} - \langle
\hat{A} \rangle ), ( \hat{B} - \langle \hat{B} \rangle
) \big]_+ \right\rangle \right|^2,
\end{eqnarray*}
\end{widetext}
where $[,]_+$\ denotes the anticommutator. The well-known
(generalized) uncertainty relation follows immediately by setting
$\hat{A} = \hat{q}$ and $\hat{B} = \hat{p}$. In the present work we
will \emph{not} assume that all {\em kinematical} moments satisfy
these inequalities, or even that their values are real. We will
instead impose (order by order in the semiclassical expansion) these
inequalities and reality on the relational observables \emph{after}
the constraint is solved. This is discussed in greater detail in
Sec.~\ref{sec:positivity} and in Appendix~\ref{positivity}.  Notice
that the generalized uncertainty relation is then the only remaining
inequality at order $\hbar$.

The effective formalism provides approximation techniques for the
evaluation of quantum dynamics. While it is motivated by the operator
algebras of standard quantum theory, it is not necessarily equivalent
to the standard theory. For instance, an expression such as
$\langle\hat{q}\rangle$ need not and cannot necessarily be interpreted
literally as the expectation value of a well-defined operator in a
Hilbert space with a specifically defined inner product.  Especially
in the context of the problem of time, a crucial new feature arises
--- local internal time and the corresponding local relational
observables, or fashionables \cite{EffTime1} --- which at present do
not have a known analog at the Hilbert-space level.  Changing one's
local time in practice additionally amounts to a gauge transformation
(see Sec.~\ref{gaugec}), and we shall see later that different choices
of gauge in the effective theory correspond to different, and in
general inequivalent, choices of a Hilbert space for the quantum
theory.
Eventually, these new notions may be used to arrive at a
generalization of quantum mechanics for situations in which time is
not idealized as a monotonic parameter without turning points. If so,
the generalization cannot be fully specified in the current effective
framework which makes use of semiclassicality for explicit evaluations
of its equations. But the examples provided in this article should
play a key role in exploring these issues.

\subsection{Example: ``Relativistic'' harmonic oscillator}

To illustrate the procedure, we consider two copies of the canonical
algebra $[\hat{t}, \hat{p}_t] = i\hbar = [\hat{\alpha},
\hat{p}_{\alpha}]$, subject to the constraint $\hat{C}=\hat{p}_t^2 -
\hat{p}_{\alpha}^2 - \hat{\alpha}^2$. This system%
\footnote{This toy model is clearly not relativistic in the standard
sense. However, here (and in the remaining models of this work) we are
not interested in the precise physical interpretation of this system
(of which there exist both relativistic and non-relativistic ones),
but rather in its structural properties. The constraints considered in
the present article, similarly to Hamiltonian constraints in
relativistic cosmology, are all quadratic in momenta.}
has been treated
in a fair amount of detail in~\cite{EffConsRel}
and~\cite{EffConsComp}, so here we only provide an outline. We
truncate the system at order $\hbar$\ of the semiclassical
expansion. Specifically, this means that in addition to the terms
explicitly proportional to $\hbar^{\frac{3}{2}}$, we discard all
moments of third order and above, products of two or more second
order moments, as well as products between a second order moment and
$\hbar$. In particular, of the infinite number of degrees of freedom
at this order, we only need to consider fourteen: four expectation
values $\langle \hat{a} \rangle$, four spreads $(\Delta a)^2$ and
six covariances $\Delta(ab)$, where $a$, $b$\ can be any of the four
basic kinematical variables.

In this model, for example, one of the constraint conditions to be
enforced is $C_{\alpha}:=\langle (\hat{\alpha} - \langle
\hat{\alpha} \rangle) \hat{C} \rangle = 0$. Here we are dealing with
low order polynomials and the corresponding condition on expectation
values and moments is straightforward to derive explicitly:
\begin{widetext}
\[
C_{\alpha} = \left\langle \left(\hat{\alpha} - \langle \hat{\alpha}
\rangle \right) \left(\hat{p}_t^2 - \hat{p}_{\alpha}^2 -
\hat{\alpha}^2 \right) \right\rangle = \left\langle
\left(\hat{\alpha} - \langle \hat{\alpha} \rangle \right)
\hat{p}_t^2 \right\rangle - \left\langle \left(\hat{\alpha} -
\langle \hat{\alpha} \rangle \right) \hat{p}_{\alpha}^2
\right\rangle - \left\langle \left(\hat{\alpha} - \langle
\hat{\alpha} \rangle \right) \hat{\alpha}^2 \right\rangle\q.
\]
This quantity should be expressed in terms of the expectation values
and moments, our phase-space coordinates.  In each of the terms in
the last expression one needs to replace powers of kinematical
operators with corresponding powers of $(\hat{O}-\langle \hat{O}
\rangle )$. For example, the middle term can be rewritten as
\begin{eqnarray*}
\left\langle \left(\hat{\alpha} - \langle \hat{\alpha} \rangle
\right) \hat{p}_{\alpha}^2 \right\rangle = \left\langle
\left(\hat{\alpha} - \langle \hat{\alpha} \rangle \right)
(\hat{p}_{\alpha} - \langle \hat{p}_{\alpha} \rangle )^2
\right\rangle + 2 \langle \hat{p}_{\alpha} \rangle \left\langle
\left(\hat{\alpha} - \langle \hat{\alpha} \rangle \right)
(\hat{p}_{\alpha} - \langle \hat{p}_{\alpha} \rangle ) \right\rangle
+  \langle \hat{p}_{\alpha} \rangle^2\left\langle \hat{\alpha} -
\langle \hat{\alpha} \rangle \right\rangle\q,
\end{eqnarray*}
where the last term vanishes as $\langle (\hat{\alpha} - \langle
\hat{\alpha} \rangle) \rangle=\langle \hat{\alpha} \rangle - \langle
\hat{\alpha} \rangle = 0$. The remaining terms need to be ordered
symmetrically in order to write them in terms of moments, which can
be accomplished with the use of the canonical commutation relations.
Continuing with the example, the above term becomes
\[
\left\langle \left(\hat{\alpha} - \langle \hat{\alpha} \rangle
\right) \hat{p}_{\alpha}^2 \right\rangle = \left\langle
(\hat{\alpha} - \langle \hat{\alpha} \rangle)(\hat{p}_{\alpha} -
\langle \hat{p}_{\alpha} \rangle )^2\right\rangle_{\rm Weyl} +
\langle \hat{p}_{\alpha} \rangle \left( 2\left\langle (\hat{\alpha}
- \langle \hat{\alpha} \rangle) (\hat{p}_{\alpha} - \langle
\hat{p}_{\alpha} \rangle)\right\rangle_{\rm Weyl} + i\hbar
\right)\q,
\]
with
\begin{eqnarray*}
\left\langle (\hat{\alpha} - \langle \hat{\alpha}
\rangle)(\hat{p}_{\alpha} - \langle \hat{p}_{\alpha} \rangle
)^2\right\rangle_{\rm Weyl}=&
 \frac{1}{3}\left\langle (\hat{\alpha} - \langle \hat{\alpha} \rangle)(\hat{p}_{\alpha} -
\langle \hat{p}_{\alpha} \rangle )^2+ (\hat{p}_{\alpha} - \langle
\hat{p}_{\alpha} \rangle ) (\hat{\alpha} - \langle \hat{\alpha}
\rangle)(\hat{p}_{\alpha} - \langle \hat{p}_{\alpha} \rangle ) +
(\hat{p}_{\alpha}-\langle \hat{p}_{\alpha} \rangle )^2(\hat{\alpha}
- \langle \hat{\alpha} \rangle)\right\rangle\,.
\end{eqnarray*}
Proceeding in this way, one can write the constraint condition using
moments as
\begin{eqnarray*}
C_{\alpha}  = 2\langle\hat{p}_\alpha\rangle \Delta(p_t\alpha) -
2\langle\hat{p}_{\alpha}\rangle \Delta(\alpha p_{\alpha}) - i\hbar
\langle\hat{p}_{\alpha}\rangle- 2\langle\hat{\alpha}\rangle(\Delta
\alpha)^2  + \Delta(\alpha p_t^2) - \Delta(\alpha p_{\alpha}^2) +
\Delta(\alpha^3)\q.
\end{eqnarray*}
\end{widetext}

Evaluating other constraints in this manner and truncating the
system at order $\hbar$, the infinite set of constraint functions
reduces to just five:
\begin{eqnarray}\label{eq:harm_constraints}
C&=& \langle\hat{p}_t\rangle^2 - \langle\hat{p}_{\alpha}\rangle^2 -
\langle\hat{\alpha}\rangle^2 + (\Delta p_t)^2 - (\Delta
p_{\alpha})^2 -
(\Delta \alpha)^2 \nn\\
C_{t}& =& 2\langle\hat{p}_t\rangle \Delta(t p_t) + i\hbar
\langle\hat{p}_t\rangle - 2\langle\hat{p}_{\alpha}\rangle \Delta(t
p_{\alpha}) - 2\langle\hat{\alpha}\rangle
\Delta(t\alpha) \nn \\
C_{p_t}& =& 2\langle\hat{p}_t\rangle (\Delta p_t)^2 -
2\langle\hat{p}_{\alpha}\rangle  \Delta(p_tp_{\alpha})
-2\langle\hat{\alpha}\rangle\Delta(p_t\alpha)\nn \\
C_{\alpha} & =& 2\langle\hat{p}_t\rangle \Delta(p_t\alpha) -
2\langle\hat{p}_{\alpha}\rangle \Delta(\alpha p_{\alpha}) - i\hbar
\langle\hat{p}_{\alpha}\rangle-
2\langle\hat{\alpha}\rangle(\Delta \alpha)^2 \nn \\
C_{p_{\alpha}}& =& 2\langle\hat{p}_t\rangle \Delta(p_tp_{\alpha}) -
2\langle\hat{p}_{\alpha}\rangle(\Delta p_{\alpha})^2
 \nn \\ && -2\langle\hat{\alpha}\rangle \Delta(\alpha p_{\alpha}) +
i\hbar \langle\hat{\alpha}\rangle\q.
\end{eqnarray}
The constraint functions are first-class to order $\hbar$\ and,
therefore, generate gauge transformations through their Poisson
brackets with the expectation values and moments.\footnote{The
Poisson brackets between the expectation values and moments
generated by two canonical pairs of operators is tabulated in
Appendix~\ref{app:PB}.} Following \cite{EffCons,EffConsRel}, we fix
the gauge that corresponds to the evolution of $\hat{\alpha}$\ and
$\hat{p}_{\alpha}$\ in $\hat{t}$, by setting fluctuations of the
latter to zero
\begin{equation}\label{eq:harm_gauge}
(\Delta t)^2 = \Delta(t \alpha) = \Delta (t p_{\alpha}) = 0\q.
\end{equation}
Through reorderings, imaginary contributions in the constraints have
arisen, which require some of the moments to take complex values.
For instance, with our gauge choice $\Delta(tp_t)=
-\frac{1}{2}i\hbar$. All these moments refer to $t$ which, when
chosen as (internal) time in this deparameterizable system, is not represented
as an operator and does not appear in physical moments. The
gauge-dependence or complex-valuedness of these moments thus is no
problem.

Moments not involving time or its momentum, on the other hand,
should have a physical analog taking strictly real values. This is,
indeed, the case. With the gauge fixed as above, a single gauge flow
remains on the
expectation values and moments
evolving in $t$.
(We need just three gauge-fixing conditions for four $o(\hbar)$-constraints
because the Poisson tensor for the moments is degenerate.)
It is
generated by the constraint function $C_{\rm H} =
\langle\hat{p}_t\rangle \mp H_Q $ with the quantum Hamiltonian
\begin{eqnarray}\label{HQ}
H_Q &=& \sqrt{\langle\hat{p}_{\alpha}\rangle^2 +
\langle\hat{\alpha}\rangle^2}  \Biggl( 1   \\ &&  + \frac{
\langle\hat{\alpha}\rangle^2(\Delta p_{\alpha} )^2 -
2\langle\hat{\alpha}\rangle \langle\hat{p}_{\alpha}\rangle
\Delta(\alpha p_{\alpha}) + \langle\hat{p}_{\alpha}\rangle^2 (\Delta
\alpha)^2}{2(\langle\hat{p}_{\alpha}\rangle^2 +
\langle\hat{\alpha}\rangle^2)^2} \Biggr) \,.\nonumber
\end{eqnarray}
Solving the Hamiltonian equations of motion for
$\langle\hat{\alpha}\rangle(t)$,
$\langle\hat{p}_{\alpha}\rangle(t)$, $\Delta(\alpha p_{\alpha})(t)$,
$(\Delta \alpha)^2(t)$, $(\Delta p_{\alpha})^2(t)$
 yields the Dirac observables of the constrained system
in relational form, on which reality can easily be imposed just by
requiring real initial values at some $t$.
At this stage, we have arrived at the usual results for a
deparameterized system with time $t$, in which evolving variables such as
$\langle\hat{\alpha}\rangle(t)$ solving equations of motion with
respect to (\ref{HQ}) would be considered physical while no physical
operator for time itself exists.

In our framework, it is gauge fixing that distinguishes one of the
original variables as time without an operator analog: Time moments
$\langle \hat{p}_t \rangle$, $(\Delta p_t)^2$, $\Delta(p_t p)$,
$\Delta( p_t \alpha)$, $\Delta(tp_t)$ are eliminated using the
constraints~(\ref{eq:harm_constraints}), while $(\Delta t)^2$,
$\Delta(t \alpha)$, $\Delta (t p_{\alpha})$\ are fixed by the gauge
condition~(\ref{eq:harm_gauge}). Generally, there may be several ways
to interpret a given quantum constraint dynamically with respect to
different choices of (internal) time. Collectively, the choice of a
time variable, the associated gauge conditions and the selection of
evolving variables within that gauge will be referred to, following
\cite{EffTime1}, as a \emph{Zeitgeist}.  Usually, the selection of
which variable to choose as clock function in which other variables
may evolve relationally does not constitute a gauge choice.  The
effective formalism as developed here, however, provides a
relationship between (the interpretation of a quantum variable as)
time and gauge: we are free to fix the independent gauge flows in a
way that describes and interprets relational evolution in the most
convenient way. We will come back to this issue in detail in
Sec.~\ref{gaugec}; for now, we warn the reader about an inherent
weakness of evolving observables, which underlies the comparison
problem of time: If transformations of internal time variables are
allowed, and if they are essentially implemented by gauge changes, the
physical nature of some variables may appear
(but is not)
gauge dependent. To avoid
apparently contradictory language, we use the term {\em fashionables}
for local relational observables, as introduced in \cite{EffTime1}.

\section{A model of a bad internal clock}\label{lt}

In this section, through the use of a toy model, we showcase an
effective semiclassical solution to the problem of defining quantum
dynamics with respect to a time variable which is non-monotonic along
a (classical) trajectory.

We introduce the model together with its classical properties in
Sec.~\ref{sec:lt_class}; its Dirac quantization is briefly discussed
in Sec.~\ref{sec:lt_dirac}.  In Sec.~\ref{sec:lt_eff} we apply the
effective scheme of~\cite{EffCons,EffConsRel} for solving constraints
to define approximate dynamics; among the many viable choices for
internal time, we elect to study the dynamics relative to a variable
that cannot be used for a global deparameterization. Evolution with
respect to such a clock variable breaks down near its turning points
and translation to a new clock variable is required. Within the
effective approach, the choice of a clock is practically incorporated
by selecting a gauge as in (\ref{eq:harm_gauge}) and, therefore,
switching a clock is achieved by a gauge transformation. Another
novelty is that the expectation value of the time variable acquires an
imaginary contribution, a feature further discussed in
Sec.~\ref{compt} and the second model in Sec.~\ref{rovmod}. The end
result of the present section is an internally consistent approximate
method for evolving initial data in a non-global clock variable
through its extremal point on the trajectory, by temporarily switching
to a different variable used as internal time.

\subsection{Classical discussion}\label{sec:lt_class}

The model we are interested in possesses a ``time potential'' $\lambda
t$ and is classically determined by the constraint \ba\label{class}
C_{\rm class}= p_t^2 - p^2 - m^2 + \lambda t\q\,.  \ea We assume
$\lambda \geq 0$\ for concreteness. This model has been briefly
discussed in \cite{EffConsRel} and structurally resembles a perturbed
free relativistic particle.%
\footnote{Although, again, the system is clearly not relativistic in
the standard sense.}
Of particular
interest to us is the fact that $t$\ exhibits a
specific trait of a bad clock, namely it is not
monotonic along a classical trajectory. As regards the
parametrization of the flow generated by $C_{\rm class}$, we infer
from \ba\label{ltclass1}
 \{t,C_{\rm class}\}=2p_t \q \text{and} \q
\{p_t,C_{\rm class}\}=-\lambda<0\q, \ea that
\begin{flalign} \label{ltclass2} t(s)=-\lambda s^2+2{p_t}_0s+t_0 \ \
\text{and} \ \ p_t(s)=-\lambda s+{p_t}_0 \ , \end{flalign} where $s$
is the parameter along the flow $\alpha^s_{C_{\rm class}}(x)$
generated by $C_{\rm class}$. We see that $t$\ has an extremum and
runs twice through each value it assumes; therefore globally it is
not a good clock function for the gauge orbits generated by $C_{\rm
class}$. Note that both $p_t$ and $q$ provide good parametrizations
of the gauge orbit and $p$ is an obvious Dirac observable. Although
this model is deparametrizable in either $q$ or $p_t$, we would like
to interpret the relational evolution of the configuration variable
$q$ with respect to the non-global clock function $t$.

\begin{center}
\begin{figure}[htbp!]
\includegraphics[width=7cm]{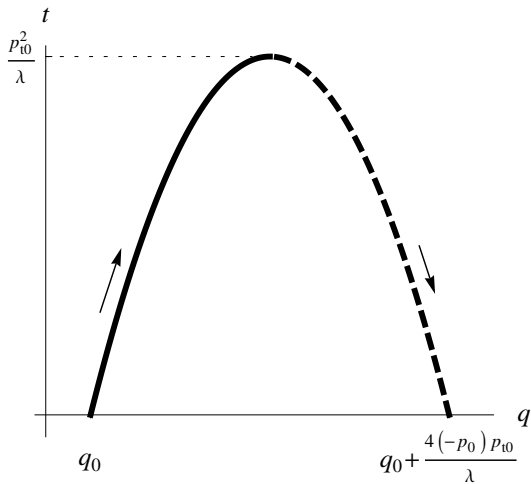}
\caption{\label{fig:classtraj} A typical classical configuration
space trajectory is a parabola with the peak value of $t$\ dependent
on ${p_t}_0$\ and the separation of branches dependent on $p_0$. The
orientation of evolution, indicated by the arrows, is consistent
with $p_0<0$\ and ${p_t}_0>0$. We refer to the left branch (solid)
as ``incoming'' or ``evolving forward in $t$'', the right branch
(dashed) as ``outgoing'' or ``evolving backward in $t$''.}
\end{figure}
\end{center}

For completeness, we also note that the Dirac observables of this
system are easy to find and they themselves form a canonical Poisson
algebra,
\begin{equation}
\cq := q - \frac{2}{\lambda} pp_t \quad {\rm and} \quad \cp:= p,
\quad {\rm satisfy \ \ } \{ \cq, \cp \} = 1 \q.
\label{eq:Dobs_lambda}
\end{equation}

\subsection{Dirac quantization} \label{sec:lt_dirac}

Following Dirac's algorithm for a constraint quantization, one would
first quantize the kinematical system in the usual way, by
representing canonical operators on the space $L^2(\mathbb{R}^2,
dtdq)$\ as
\[
\hat{t} = t \q,\q \hat{p}_t = \frac{\hbar}{i}
\frac{\partial}{\partial t} \q,\q \hat{q} = q \q,\q \hat{p} =
\frac{\hbar}{i} \frac{\partial}{\partial q}\q.
\]
The constraint function~(\ref{class}) can be straightforwardly
quantized as $\hat{C}=\hat{p}_t^2 - \hat{p}^2-m^2 + \lambda\hat{t}$\
and the physical state condition $\hat{C} \psi_{\rm phys} = 0$\
becomes a partial differential equation
\begin{equation}
\left(-\hbar^2 \frac{\partial^2}{ \partial t^2} + \lambda t - m^2 +
\hbar^2 \frac{\partial^2}{ \partial q^2} \right) \psi(t, q) = 0 \q.
\label{eq:lt_cons_pde}
\end{equation}
The operators $\hat{p}^2$\ and $\hat{p}_t^2+\lambda \hat{t}$\
commute and thus can be simultaneously diagonalized. The solution to
the constraint equation can be constructed from their simultaneous
eigenstates. The general solution has the
form
\begin{equation}
\psi_{\rm phys} (t, q) = \int dk\, f(k) {\rm Ai}\left[ \left(
\frac{\lambda}{\hbar} \right)^{\frac{2}{3}} \left( \lambda t - k^2 -
m^2 \right) \right] e^{\frac{-ikq}{\hbar}} ,
\end{equation}
where ${\rm Ai}[x]$\ is the bounded and integrable Airy-function. As
it often happens, none of the solutions are normalizable with
respect to the kinematical inner product and a separate
\emph{physical} inner product must be defined on the solutions. A
common way to proceed in the context of quantum cosmology is to
deparameterize the system with respect to a suitable time variable.
The simplest option is to formulate the constraint equation as a
Schr\"odinger equation giving evolution of wavefunctions of $q$\ in
the time-parameter $p_t$
\begin{equation}
i\hbar \frac{\partial}{\partial p_t} \tilde{\psi}(p_t, q) =
\frac{1}{\lambda} \left( -\hbar^2 \frac{\partial^2}{ \partial q^2} -
p_t^2+m^2 \right)\tilde{\psi}(p_t, q) \q,
\end{equation}
where $\tilde{\psi}(p_t, q) := \int dt\, \psi(t, q) e^{-itp_t/
\hbar}$. We then define the physical inner product by integrating
over $q$\ at a fixed value of $p_t$
\begin{equation}
\langle \psi, \phi \rangle_{\rm phys} := \int_{p_t = {p_t}_0} dq\,
\bar{\tilde{\psi}}(p_t, q) \tilde{\phi}(p_t, q)\q.
\end{equation}
For solutions to~(\ref{eq:lt_cons_pde}), the result is independent
of the value of ${p_t}_0$ and finite. A similar construction, one
that is more complicated due to taking square roots of operators,
can be performed if one chooses $q$\ to act as time. However, to our
knowledge, there is no exact way to deparameterize this constraint
using $t$. Here we are specifically interested in the situations
where there is no obvious time variable available to perform
deparameterization. For that purpose, in this toy model we choose a
time variable which we know to be bad in a particular way and
construct an effective initial value formulation with respect to
that variable.

Specifically, we would like to evolve initial data given at a fixed
value of $t$\ on the incoming branch onto the outgoing branch (see
FIG.~\ref{fig:classtraj}). In order to do that, one inevitably has
to find a way to evolve data through the extremum of $t$. Such an
evolution can be easily performed in the classical limit and,
therefore, should also be well-posed at least semiclassically.

\subsection{Effective treatment} \label{sec:lt_eff}

Following the procedure outlined in Sec.~\ref{sec:effcons}, we write
the constraint functions $C_{\rm pol}=0$\ in terms of moments and
truncate the system by discarding terms of order
$\hbar^{\frac{3}{2}}$\ and higher in the semiclassical
approximation. As for the ``relativistic harmonic oscillator", we
have fourteen kinematical degrees of freedom to this order, subject
to the five effective constraints
\begin{align}\label{ex1}
C& = p_t^2 - p^2 - m^2 + (\Delta p_t)^2 - (\Delta p)^2+\lambda t = 0
\nonumber \\ C_{t}& = 2p_t \Delta(t p_t) + i\hbar p_t - 2p \Delta(t
p)+\lambda (\Delta t)^2 = 0 \nonumber \\ C_{p_t}& = 2p_t (\Delta
p_t)^2 - 2p \Delta(p_tp) + \lambda\Delta(tp_t) -
\frac{1}{2}i\lambda\hbar = 0 \nonumber
\\ C_q& = 2p_t \Delta(p_tq) - 2p \Delta(qp) - i\hbar
p+\lambda\Delta(qt) = 0 \nonumber \\ C_p& = 2p_t \Delta(p_tp) -
2p(\Delta p)^2+\lambda\Delta(tp) = 0\q.
\end{align}
The five effective constraints generate only four linearly
independent flows due to a degenerate Poisson structure to order
$\hbar$. Consequently, the 14-dimensional Poisson manifold may be
reduced to a 5 dimensional surface describing the five physical
degrees of freedom to semiclassical order. Note that both $p$ and,
as a result of (\ref{effobs}), $(\Delta p)^2$ commute with all five
constraints and are, therefore, two obvious constants of motion of
this effective system. We want to find the remaining three physical
degrees of freedom as relational Dirac observables.

\subsubsection{Evolution in complex $t$ and breakdown of the corresponding gauge}\label{t-gauge}

Choosing $t$ as our clock function, it is helpful to fix three out
of the four independent gauge flows in order to facilitate explicit
calculations and avoid keeping track of three further order $\hbar$
clocks\footnote{Note that this gauge fixing occurs after
quantization.}. The system, certainly, does not single out a particular gauge
for us; nevertheless,
with our choice of clock
we can motivate certain
gauges. Once a choice of time has been implemented, the clock
function should not correspond to an operator and, hence, should not
appear in
evolving
moments; it should be ``as classical as
possible'', implying that the gauge conditions
\begin{align}\label{ex2}
\phi_1&=(\Delta t)^2=0\nn\\
\phi_2&=\Delta(tq)=0\nn\\
\phi_3&=\Delta(tp)=0
\end{align}
seem reasonable. We will refer to these conditions as {\it $t$-gauge}
or the Zeitgeist associated to $t$. At the state level, this would be
closest in spirit to an inner product evaluated on $t=const$ slices in
some kinematical representation.  Since $t$ is not a global time, this
would lead to an apparent non-unitarity in the quantum theory, which
by analogy suggests that this gauge should not be globally valid,
simply because $t$ is not a global clock. We will come back to this
issue below.

Imposing the gauge conditions renders the
combined system of (\ref{ex1}) and (\ref{ex2}) a mixture of first
and second class constraints. Since there were originally four
independent gauge flows, we expect at least one first class
constraint among the eight conditions given by~(\ref{ex1})
and~(\ref{ex2}). One additional independent first class constraint
may arise, but this constraint must generate a vanishing flow on the
variables which we choose after solving the constraints and gauge
conditions. It is easily verified that the first class constraint
with the vanishing flow on the variables $q,p,t,p_t,(\Delta
q)^2,(\Delta p)^2,\Delta (qp)$ must be directly proportional to
$C_t$ in this gauge. Solving this constraint
\begin{flalign}\label{Ct} C_t\approx 2p_t\Delta(tp_t)+i\hbar p_t=0 \q
\Rightarrow \q \Delta(tp_t)=-\frac{i\hbar}{2}\ , \end{flalign}
implies a saturation of the (generalized) uncertainty relation for $t$ and $p_t$
in this system. Here and throughout the rest of the present work `$\approx$' denotes equality restricted to the region where both constraint functions and the gauge conditions of the relevant Zeitgeist are satisfied.

The remaining first class constraint with non-vanishing flow on the
chosen variables will generate our relational evolution in $t$;
therefore, we refer to it as the ``Hamiltonian constraint'' in the
$t$-gauge. It has the form $C_H\propto C_eV^{e}$, where $V^e$ is the
solution to $\{\phi_i,C_e\}V^e=0$ and $i=1,2,3$ and the $C_e$ denote
the constraints of~(\ref{ex1}), except~$C_t$. The matrix
$\{\phi_i,C_e\}$ is generically of rank 3 from which we infer that
there is only one independent $C_H$. The coefficients of this matrix
are given in Tab.~\ref{tab:pbgc}, and, up to an overall factor, we
find
\ba\label{ex8a} C_H=C+\alpha C_{p_t}+\beta C_q+\gamma C_p\q, \ea
where, on the constraint surface, the coefficients read
\ba\label{ex8b} \alpha=-\frac{1}{2p_t} \q, \q \beta= 0 \q \q
\text{and} \q \q
  \gamma=-\frac{p}{2p_t^2}\q.
  \ea
Four non-physical moments in this gauge may be solved for via $C_t$,
$C_{p_t}$, $C_q$ and $C_p$. Equation~(\ref{Ct}) gives
$\Delta(tp_t)$, the rest are given by \ba\label{depmoms} (\Delta
p_t)^2&=\frac{2p^2(\Delta p)^2+i\hbar\lambda p_t}{2p_t^2}\q, \q
\Delta(p_tp)=\frac{p(\Delta p)^2}{p_t}\nn \\ \text{and}& \q \q
\Delta(qp_t)=\frac{i\hbar p+2p\Delta(qp)}{2p_t}\q. \ea

\begin{table*}[h] \centering \caption{Poisson
algebra of gauge conditions (\ref{ex2}) with the constraints
(\ref{ex1}). First terms in the bracket are labeled by rows, second
terms are labeled by columns. Note that these results only hold on
the gauge surface defined in (\ref{ex2}).}
\begin{tabular}{|c|| c|c|c||} \hline & $\phi_1$
& $\phi_2$ & $\phi_3$ \\ \hline\hline $C$ & $2i\hbar$ &
$-2\Delta(qp_t)$ & $-2\Delta(p_tp)$\\ \hline $C_{p_t}$ & $4i\hbar
p_t$ & $-2p_t\Delta(qp_t)-2i\hbar p$ &$-2p_t\Delta(p_tp)$ \\ \hline
$C_q$ & $0$ & $-2p_t(\Delta q)^2$ & $-2p_t\Delta(qp)-i\hbar p_t$\\
\hline $C_p$ & $0$ & $i\hbar p_t-2p_t\Delta(qp)$ & $-2p_t(\Delta
p)^2$\\ \hline
\end{tabular}
\label{tab:pbgc}
\end{table*}

When these relations are used together with the $t$-gauge
conditions~(\ref{ex2}), the equations of motion generated by $C_H$\
on the remaining variables read (recall that $p$\ and $(\Delta
p)^2$\ are constants of motion) \ba\label{eom2} \dot{t} &= \{t,C_H\}
&= 2p_t-\frac{2p^2(\Delta
p)^2}{p_t^3}-\frac{i\hbar\lambda}{2p_t^2}\q, \nn\\
\dot{p_t} &= \{p_t,C_H\} &=-\lambda\q,\nn\\
\dot{q} &= \{q,C_H\} &=  -2p\left(1-\frac{(\Delta p)^2}{p_t^2}\right)\q,\nn\\
\dot{(\Delta q)^2} &= \{(\Delta q)^2, C_H\} &= -4\Delta(qp)\left(1-\frac{p^2}{p_t^2}\right)\q,\nn\\
\dot{\Delta(qp)} &= \{\Delta(qp), C_H\} &= -2(\Delta
p)^2\left(1-\frac{p^2}{p_t^2}\right)\q. \ea These can be solved
analytically by \ba\label{solq} t(s) &=&
-\frac{p_t(s)^2}{\lambda}-\frac{p^2(\Delta p)^2}{\lambda p_t(s)^2} -
\frac{i\hbar}{2p_t(s)} + c\q,
\nn \\ p_t(s) &=& -\lambda s+{p_t}_0 \q, \nn \\
q(s) &=& 2\frac{pp_t(s)}{\lambda}
\left(1 + \frac{(\Delta p)^2}{p_t(s)^2}\right) + c_1\q,\nn\\
(\Delta q)^2(s) &=& 4(\Delta p)^2 \frac{\left( p^2 + p_t(s)^2
\right)^2}{\lambda^2p_t(s)^2} \nn \\ && + \frac{4
\left(p^2+p_t(s)^2\right)}{\lambda p_t(s)}c_2 + c_3\q,\nn\\
\Delta(qp)(s) &=& 2(\Delta p)^2 \frac{p^2+p_t(s)^2}{\lambda p_t(s)}
+ c_2\q\,, \label{RelOb} \ea where $c$, ${p_t}_0 $\ and
$\{c_i\}_{i=1,2,3}$\ are integration constants related to the
initial conditions. (These solutions, expressed via $p_t$, provide
relational observables of the system. A comparison with
(\ref{eq:Dobs_lambda}) shows that the classical observables receive
quantum corrections via the moments.) In particular, we note that to
this order $p_t$\ experiences no quantum back-reaction and evolves
entirely classically, which is due to the fact that the only
constraint function that has non-trivial bracket with $p_t$\ is $C$.

Neither $p_t$, nor $t$\ is a Dirac observable and one of them can be
eliminated by using $C$. Combining relations~(\ref{depmoms}) and the
gauge conditions~(\ref{ex2}) with $C=0$, we obtain \ba\label{pt} 0 =
p_t^4&-&\left(p^2+m^2-\lambda t+(\Delta p)^2\right)p_t^2 \nn
\\ &+&\frac{i\hbar\lambda}{2}p_t+p^2(\Delta p)^2\q. \ea It is not
difficult to see that, if we want to keep the variables $q,p,(\Delta
q)^2,(\Delta p)^2,\Delta (qp)$\ real (see
Sec.~\ref{sec:positivity}), the above relation necessarily forces
either $t$\ or $p_t$\
 to be complex. When we look at the equations of motion~(\ref{eom2}) and
their solutions~(\ref{solq}), the choice is almost obvious. The
equation of motion for $p_t$\ has no imaginary component and hence
equipping it with a constant imaginary part appears somewhat
artificial. More importantly, $p_t$\ features prominently in the
solutions for $q,p,(\Delta q)^2,(\Delta p)^2,\Delta (qp)$, in order
to keep all these real, we are forced to keep $p_t$\ real and,
consequently, $t$\ must be complex-valued.

Let us quantify the imaginary contribution to $t$. We determine $c$\
by substituting both $p_t(s)$ and $t(s)$\ from~(\ref{solq}) into the
constraint (\ref{pt}) which yields the real-valued result
\ba\label{c} c=\frac{p^2+m^2+(\Delta p)^2}{\lambda}\q. \ea The
imaginary contribution to the clock $t$ is, therefore, a quantum
effect of order $\hbar$ and given by \ba\label{imagt}
\Im[t(s)]=-\frac{\hbar}{2{p_t}(s)}\q. \ea A more thorough analysis
of the complex nature of the effective non-global clocks will be
explored in Sec.~\ref{compt} and its general features have been
discussed in~\cite{EffTime1}.

\begin{center}
\begin{figure}[htbp!]$
\begin{array}{c}
\includegraphics[width=7.2cm]{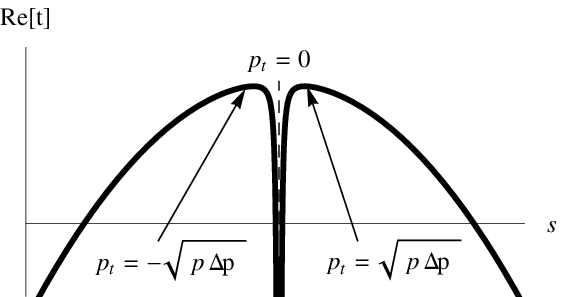} \\
\includegraphics[width=7.2cm]{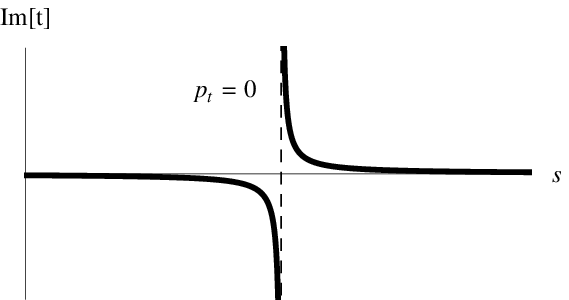}
\end{array}$
\caption{\label{fig:ts} Schematic plots of the real part of $t$
(top) and the imaginary part of $t$ (bottom) against the flow
parameter $s$.}
\end{figure}
\end{center}

We have previously stated that the gauge defined by the
conditions~(\ref{ex2}) is related to choosing $t$\ as time. However,
the equations of motion, as well as their solutions are written in
terms of the gauge parameter $s$\ that parameterizes the flow
generated by $C_H$. Since $t$\ is a complex variable we can relate
$s$\ to its real and imaginary parts separately. In
FIG.~\ref{fig:ts}, we plot the real and imaginary parts of $t(s)$,
deduced directly from~(\ref{solq}) and~(\ref{c}).

{}From the plot we see that away from $p_t=0$, $\Re[t]$ is monotonic
in $s$ on each of the two branches and, asymptotically far away from
$p_t=0$, they become proportional. On the forward moving branch,
$\Re[t]$\ is increasing with $s$; on the backwards moving branch
$\Re[t$] is decreasing with $s$. From the plot we can also see that
$\Re[t]$\ reaches its peak value at $p_t = \pm \sqrt{p \Delta p}
\neq 0$. However, at this point we can no longer trust the
semiclassical approximation as the small value of $p_t$\ in the
denominators in the equations of motion~(\ref{solq}) will result in
values of the moments that no longer satisfy the assumed drop-off.

Figure~\ref{fig:ts} also shows that $\Im[t]$ is monotonic in $s$ in
the same regimes.
Thus, when it comes to parameterizing dynamics using $t$, we have
the option of using either $\Im[t]$\ or $\Re[t]$. We opt to refer to
the real part of $t$\ as ``time'', for several reasons: 1) in the
classical limit the imaginary part vanishes and it is, indeed, the
real part of $t$\ that matches the classical internal time; 2) for large
$p_t$ or small $\lambda$ when the time-dependent term in the
constraint becomes insignificant, the imaginary part of $t$\ is
small and approximately constant; 3) finally, as we will see later,
the expectation value that reproduces $\Im[t]$\ in the case of a
free relativistic particle is based on integrating at a fixed value
of (parameter) $t$\ equal to precisely the real part of the
expectation value.

\begin{center}
\begin{figure}[htbp!]$
\begin{array}{c}
\includegraphics[width=7.2cm]{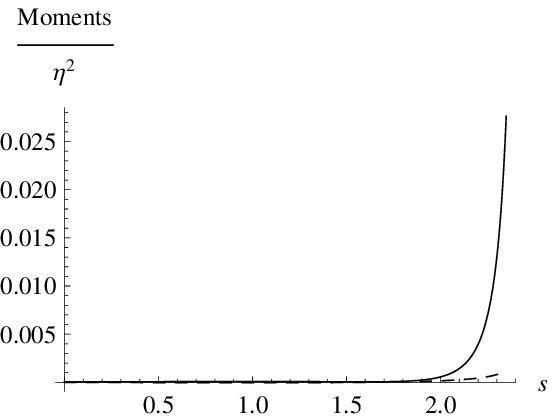} \\
\includegraphics[width=7.2cm]{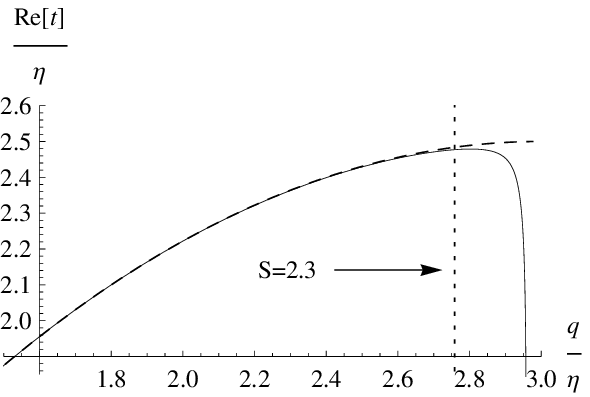}
\end{array}$
\caption{\label{fig:ltbreak}  Top: evolution of moments $(\Delta
q)^2$\ (solid) and $\Delta(qp)$\ (dashed) in $t$-gauge ($(\Delta
p)^2 = {\rm const}$). Somewhere after $s=2.3$\ the spread $\Delta
q:=\sqrt{(\Delta q)^2}$ becomes comparable to the expectation
values, as $\Delta q/\eta
>.1$, and the semiclassical approximation breaks down in $t$-gauge. Bottom: corresponding
effective trajectory (solid) and the related classical trajectory
(dashed); the effective trajectory quickly diverges after $s=2.3$.}
\end{figure}
\end{center}

As one would expect from the classical behavior of $t$, this gauge
is not valid for the whole ``quantum trajectory''. In particular, we
noted that $p_t$\ evolves entirely classically, so that its solution
is simply given by (\ref{ltclass2}). As a result $p_t$ passes
through zero for a finite value of the evolution parameter $s$,
which immediately implies the breakdown of the $t$-gauge: the
coefficients in (\ref{ex8b}) and in~(\ref{solq}) become singular,
the magnitudes of the moments $(\Delta q)^2$\ and $\Delta(qp)$\ blow
up, thereby violating semiclassicality. An example of this
divergence is shown in FIG.~\ref{fig:ltbreak}. Here
$\eta:=\sqrt{p^2+m^2}$\ provides us with a classical length-scale on
the phase space, and the quantum length-scale is set to
$\sqrt{\hbar}=.01\eta$. Classical quantities such as $p$, $m$,
$\lambda$ are all of order $\eta$, while the values of second order
moments are initially of order $\hbar$. Qualitative features of the
plot are insensitive to the precise values chosen so long as the
relative scales are preserved.

Due to the non-global nature of the relational clock $t$, this breakdown does
not come unexpected. In order to evolve a semiclassical state
through the turning point of the clock, we, therefore, need to
switch the gauge and --- unlike in the classical case
--- the clock (see also Sec.~\ref{gaugec} on this issue). A more
complete discussion of the breakdown of the gauge and its
counterpart on the exact side of the quantum theory will be
discussed in the second model in Sec.~\ref{rovmod}, while the
transformation to {\it $q$-gauge} and the evolution through the
turning point will be discussed in Sections~\ref{q-gauge} and
\ref{sec:gaugec_lt} below.

\subsubsection{Evolution through the extremal point of $\Re[t]$ in a new gauge}\label{q-gauge}

Based on the evidence that the $t$-gauge (\ref{ex2}) fails globally
due to the fact that $t$ is a non-global time function, we can,
instead, make use of the fact that, e.g., $q$ {\it is} a good clock
variable for the entire trajectory. For the evolution through the
$t$-turning point we could, therefore, simply choose the following
{\it $q$-gauge} (``as if we chose $q$ as time") \ba\label{gaq}
\tilde{\phi}_1&=&(\Delta q)^2=0\nn\\
\tilde{\phi}_2&=&\Delta(tq)=0\nn\\
\tilde{\phi}_3&=&\Delta(qp_t)=0\q. \ea This gauge is closest in
spirit to choosing a $q={\rm const}$-slicing in an analogous
treatment of the model at the Hilbert space level and since $q$ is
a good clock, in this gauge we expect to be able to evolve through
the extremum in $\Re[t]$ without difficulty. Such a procedure of
adapting the gauge to a good local clock should work in general even
if no global clock functions exist, since generically we expect the
existence of some degree of freedom which may serve as a good local
clock where other clock degrees of freedom fail. To evolve through
the whole trajectory one would in general need to switch gauges,
which we discuss in Sec.~\ref{sec:gaugec_lt} below.

We immediately notice that this gauge is inconsistent with treating
the moments of $\hat{p}$\ and $\hat{q}$\ as
independent phase-space
degrees of
freedom, since several of them are completely fixed by the gauge
conditions. We, therefore, interpret $q$ as a clock in this gauge
(see also Sec.~\ref{gaugec} on this issue) and eliminate the
remaining moments of $\hat{p}$\ and $\hat{q}$\ through constraints
leaving the free variables $t$, $p_t$, $q$, $p$, $(\Delta t)^2$,
$(\Delta p_t)^2$, $\Delta(tp_t)$. The first class constraint with
vanishing flow on these variables is now given by $C_q$. Solving
this constraint then implies $\Delta(qp)=-\frac{i\hbar}{2}$ and,
together with (\ref{gaq}), the saturation of the uncertainty
relation between $\hat{q}$ and $\hat{p}$. The ``Hamiltonian
constraint'' of the {\it q-gauge} reads \ba\label{dq1}
\tilde{C}_H=C+\tilde{\alpha}C_t+\tilde{\beta}C_{p_t}+\tilde{\gamma}C_p\q,
\ea where the coefficients are given on the constraint surface by
\ba\label{dq2} \tilde{\alpha}=-\frac{\lambda}{4p^2}   \q,\q
\tilde{\beta}= -\frac{p_t}{2p^2}   \q\q\text{and}\q\q
\tilde{\gamma}=-\frac{1}{2p}\q. \ea These coefficients are clearly
well-behaved along the entire trajectory, as long as the constant of
motion $p\neq0$. In addition to $\Delta(qp)$, we eliminate the three
remaining unphysical moments through constraints
\begin{eqnarray}
(\Delta p)^2 &=& \frac{p_t^2}{p^2} (\Delta p_t)^2 + \frac{\lambda
p_t}{p^2} \Delta(t p_t) + \frac{\lambda^2}{4p^2} (\Delta t)^2 \nn
\q, \\ \Delta(p_tp) &=& \frac{p_t}{p} (\Delta p_t)^2 +
\frac{\lambda}{2p} \left( \Delta(tp_t) - \frac{i\hbar}{2} \right)
\nn \q, \\ \Delta(tp) &=& \frac{p_t}{p} \left( \Delta(tp_t) +
\frac{i\hbar}{2} \right) + \frac{\lambda}{2p} (\Delta t)^2 \q.
\label{eq:qg_const_lt}
\end{eqnarray}
The dynamical equations generated by this Hamiltonian constraint on
the $q$-gauge surface are
\begin{eqnarray}
\dot{t} &=& 2p_t - \frac{2p_t (\Delta p_t)^2 + \lambda
\Delta(tp_t)}{p^2} \q, \nn \\ \dot{p}_t &=& -\lambda \q, \nn  \\
\dot{q} &=& -2p + \frac{\lambda^2 (\Delta t)^2 + 4 p_t^2 (\Delta
p_t)^2 + 4 \lambda p_t \Delta(tp_t)}{2p^3} \q, \nn \\ \dot{(\Delta
t)^2} &=& \frac{4(p^2 - p_t^2) \Delta(tp_t) - 2\lambda p_t (\Delta
t)^2}{p^2} \nn \\ \dot{\Delta(tp_t)} &=& \frac{4(p^2 - p_t^2)
(\Delta p_t)^2 + \lambda^2(\Delta t)^2}{2p^2} \q,\nn \\ \dot{(\Delta
p_t)^2} &=& \frac{2\lambda p_t (\Delta p_t)^2 + \lambda^2
\Delta(tp_t)}{p^2} \q. \label{eq:lt_qgauge_eom}
\end{eqnarray}
As in the $t$-gauge before, $p_t$\ evolves classically
$p_t(\tilde{s}) = -\lambda \tilde{s} + {p_t}_0$. The moments evolve
according to
\begin{widetext}
\begin{eqnarray}
(\Delta t)^2(\tilde{s}) &=& \frac{p_t(\tilde{s})^2}{p^2} \tilde{c}_1
+ \frac{4 \left( p_t(\tilde{s})^2 + p^2 \right)^2}{\lambda^2p^2}
\tilde{c}_2 + \frac{4p_t(\tilde{s}) \left( p_t(\tilde{s})^2 + p^2
\right)}{\lambda p^2} \tilde{c}_3 \q, \q (\Delta p_t)^2 (\tilde{s})
= \frac{p_t(\tilde{s})^2}{p^2} \tilde{c}_2 + \frac{\lambda
p_t(\tilde{s})}{p^2} \tilde{c}_3 + \frac{\lambda^2}{p^2} \tilde{c}_1
 \q, \nn \\ \Delta(tp_t) (\tilde{s}) &=& - \frac{2 p_t(\tilde{s})^2 + p^2}{p^2}
\tilde{c}_3 - \frac{2 p_t(\tilde{s}) \left( p_t(\tilde{s})^2 + p^2
\right)}{\lambda p^2} \tilde{c}_2 - \frac{\lambda
p_t(\tilde{s})}{p^2} \tilde{c}_1 \q. \label{eq:sol_lt_qgauge}
\end{eqnarray}
The above solutions can be substituted into the equations of motion
for $q(\tilde{s})$\ and $t(\tilde{s})$, which can then be integrated
separately.

Once again, we can eliminate yet another variable. By using $C=0$\
combined with~(\ref{eq:qg_const_lt}), we obtain an equation for $p$,
\begin{equation}
p^4 - \left(p_t^2 - m^2 + (\Delta p_t)^2 + \lambda t \right) p^2 +
p_t^2 (\Delta p_t)^2 +\lambda p_t \Delta(tp_t) + \frac{\lambda^2}{4}
(\Delta t)^2 = 0 \q. \label{eq:p}
\end{equation}
\end{widetext}
We see that there is no need to make either $p$\ or $q$\ complex to
satisfy this equation. Nor are there any explicitly imaginary terms
in the equations of motion or their solutions. Nevertheless, in
order to consistently switch between $t$-gauge and $q$-gauge, we
will require $q$\ to carry an imaginary contribution in this gauge
analogous to~(\ref{imagt})
\begin{equation}\label{imagq}
\Im[q(\tilde{s})] = -\frac{\hbar}{2p} \q,
\end{equation}
which in this case is constant, since $p$\ is a constant of motion.

Finally, we note that --- as expected --- the evolution in this
gauge encounters no difficulty near the extremal point of $t$ when
$p_t=0$. The coefficients in~(\ref{dq2}) stay finite and we can see
from~(\ref{eq:sol_lt_qgauge}) that the moments of $\hat{p}_t$\ and
$\hat{t}$\ remain well-behaved as we go through $p_t=0$. In the next
section we describe a method for switching between the two gauges.

\subsubsection{Switching gauges}\label{sec:gaugec_lt}

The two gauges discussed in Sections~\ref{t-gauge} and~\ref{q-gauge}
describe evolution of two different sets of
degrees of
freedom. If we switch from one gauge to another, for example, to
evolve through the turning point of a time function, we need to be
able to translate between the two sets of variables. We recall that
the original gauge orbit for the truncated system of
constraints~(\ref{ex1}) is, in general, four-dimensional. The three
gauge-fixing equations of either~(\ref{ex2}) or~(\ref{gaq}) restrict
us to a one-dimensional flow on this gauge orbit generated by the
remaining first-class constraint~(\ref{ex8a}) or~(\ref{dq1}),
respectively. In order to ensure that the two sets of variables lie
on the same four-dimensional gauge orbit we need to find
a gauge transformation which takes us from the surface defined
by~(\ref{ex2}) to the one defined by~(\ref{gaq}) and vice versa.

In other words, to transform from $t$-gauge to $q$-gauge we need to
find a combination of the constraint functions $G = \sum_i \xi_i
C_i$, such that a (possibly finite) integral of its flow transforms
the variables as
\begin{equation}
\left\{ \begin{array}{l} (\Delta q)^2 = (\Delta q)^2_0 \\ \Delta(tq)
= 0 \\ \Delta(p_tq) = \Delta(p_tq)_0 \end{array} \right. \rightarrow
\left\{ \begin{array}{l} (\Delta
q)^2 = 0 \\ \Delta(tq) = 0 \\
\Delta(p_tq) = 0 \end{array} \right. \q, \label{eq:tgauge_to_qgauge}
\end{equation}
where the subscript $0$ labels the value of the corresponding
variable prior to the gauge transformation. In general, one would
expect such a transformation to be unique up to the flows generated
by $C_H$\ and $\tilde{C}_H$, since they preserve the corresponding
sets of gauge conditions (see Sec.~\ref{gtmoment} for additional
discussion). To get a unique answer, and to make the transformation
induced on the expectation values small, we fix the multiplicative
coefficient of $C$\ in $G$\ to zero. 

For convenience, we only present and work with the flows generated
by the constraint functions rather than displaying the generators
themselves whose explicit expressions turn out to be rather
complicated and less well-behaved than their flows. The flow
generated by a generator $G$ will be denoted by $\alpha_G^s(x)$,
$x\in\mathcal{C}$, where $\mathcal{C}$ denotes the constraint
surface and $s$ is the gauge parameter along the flow. Its (finite)
action on a quantum phase space function $f$ can be computed via a
derivative expansion \ba\label{gaugeaction}
\alpha^s_G(f)(x):=f(\alpha_G^s(x))=\sum^\infty_{n=0}\frac{s^n}{n!}\{f,G\}_n(x)\q,
\ea where $\{f,G\}_n:=\{\{f,G\}_{n-1},G\}$ and $\{f,G\}_0=f$. The
 Hamiltonian vector field of the generator $G$ is denoted by $X_G$ and
 we have $X_G(f):=\{f, G\}$. The required flows for the transformation
 may be computed explicitly with the aid of the table in
 Appendix~\ref{app:PB}. There is still some freedom in choosing a path
 for the gauge transformation: as mentioned at the beginning of
 Sec.~\ref{sec:lt_eff}, the five constraints generate only four
 independent flows. Removing $C$ still leaves us with three
 independent flows which we can combine. At this point we construct
 the gauge transformation in two steps. First we search for a flow that
 satisfies $X_{G_1}\left(\Delta(qp) \right) = X_{G_1}\left( \Delta(tq)
 \right)=0$\ on the constraint surface and re-scale the flow such that
 $X_{G_1}\left( (\Delta q)^2 \right) =1$. The second step involves
 finding the flow that satisfies $X_{G_2}\left((\Delta q)^2 \right) =
 X_{G_2}\left( \Delta(tq) \right)=0$ and re-scaling this flow such
 that $X_{G_2}\left( \Delta(qp) \right) = 1$.
 The required gauge transformation will then be given by the
flow\footnote{This expression
 might appear surprising at a first glance since gauge parameters are
 real-valued. However, the flow of $G_2$ can be understood via
 $\alpha_{G_2}^{-(\Delta(qp)_0+i\hbar/2)}=
\alpha^{-\Delta(qp)_0}_{G_2}\circ\alpha^{-\hbar/2}_{iG_2}$
 which directly follows from (\ref{gaugeaction}).}
 $\alpha^s_G(f)(x):=\alpha_{G_2}^{-(\Delta(qp)_0+i\hbar/2)}
\circ\alpha_{G_1}^{-(\Delta
 q)_0^2}(f)(x)$ if we can argue that the second and higher derivative
 terms in the respective expansion via (\ref{gaugeaction}) can be
 consistently neglected to order $\hbar$. Equation (\ref{gaugeaction})
 implies that to linear order in the derivative expansion we also have
 $\alpha^u_{G_2}\circ\alpha^v_{G_1}=\alpha^v_{G_1}\circ\alpha^u_{G_2}$
 for fixed values of $u,v$. Note that this
 composition of the $G_1$ and $G_2$ flows only determines $\alpha^s_G$
 up to re-scalings of $G$ and, consequently, the value of $s$ where
 the new $q$-gauge is reached, but any such $\alpha^s_G$ will be
 suitable.

For the particular system at hand, the procedure simplifies if we
impose, in addition to the constraint functions, the gauge condition
$\Delta(tq)=0$, which is shared by both $t$-gauge and $q$-gauge and
is preserved by $\alpha_{G_1}$ and $\alpha_{G_2}$ by construction;
we then find for the other variables
\begin{eqnarray*}
X_{G_1}(t)= \frac{\lambda}{4p^2} \quad&,&\quad X_{G_2}(t)= -\frac{1}{p_t}\,,\\
X_{G_1}(q) = 0 \quad&,&\quad X_{G_2}(q) = \frac{1}{p}\,,\\
X_{G_1}
\left( (\Delta t)^2 \right) = -\frac{p_t^2}{p^2} \quad&,&\quad X_{G_2} \left( (\Delta t)^2 \right) = 0 \,,\\
X_{G_1} \left( (\Delta p_t)^2 \right) = -\frac{\lambda^2}{4p^2}
\quad&,&\quad X_{G_2} \left( (\Delta p_t)^2
\right) = \frac{\lambda^2}{p_t}\,,\\
X_{G_1} \left( \Delta(tp_t) \right) = \frac{\lambda p_t}{2p^2}
\quad&,&\quad X_{G_2} \left( \Delta(tp_t) \right) = -1\,.
\end{eqnarray*}
Noting that $p$\ has a vanishing bracket with all constraints and
$p_t$\ with all constraints except for $C$,
whose flow is neither contained in $\alpha_{G_1}$\ nor in
$\alpha_{G_2}$, we see that all of the derivatives are constant, and
thus the gauge transformation is infinitesimal and, indeed, simply
given by the terms up to linear order in the derivative expansion
(\ref{gaugeaction}) of
$\alpha^s_G(f)(x):=\alpha_{G_2}^{-(\Delta(qp)_0+i\hbar/2)}\circ\alpha_{G_1}^{-(\Delta
q)_0^2}(f)(x)$. Without this simplification, one may, in general,
have to integrate the flows numerically.\footnote{In general, the
Poisson structure of the quantum phase space is such that the
Poisson bracket of the $o(\hbar)$-quantum constraint functions with
a quantum phase space function of a certain order preserves or
increases the order in $\hbar$, while, for instance, Poisson
brackets of ratios of moments can actually decrease the order in
$\hbar$. This follows from the Poisson algebra of moments in
Appendix~\ref{app:PB}. Now the rescaling of the flow such that,
e.g., $X_{G_1}\left( (\Delta q)^2 \right) =1$ has the consequence
that $G_1$ will be of order $\hbar^0$, consisting of ratios of
moments which, in general, may lead to negative orders of $\hbar$
when taking higher derivatives of moments along the flow. It is then
not consistent anymore to neglect the higher derivative terms in the
expansion (\ref{gaugeaction}) of the flow action even if one
multiplies with $o(\hbar)$ values of the flow parameter. In such
situations one must numerically integrate the flow. However, in
general, we expect the gauge transformation between $t$- and
$q$-gauge to be infinitesimal to order $\hbar$.} The initial value
for $(\Delta t)^2$\ is zero as we are starting with the $t$-gauge,
initial values of $\Delta(tp_t)$\ and $(\Delta p_t)^2$\ can be
deduced from~(\ref{Ct}) and~(\ref{depmoms}), respectively. We find
the complete transformation of $t$-gauge variables into the
$q$-gauge variables to order $\hbar$ given by
\begin{eqnarray}
t &=& t_0 + \frac{i\hbar+2\Delta(qp)_0}{2p_t}-\frac{(\Delta q)^2_0
\lambda}{4 p^2} \nonumber \\
q &=& q_0 -\frac{i\hbar+2 \Delta(qp)_0}{2 p} \nonumber \\
(\Delta t)^2 &=& (\Delta q)^2_0\frac{ p_t^2}{p^2} \nonumber \\
(\Delta p_t)^2 &=& \frac{p^2 (\Delta p)^2_0 - \Delta(qp)_0 \lambda
p_t}{p_t^2} + \frac{\lambda^2}{4 p^2}(\Delta q)^2_0 \nonumber \\
\Delta(tp_t) &=& \Delta(qp)_0 - \lambda\frac{ p_t}{2p^2}(\Delta
q)^2_0\q. \label{eq:time_pot_ftransf}
\end{eqnarray}
No gauge transformations for $p_t$\ and $p$\ are listed since
these variables are invariant along the flow of $G$.
The reverse transformation can be obtained in an identical manner,
or simply by inverting~(\ref{eq:time_pot_ftransf})
\begin{eqnarray}
t &=& t_0 - \frac{2 p_t \left(i\hbar+2 \Delta(tp_t)_0\right)+
(\Delta t)^2_0 \lambda }{4 p_t^2} \nonumber \\
q &=& q_0 + \frac{p_t \left(i\hbar+2 \Delta(tp_t)_0 \right)+ (\Delta
t)^2_0 \lambda }{2 p p_t} \nonumber \\
(\Delta q)^2 &=& (\Delta t)^2_0\frac{p^2}{p_t^2} \nonumber \\
(\Delta p)^2 &=& \frac{4p_t^2 (\Delta p_t)^2_0 + 4\lambda  p_t
\Delta(tp_t)_0 + \lambda^2(\Delta t)^2_0 }{4 p^2}\nonumber \\
\Delta(qp) &=& \frac{\lambda }{2 p_t}(\Delta t)^2_0  +
\Delta(tp_t)_0\q. \label{eq:time_pot_btransf}
\end{eqnarray}
In particular, both $q$\ and $t$\ acquire imaginary contributions
during these transformations. We point out that these contributions
exactly cancel out the imaginary terms~(\ref{imagt})
and~(\ref{imagq}), so that upon transformation from $t$-gauge to
$q$-gauge $t$ becomes real and $q$\ acquires the imaginary
term~(\ref{imagq}) and vice versa. Observe that in the case of the
global clock function $q$ in the $q$-gauge, its imaginary part is a
constant of motion and, therefore, does not play any role for
evolution, while in the case of the non-global clock $t$ in the
$t$-gauge, its imaginary part is actually dynamical. We return to this
characteristic in Sec.~\ref{genimagcomment}.  For more discussion of
gauge switching and an argument for the irrelevance of the precise
instant of the gauge change see Sec.~\ref{gaugec} and \ref{gtmoment}.

\begin{center}
\begin{figure}[htbp!]
\includegraphics[width=7.5cm]{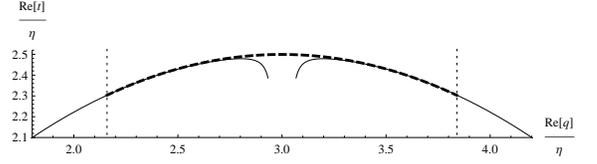}
\caption{\label{fig:fullevollt} Plot of the semiclassical trajectory
evolved past the extremal point in $t$-gauge (solid part of the
trajectory), by temporarily switching to the $q$-gauge (dashed part
of the trajectory). Dotted vertical lines indicate the points where
gauges were switched.}
\end{figure}
\end{center}

FIG.~\ref{fig:fullevollt} gives a segment of a semiclassical
trajectory that has been evolved through the extremal point of $t$\
by temporarily switching to $q$-gauge. The initial conditions and
the values of parameters used here are identical to the ones used to
generate FIG.~\ref{fig:ltbreak}. We switch to $q$-gauge before the
moments have a chance to become large (at $s=1.8$). The evolution
in $q$-gauge stays semiclassical through the turning point in $t$
and sufficiently far away from the extremum ($\tilde{s}$ evolved
from $0$ to $1.4$); the reverse gauge transformation yields a
semiclassical outgoing state in $t$-gauge. Incoming and outgoing
trajectories in $t$-gauge were continued into the region where the
$q$-gauge was used in order to demonstrate their divergence.  We
note that, although the quantities $q(\Re[t])$ in the $t$-gauge and
$t(\Re[q])$ in the $q$-gauge refer to different pairs of objects
(two examples of fashionables in the terminology of \cite{EffTime1})
from the point of view of quantum mechanics, their classical limits
correspond to the same correlations between $q$ and $t$ and plotting
one trajectory as following the other (with jumps of $o(\hbar)$
between the trajectories as a consequence of the gauge changes
above) makes sense for a semiclassical state. The resulting
composite trajectory agrees extremely well with its classical
counterpart, which is why the latter is not present in
the plot.

\subsubsection{Effective positivity conditions and physical states}\label{sec:positivity}

In the discussion of dynamics in the $t$-gauge, we implicitly
interpreted the variables $q(s)$, $p(s)$, $(\Delta q)^2(s)$,
$\Delta(qp) (s)$, $(\Delta p)^2(s)$\ as expectation values and
moments of a canonical pair of evolving operators, with $t$ keeping
track of the ``flow of (internal) time''. In order to make this interpretation
consistent, these variables must have the correct Poisson algebra,
which follows directly from the canonical commutation relation
(CCR). The non-trivial brackets of this algebra are
\begin{flalign}
&\{q, p\} = 1, \quad \{ (\Delta q)^2, (\Delta p)^2\} = 4\Delta(qp)
\\ &\{ (\Delta q)^2,  \Delta(qp) \} = 2 (\Delta q)^2, \quad \{
\Delta(qp), (\Delta p)^2 \} = 2 (\Delta p)^2 \q. \nn
\end{flalign}
In particular, $t$\ must have a vanishing bracket with the rest of
the above variables. These relations are, of course, satisfied
kinematically simply by construction. However, when we introduce
gauge conditions the Poisson bracket on the gauge surface is defined
with the use of the Dirac bracket~\cite{HenTeit}.  It is an
important feature of the gauge conditions~(\ref{ex2}) that the Dirac
brackets between precisely the free variables in the $t$-gauge are
the same as their kinematical counterparts. For the details we refer
the interested reader to~\cite{EffConsRel}.

The above result ensures that the dynamics is consistent with that
of a pair of operators subject to the CCR. However, if we are to
interpret these operators as self-adjoint (which is required for
well-behaved observables), we have to impose additional conditions
on their expectation values and moments:
\begin{eqnarray}
&& q, p, (\Delta q)^2, (\Delta p)^2, \Delta(qp) \in \mathbb{R} \nonumber \\
&& (\Delta p)^2, (\Delta q)^2 \geq  0 \nonumber \\ && (\Delta q)^2
(\Delta p)^2 - \left(\Delta(qp)\right)^2 \geq \frac{1}{4} \hbar^2\q.
\label{eq:pos_conditions1}
\end{eqnarray}
These conditions, in particular, guarantee similar conditions
holding to order $\hbar$ for any polynomial constructed out of
symmetrized products of $\hat{q}$ and $\hat{p}$ (see
Appendix~\ref{positivity}). There is, of course nothing that would
prevent us from imposing these conditions on the initial values of
the variables. However, it is \emph{a priori} not clear whether such
conditions will be preserved by the dynamics in either gauge or by
the gauge transformations. Below we list the specific results that
ensure the consistency of the effective dynamics with the
interpretation of the variables we have chosen as observable
expectation values and moments. The details of the calculations may
be found in Appendix~\ref{positivity}. We find that
\begin{itemize}
\item the conditions~(\ref{eq:pos_conditions1}) are preserved by the
dynamics of the $t$-gauge,
\item the conditions on the expectation values and moments of $\hat{t}$\
and $\hat{p}_t$\ analogous to~(\ref{eq:pos_conditions1}) are
preserved by the dynamics in the $q$-gauge,
\item if the variables in the $t$-gauge
satisfy~(\ref{eq:pos_conditions1}), then the gauge transformed
variables satisfy the $q$-gauge analog
of~(\ref{eq:pos_conditions1}).
\end{itemize}

\section{Complex internal time and relational observables}\label{compt}

In this section we reflect on some of the general features of the
effective analysis performed on the model of Sec.~\ref{lt}. We focus
on the interpretation of the imaginary contribution to internal time,
transformations between local choices of clocks (Zeitgeist) and the
status of relational observables in a system without global
time. Complex internal time arising in the effective approach to local
clocks and in local deparametrizations at the state level has been
discussed in detail in~\cite{EffTime1}, along with general issues
related to relational evolution and observables and we refer the
interested reader to that work. However, the results concerning
complex internal time are worth summarizing in the context of the
concrete examples provided within the present manuscript, which we do
in Sec.~\ref{sec:imtime}. Considerations of this section are general,
and hence equally applicable to the second model studied in
Sec.~\ref{rovmod},
for which some of the general discussions of this
section will be helpful.

\subsection{Imaginary contribution to internal time} \label{sec:imtime}

At this moment, it is useful to pause and ask how meaningful an
imaginary contribution to time can be. First, we would like to
acquire some intuition regarding its origin. From a certain point of
view this feature is not entirely surprising --- after all, there are
old and well-known arguments in quantum mechanics saying that time
cannot be a self-adjoint operator. Otherwise, it would be conjugate
to an energy operator bounded from below for stable systems. Since a
self-adjoint time operator would generate unitary shifts of energy
by arbitrary values, a contradiction to the lower bound would be
obtained. The result of complex expectation values for local
internal times obtained here looks similar at first sight --- a
non-self-adjoint time operator could, certainly, lead to complex time
expectation values --- but it is more general. In the model of
Sec.~\ref{lt}, we are using a linear potential which does not
provide a lower bound for energy. The usual arguments about time
operators thus do not apply; instead our conclusions are drawn
directly from the fact that we are dealing with a time-dependent
potential. (For time-independent potentials, $\langle\hat{t}\rangle$
does not appear in the effective constraints and can consistently be
chosen real. The time dependence is thus crucial for the present
discussion.)

Rather, the imaginary contribution to internal time may be regarded in the
same vein as the imaginary contributions to the various unphysical
moments (see e.g.\ Eq.~(\ref{Ct})) --- as an artifact of assigning
expectation values to all kinematical observables, which typically
do not project in any natural way to self-adjoint operators on the
physical Hilbert space. We recall a simple example given
in~\cite{EffTime1} of a physical inner product, which in a
deparameterizable system assigns a complex ``expectation value" to
internal time. A free relativistic particle in 1+1 Minkowski spacetime,\footnote{In this example, $t$ has the usual notion of proper time as
  experienced by inertial observers in addition to the more general notion of internal time as a phase-space degree of freedom of the
  cotangent bundle of Minkowski space. In this context, as in our
  other examples, we are interested only in the phase-space notion of
  internal times.}
is subject to the constraint
\begin{equation}
\left( -\hbar^2 \frac{\partial^2}{\partial x_0^2} + \hbar^2
\frac{\partial^2}{\partial x_1^2} - m^2 \right) \psi(x_0, x_1) = 0
\q. \label{eq:free_rel_PDE}
\end{equation}
The standard inner product used for positive frequency solutions has
the form
\begin{eqnarray}
\begin{split}\label{eq:KG_IP}
\left( \phi, \psi \right) :=& i\hbar \int_{-\infty}^{\infty}
\left(\bar{\phi}(x_0, x_1) \frac{\partial}{\partial x_0} \psi(x_0,
x_1) \right.\\&\left.\left.- \left( \frac{\partial}{\partial x_0}
\bar{\phi}(x_0, x_1) \right) \psi(x_0, x_1) \right) dx_1
\right|_{x_0=t}\q. \label{eq:KG_prod}
\end{split}
\end{eqnarray}
Evaluating the ``expectation value" of the kinematical internal time operator, using
a positive frequency solution with this inner product,\footnote{Strictly speaking, this is clearly not a true expectation value, since the kinematical internal time operator does not preserve the (physical) positive frequency Hilbert space. Nevertheless, we can use this inner product as a well-defined bilinear form in this case.} yields
\begin{equation} \label{tOp}
\langle \hat{t} \rangle = \left( \phi, x_0 \phi \right) = t -
\frac{i\hbar}{2} \left\langle \widehat{\frac{1}{p_t}}
\right\rangle\,.
\end{equation}
To order $\hbar$\ the imaginary part is identical
to Eq.~(\ref{imagt}), and, indeed, to the analogous result in
Sec.~\ref{rovmod} given in Eq.~(\ref{imqi}). The key ingredient in
this result is the use of both $\phi$\ and $\partial\phi /\partial
x_0 $\ in the construction of the inner product, which is
ultimately related to the fact that the constraint equation is
second order in the time derivative, so that locally both $\phi$\
and $\partial\phi /\partial x_0$\ are independent degrees of
freedom. This suggests a generalization of the form of the imaginary
contribution to $\langle \hat{t} \rangle$, to all constraints where
$\hat{p}_t$\ appears quadratically. One may then ask whether the
effective procedure supports such a generalization. It was, indeed,
demonstrated in~\cite{EffTime1}, that for any constraint of the form
\[
\hat{C} = \hat{p}_t^2 - \hat{p}^2 + V(\hat{q}, \hat{t})\,,
\]
the imaginary contribution at order $\hbar$\ is precisely the same in
the effective framework, $\Im [t]= -\hbar/2\langle\hat{p}_t\rangle$.

One choice was made at the beginning of the effective analysis,
namely the gauge-fixing of the effective constraints. We used the
gauge-fixing that worked well for deparameterizable systems, but it
may not be suitable for non-deparameterizable ones. One could then
try to change the gauge-fixing conditions and perhaps move the
complex-valuedness to some of the kinematical moments rather than
the internal time expectation value. It is, however, unlikely that this would
give a general procedure  because the form of the constraints would
require gauge-fixing conditions adapted to the system under
consideration, and, in particular, to the potential. The gauge-fixing
conditions used here, on the other hand, work for arbitrary
potentials and are specifically motivated by and associated to our choice of clock and corresponding relational time
(see also Sec.~\ref{gaugec}).

Finally, there is concrete evidence, that this imaginary
contribution is a generic feature associated with local
deparameterizations of a Dirac constraint of the form
\ba\label{WdW}
\left(\hat{p}_t^2-\hat{H}^2(\hat{t},\hat{q},\hat{p})\right)\psi(q,t)=0\q,
\ea where $\hat{H}^2$ is a positive operator at least on some set of
states. For example, such a constraint features in the Wheeler-DeWitt (WDW)
equation in homogeneous and isotropic cosmology. In
general, Eq.~(\ref{WdW}) is not equivalent to a Schr\"odinger equation
\ba\label{schrod1}
\left(-i\hbar\partial_\tau+\hat{H}(\tau,\hat{q},\hat{p})\right)\psi(q,\tau)=0\q,
\ea since the solutions to the latter satisfy
\ba\label{doubleschrod}
-\hbar^2\partial_\tau^2\psi=\hat{H}^2\psi+i\hbar\partial_\tau\hat{H}\psi\q.
\ea The inequivalence formally appears to be of order $\hbar$\ and
is based in part on erroneously identifying the kinematical operator $\hat{t}$\
of Eq.~(\ref{WdW}) with the time parameter $\tau$\ of Eq.~(\ref{schrod1}).
In~\cite{EffTime1} it was shown, however, that Eq.~(\ref{WdW}) and
an internal time version of
Eq.~(\ref{schrod1})
are both solved by the same state (in the sense that their
expectation values vanish) at order $\hbar$, if one defines
\begin{equation}\label{clockop}
\hat{t} = \hat{\tau} - \frac{i\hbar}{2} \widehat{p_\tau^{-1}} \q ,
\end{equation}
(for states outside the zero-eigenspace of $\hat{p}_\tau$) where the
(continuous) eigenvalues of the kinematical internal time operator
$\hat{\tau}$ assume the role of the parameter $\tau$ of the
Schr\"odinger equation. The internal time Schr\"odinger equation
represents a local deparametrization of Eq.\ (\ref{WdW}) and arises
from a kinematical quantization of one of the two factors of a
classical factorization of the quadratic constraint,
$C=(p_\tau-H(\tau,q,p))(p_\tau+H(\tau,q,p))$, where both internal time
$\tau$ and $p_\tau$ are dynamical phase space variables. The result
once again agrees with the general form of the imaginary contribution
obtained effectively. This comparison of the quadratic relativistic
constraint with a local (internal time) Schr\"odinger equation at the
state level is demonstrated on a concrete example in
Sec.~\ref{schrodreg}. We also compare the corresponding semiclassical
dynamics of local deparametrization to the effective evolution in
Sec.~\ref{effschrodcomp}.

\subsection{Dynamics with a complex relational clock}\label{genimagcomment}

As we saw in the previous section, the expectation value of internal time can
acquire an imaginary contribution even in the standard treatments of
deparameterizable systems. The difference is only that
deparameterizable systems with a global internal time do not force
us to include the imaginary part, while systems with local internal
times do. This can also be seen from the shape of the generic
imaginary contribution $\Im[t]=-\hbar/2\langle \hat{p}_t\rangle$:
While in the presence of a ``time potential'', $p_t$ will fail to
be a constant of motion and, consequently, $\Im[t]$ will actually be dynamical,
in the absence of a ``time potential'' in the
constraint $p_t$ is automatically a Dirac observable and,
therefore, $\Im[t]$ a constant of motion.
But a constant imaginary contribution, in contrast to a dynamical one, is not
needed in order to avoid a violation of the constraints since it
can be interpreted as an integration constant at the effective level
and does not even appear in the constraints in the absence of a
``time potential''. Indeed, the WDW and
 (the internal time version of the) Schr\"odinger equation,
Eqs.\ (\ref{WdW}) and (\ref{schrod1}), are automatically equivalent in
this case. The imaginary contribution to internal time may, therefore, be
disregarded altogether for relational evolution in the absence of
a ``time potential", but it cannot be neglected otherwise.

We emphasize that a non-global clock necessarily implies a ``time
potential,'' while a time-dependent potential does not automatically
imply a non-global clock.\footnote{For instance, in a relativistic
system governed by a constraint $C=p_t^2-H^2(q,p,t)$, where $H^2>0$
$\forall\, t$, the clock $t$ will be global.} The dynamical imaginary
contribution is, therefore, more general than a pure consequence of
non-unitarity following from non-global clocks. Nevertheless, the
imaginary contribution becomes more prominent where the momentum of
the clock variable becomes small and is, thus, especially relevant
near turning points of non-global clocks. In fact, the dynamical
imaginary contribution, being inversely proportional to the kinetic
energy of the clock variable, can be interpreted as a measure for the
quality of the relational clock: the higher the clock's momentum,
i.e., the further away it is from a turning point where quantum
effects restrict its applicability, the smaller the imaginary term and
the better behaved the clock. This coincides with the intuition that,
the faster the clock, the better its time-resolution. The inverse
kinetic energy also appears in other discussions of the qualities of
clocks. A brief comparison of this and further references may be found
in \cite{EffTime1}.

Facing a dynamical imaginary part, we ought to make sense out of such
a ``vector time'' with two separate degrees of freedom. (Relational)
time is commonly understood as a single (scalar) degree of freedom
and, in principle, we may choose any (real) phase space function which
is reasonably well-behaved. In this light, we appoint the real part of
the clock function for relational time, for several reasons: 1) it
gives the correct classical internal time in the classical limit; 2)
for small ``time potentials'', or in the absence thereof, the
imaginary contribution is approximately, or exactly constant,
respectively; 3) the ``expectation value", Eq.~(\ref{tOp}),
reproducing the specific imaginary term for the free relativistic
particle is based on a constant real parameter time slicing; 4) the
Schr\"odinger regime (obtained from a local deparametrization of the
relativistic constraint) which, at least locally, should give a
conventional quantum time evolution, is based on a real-valued time,
and 5) as we will see in an example in FIG.\ \ref{reqiimqi} in Sec.\
\ref{effschrodcomp} below, the dynamical imaginary contribution for
non-global clocks can fail to be monotonic where the real part serves
as a suitable local clock.

\subsection{Switching clocks is equivalent to changing gauge}\label{gaugec}

From the point of view of the Poisson manifold of the effective
framework no variables or gauges are preferred over others and we
could, in principle, choose a $q$-gauge like (\ref{gaq}) and still use
$t$ as our clock for relational evolution. However, as we will see in
the second model in Sec.~\ref{rovmod}, the effective evolution in a
given $\tau$-gauge is matched by a Schr\"odinger type state evolution
(\ref{schrod1}) in internal time $\tau$, where the conventional
Schr\"odinger type inner product is defined on constant-$\tau$
slicings. This Schr\"odinger regime analog can, thus, only be
meaningfully interpreted as local evolution in $\tau$.  Moreover, when
nevertheless using, e.g., $t$ as a local clock in the $q$-gauge in
Sec.\ \ref{q-gauge}, one faces the undesirable consequence that
moments involving $t$ or $p_t$ become evolving degrees of freedom,
while the moments of our actual variables of interest, $(q,p)$, are
(at least partially) gauge fixed, essentially leaving only an
evolution parameter $q$.  The resulting moments would no longer be
associated to a canonical pair, which has an impact on Dirac brackets
and unnecessarily complicates the physical relational interpretation
of such moments relative to $t$.  Consequently, it is unavoidable to
switch the local clock in the effective procedure when choosing a new
gauge; the choice of gauge is
intimately intertwined with the
choice of
(internal) time and changing the clock and corresponding time is
practically tantamount to changing gauge and Zeitgeist.  Accordingly,
certain questions about (physical) correlations of variables are best
described in certain gauges and in each gauge we evolve a {\it
different} set of relational observables which is associated to the
chosen relational clock.

The peculiar circumstance that the set of degrees of freedom that
evolve in relational time appears to depend on the gauge has its roots
in the fact that, by the choice of Zeitgeist, local relational
observables considered here describe the system in partially gauge
fixed form.  While the physical information computed for the system
is, certainly, gauge independent, its presentation in gauge fixed form
depends on the gauge chosen. One can illustrate this feature also with
the standard notions of partial and complete observables.  Complete
relational observables (invariant under all gauge flows) can be
understood as gauge invariant extensions of gauge restricted
quantities \cite{Bianca1,Haj1,HenTeit}; when restricting a complete
observable to certain fixed values of some clock functions
(parametrizing the full gauge orbit), it is reduced to a ``partial''
observable, evaluated on a gauge-fixing surface. In such a gauge not
all correlations between the phase-space degrees of freedom are
accessible and, hence, not all questions about correlations
meaningful. (The choice of clock functions along full gauge orbits, of
course, does not constitute gauge fixing.)  Evolving partial
observables along the (full) gauge orbits results in complete
relational observables that clearly depend on the choice of the
relational clock functions,\footnote{Different choices of clocks
parametrizing the full gauge orbits will yield different parameter
families of observables, although still describing the correlations on
the same gauge orbits (albeit along different flow lines).} just as
the gauge-fixing surfaces corresponding to constant values of (some
of) the clock functions and the associated partial relational
observables do.

In the effective framework as well one could gauge invariantly extend
the local relational observables of the different Zeitgeister to
complete observables by, apart from the $o(\hbar^0)$-clock $t$ or $q$,
taking three further $o(\hbar)$-clock functions into account to keep
track of the remaining three gauge flows on quantum phase
space.\footnote{In general, global obstructions may prevent the clock
functions from globally parametrizing the full gauge orbit.}  However,
for practical reasons, it is advantageous to gauge fix these three
$o(\hbar)$-clocks such that the relational evolution we want to
describe in the $o(\hbar^0)$-clock can be expressed and compared to
Hilbert-space approaches in the most convenient way.  One possibility
is by using the mentioned relationship of the effective framework with
a (local) deparametrization in an internal time Schr\"odinger regime.
To define a Schr\"odinger type evolution, one can choose which slicing
to employ (where the constant-$t$-slicing is the most convenient one
when choosing $t$ as internal time and corresponds to the
deparametrization given by (\ref{schrod1})). The choice of the slicing
and corresponding inner product determines how the spreads of the
states solving the internal time Schr\"odinger equation are
measured. For instance, in standard constant-$\tau$-slicing for
(\ref{schrod1}) (corresponding to constant-$t$-slicing and evolution
in $t$ in the relativistic system), not all the fluctuations of
$\hat{q}$ can vanish and the variable appears to be of quantum nature,
while $\hat{\tau}$ is projected to the role of a classical parameter
$\tau$\ since the spreads related to $\hat{\tau}$ will vanish.
In constant $q$-slicing the situation is reversed. Note, however, that
deparametrizations with respect to different internal time variables
will, in general, yield different quantum theories with inequivalent
Hilbert spaces.

Alternatively, we could use a tilted slicing that corresponds to
neither configuration coordinate. For a concrete example recall the
free relativistic particle, which is subject
to~(\ref{eq:free_rel_PDE}).  This constraint equation is
Lorentz--invariant and we can construct a physical inner product on
its solutions of the same form as~(\ref{eq:KG_IP}) but evaluated in a
different Lorentz frame on surfaces of constant $x'_0$, where
$x'_{\mu} = \Lambda_{\mu}^{\ \nu} x_{\nu}$\ are the boosted
coordinates; the corresponding multiplicative kinematical operators
will be denoted by $\hat{x}'_{\mu}$. Kinematical expectation values
and moments of $\hat{t}$\ and $\hat{q}$\ are linear combinations of
the expectation values and moments of $\hat{x}'_{\mu}$. For instance,
by linearity of the expectation values, the correlation $\Delta(tq)=
\Lambda_{ \ 0}^{ \mu} \Lambda_{ \ 1}^{\nu} \Delta(x'_{\mu} x'_{\nu}) =
\Lambda_{ \ 0}^{1} \Lambda_{ \ 1}^{1} (\Delta x'_1)^2$, which is
non-zero unless the boost is trivial. (Here the last equality follows
as fluctuations of $\hat{x}'_0$\ vanish to order $\hbar$, when
evaluated in this inner product.)  In this tilted slicing one can
construct a local Schr\"odinger evolution and still use $\langle
\hat{t} \rangle$\ as internal time, though unfamiliar non-vanishing
moments (involving $\hat{t}$) severely complicate the interpretation
of $\hat{t}$\ and $\hat{q}$ as a relational time reference and an
evolving variable, respectively.

On the other hand, the quantum phase space of the effective framework,
being representation independent, must contain information about a
general class of slicings in a (local) deparametrization.  This is the
reason why unusual (time) moments such as $\Delta(qt)$ do not
necessarily vanish in the effective formalism.  The three
$o(\hbar)$-clocks do not represent true internal coordinates, but
parametrize the slicings
and thereby the (in general inequivalent) corresponding Hilbert-space
representations.
Hence, the three conditions fixing the three
$o(\hbar)$-flows will fix the slicing
and Hilbert-space representation
to which the effective
relational evolution will correspond. Certainly, when choosing $t$ as
the relational $o(\hbar^0)$-clock, we could choose gauge conditions
differing from the $t$-Zeitgeist; however, these would correspond to
tilted slicings and are, consequently, less convenient for
calculations as well as interpretations. Furthermore, the
$q$-Zeitgeist can be interpreted in terms of slicings parallel to the
$t$-axis and is, thus, not useful for describing evolution in $t$.

In the light of the present discussion, one may interpret the
evolution generated by the remaining first class (Hamiltonian)
constraint in a given Zeitgeist (e.g., (\ref{ex8a}) in $t$-Zeitgeist
in Sec.~\ref{t-gauge}) which preserves this gauge and the effective
positivity (see Sec.~\ref{sec:positivity}) as describing an
approximate, locally unitary evolution for semiclassical states in a
given (preserved) slicing in a local deparametrization. In addition,
the imaginary contribution to internal time is clearly dependent on
the chosen Zeitgeist at the effective level and the slicing in a local
deparametrization; when employing tilted slicings or gauges differing
from the Zeitgeist, the imaginary contribution to the internal clock
will take a different form.

In conclusion, certain questions about correlations are best addressed
in certain gauges and we are, indeed, evolving different sets of
(partial) relational observables in different Zeitgeister. The
presence of additional gauge flows and slicings also explains the
observation that $\langle\hat{t}\rangle(\langle\hat{q}\rangle)$ and
$\langle\hat{q}\rangle(\langle\hat{t}\rangle)$ are not in one-to-one
correspondence, while the analogous statement (at least locally) holds
in the classical system.

\subsection{The moment of gauge and clock change}\label{gtmoment}

Here we argue that the precise instant of the gauge change is
irrelevant, as long as the semiclassical approximation is valid before
and after the gauge transformation. The instant when to perform the
change of the clock then becomes a matter of convenience.

Let $q_1$ and $q_2$ be two configuration variables, which we use as
local clocks, and let $\cc$ be the constraint surface, $\cg_{1}$ the
$q_1$-gauge surface and $\cg_{2}$ the $q_2$-gauge surface (in
$\cc$). Denote by $\alpha^s_{C_{H_1}}(x)$ ($x\in\cg_1$) the flow of
the ``Hamiltonian constraint'' in $q_1$-gauge (i.e., the
$\cg_1$-preserving first class flow) and by
$\alpha^{u}_{C_{H_2}}(y)$ ($y\in\cg_2$) the flow of the
``Hamiltonian constraint'' in $q_2$-gauge, where $s,u$ are gauge
parameters along the flows. Furthermore, denote by $\alpha^t_G(x)$
the flow of the generator $G$ of some fixed gauge transformation
which maps between the $q_1$- and $q_2$-gauge for certain values of
$t$ and which, for the sake of avoiding ordering ambiguities, we
assume to be free of caustics (see Secs.\ \ref{sec:gaugec_lt} and
\ref{rovgt} for explicit constructions of such transformations in
the examples).

For the moment, assume that both $\cg_1$ and $\cg_2$ provide
complete submanifolds of $\cc$ and that there are no global
obstructions to either the $q_1$- or the $q_2$-gauge. Recall that
the first class nature of a constraint algebra with $n$ independent
flows ensures that the flows are integrable to an $n$-dimensional
submanifold in $\cc$, the gauge orbit $\fg$ \cite{HenTeit}.

For simplicity, consider a classical constraint $C(q_1,q_2,p_1,p_2)$
on a four-dimensional phase space. Then the quantum phase space to
semiclassical order will be 14-dimensional and governed by five
quantum constraint functions which generate four independent flows
\cite{EffCons,EffConsRel}. Hence, $\dim\cc=9$ and $\dim\fg=4$.
$\cg_1$ and $\cg_2$ are each described by three independent
conditions, thereby fixing three of the four independent flows in
$\fg$. $C_{H_1}$ ($C_{H_2}$) generates the only independent gauge
flow which preserves $\cg_1$ ($\cg_2$), implying
$\dim\fg\cap\cg_1=\dim\fg\cap\cg_2=1$, where the sets $\fg\cap\cg_1$
and $\fg\cap\cg_2$ are the curves $\alpha^s_{C_{H_1}}(x)$
($x\in\cg_1$) and $\alpha^u_{C_{H_2}}(y)$ ($y\in\cg_2$). Now
$\alpha^t_G(x)\in\fg\,\,\,\forall \,\,t$ and
$\alpha^{t=t^*}_G(x)\in\fg\cap\cg_2$ for some $t^*$ and $x\in\cg_1$.
This map obviously has an inverse, namely $\alpha_{-G}$, since the
flow lines of a single generator form a congruence in $\fg$, and,
thus, no point lies on two different such flow lines. Therefore,
points along $\alpha^s_{C_{H_1}}$ are mapped 1-to-1 to points along
$\alpha^u_{C_{H_2}}$ via $\alpha_G$, and we must have \ba\label{ind}
\alpha^{t=t_1^*}_G\circ\alpha^s_{C_{H_1}}(x)=\alpha^u_{C_{H_2}}\circ\alpha^{t=t_2^*}_G(x)\q,
\ea for some $x\in\cg_1$, some $s,u\in\mathbb{R}$ and fixed
$t_1^*,t^*_2$ determined via the conditions
$\alpha^{t=t_2^*}_G(x)\in\cg_2$ and
$\alpha^{t=t_1^*}_G\circ\alpha^s_{C_{H_1}}(x)\in\cg_2$.

Since the gauge transformation $\alpha_G$ maps the points along the
$C_{H_1}$-generated trajectory in $\cg_1$ bijectively to points
along the $C_{H_2}$-generated trajectory in $\cg_2$ we always map
between the same two trajectories and, therefore, it does not matter
when precisely the gauge and the clock are switched.

Locally, this argument also holds in systems without global clocks
and which suffer from global obstructions to the $q_1$- and
$q_2$-gauges, as long as one works in a regime in which the
respective gauges are valid before and after the gauge
transformation and are consistent with the semiclassical
approximation. In this regime, it should also be irrelevant when
precisely the gauge and the clock are changed. In Sec.\
\ref{fullorbit}, we numerically demonstrate this argument and its
consistency with the semiclassical approximation in an example.

\subsection{Relational observables as ``fashionables"}

As can be seen explicitly in the models studied in the present work,
relational observables of the type
$\langle\hat{q}\rangle(\langle\hat{t}\rangle)$\ can be given meaning
even if $\langle\hat{t}\rangle$\ is not used as an internal time
throughout the evolution.  This feature is implemented by switching
gauges for non-global clocks. Such gauge transformations imply shifts
of the order $\hbar$\ in correlations of expectation values and
moments as one changes clocks. This is not surprising; it merely
underlines the fact that expectation values of the same kinematical
variable taken in different Zeitgeister translate into different
relational observables. Semiclassically, however, the differences are
only of order $\hbar$.

We see that relational observables appear to be only of local
nature:\footnote{Relational observables have perhaps been understood
as a local concept in the formulations provided before, but so far
they have been made sense of in a quantum setting only in the
effective framework as developed in \cite{EffTime1}. For a discussion
of difficulties in the Hilbert-space picture, see the comment by
H\'aj\'{\i}\v{c}ek cited in \cite{Rovmod}.} a Zeitgeist comes with its
own set of relational observables and since a Zeitgeist is typically
only temporary, one is forced to use different relational observables
to describe the full evolution. Just as with local coordinates on a
manifold, we cover a semiclassical evolution trajectory by patches of
local internal times and translate between them. We, therefore,
follow~\cite{EffTime1} and refer to the correlations of the evolving
expectation values and moments with the (real part of) the expectation
value of a local internal clock in its corresponding Zeitgeist as
\emph{fashionables}. An explicit examples of a fashionable is the
correlation of $q(s)$\ and $\Re[t(s)]$ of Eq.~(\ref{solq}) (see
FIGs.~\ref{fig:ltbreak} and \ref{fig:fullevollt}).  These quantities
are only defined so long as the corresponding Zeitgeist is valid and
may subsequently ``fall out of fashion'' when the Zeitgeist
changes. By analogy, we also use the term fashionables to denote the
expectation values of operators obtained via local deparametrizations
(for example $\langle \hat{q}_2 \rangle (q_1)$\ and $\langle \hat{p}_2
\rangle (q_1)$\ of Eq.~(\ref{expec})).

It should be noted that the notion of {\it fashionables} is, in fact,
state-dependent, in contrast to usual operator versions of quantum
relational Dirac observables. Fashionables are associated to a choice
of Zeitgeist and different Zeitgeister are valid for ranges depending
on the semiclassical states considered. A fashionable breaks down
together with the corresponding Zeitgeist when it is rendered invalid,
e.g., at a turning point of the corresponding clock. Fashionables,
therefore, reflect the local nature of quantum relational evolution
and are somewhat closer to a physical interpretation by being
state-dependent. Thereby, they also avoid certain technical and
interpretational problems of operator versions of quantum relational
observables, such as non-self-adjointness issues in the presence of a
purely local time (see also the general discussion concerning
fashionables in \cite{EffTime1}).
In practice, the local nature of observables does not prevent one from
computing physically meaningful predictions, as these typically refer
to finite ranges of time. Moreover, since data is consistently
transferred between local choices of a clock, one can evolve them
through the turning point by temporarily switching to a new Zeitgeist
and employing the old Zeitgeist before and after the turning point.

Apart from being generally of merely local nature, it appears that the
standard concept of relational evolution has only semiclassical
meaning and that the standard notion of (locally unitary) relational
time evolution breaks down together with complex relational time in a highly
quantum state of a system without a global clock. For a discussion of
this issue, we again refer the interested reader to \cite{EffTime1}.

Unlike a conventional Hilbert-space representation, the effective
approach in its present form does not by itself rigorously define a
quantum theory, but rather provides a tool for evaluating quantum
dynamics.  In deparameterizable models, a close relationship between
these two formulations has been found and discussed
\cite{EffConsComp}.  On the other hand, when going beyond
deparameterizable systems, the effective method can still be used to
evaluate quantum dynamics, while local internal times and fashionables
have not been made sense of in the Hilbert-space picture, which
indicates that the effective constructions presented here already go
somewhat beyond usual formulations of quantum physics. At this stage,
we are not entitled to formulate effective dynamics as a true
alternative to quantum mechanics because mainly the semiclassical
setting has been developed so far. Given the enormous difficulties of
dealing with time at the Hilbert-space level of non-deparameterizable
systems, some non-truncated form of effective equations may be a more
suitable
setting and eventually be independent of Hilbert-space
constructions.

\section{A timeless model: the 2D isotropic harmonic oscillator with fixed total energy}\label{rovmod}

The previous example in Sec.\ \ref{lt} was deparametrizable, even
though one could locally employ a non-global clock which already
revealed a number of consequences of the {\it global time problem}, in
particular for the effective approach. Some of these features were
subsequently discussed in more generality in Sec.~\ref{compt},
complementing \cite{EffTime1}. Now we explore all this in
detail in a truly timeless, non-deparametrizable system comprised of
the 2D isotropic harmonic oscillator with prescribed total energy.
This toy model, previously discussed
by Rovelli in \cite{Rovbook,Rovmod}, leads to closed orbits in the
classical phase space and, consequently, does not admit global
clocks. The issue of changing clocks/gauges becomes inevitable. In
our discussion we will compare the classical, effective and
Hilbert-space approaches
to this model.

\subsection{Classical discussion}

Classically, the model is governed by the constraint
\ba\label{cl-rov}
C_{\rm class}= p_1^2+p_2^2+q_1^2+q_2^2-M\q
\ea
with a constant $M$.
The dynamical equations are given by
\ba\label{cl-eom}
\{q_i,C_{\rm class}\}=2p_i \q  \text{and} \q \{p_i,C_{\rm class}\}=-2q_i\q,
\ea
($i=1,2$) and straightforwardly solved by
\ba\label{cl-sol}
{q_1}_{{\rm cl}}(s)=\sqrt{A}\sin(2s) \,, \q {q_2}_{{\rm cl}}(s)=\sqrt{M-A}\sin(2s+\phi)\q, \\
{p_1}_{{\rm cl}}(s)=\sqrt{A}\cos(2s) \,, \q {p_2}_{{\rm cl}}(s)=\sqrt{M-A}\cos(2s+\phi)\q,
\ea
where $s$ is the parameter along $\alpha^s_{C_{\rm class}}(x)$ and $0\leqslant A\leqslant M$, $0\leqslant\phi\leqslant 2\pi$. The canonical pair of Dirac observables $\phi$ and $A$ satisfies
\ba\label{aphi}
2A=M+p_1^2-p_2^2+q_1^2-q_2^2 \,, \q \tan\phi=\frac{p_1q_2-p_2q_1}{p_1p_2+q_1q_2}\q,
\ea
and completely coordinatizes the reduced phase space, which is
topologically a sphere and, thus, no cotangent bundle
\cite{Rovmod}. The classical system clearly does not possess any
global clock functions; indeed, if we choose one of the $q_i$ as a
clock, we see that this function will encounter a sequence of turning
points along a classical trajectory. The classical trajectories are
ellipses in configuration space, periodic and, therefore closed.

Due to this periodicity of the orbits, states which are related by an
integer number of revolutions around such an ellipse are described by
identical phase space information. One could only distinguish these
states via the gauge parameter $s$ which, however, is not a physical
degree of freedom. In order to distinguish states related by complete
numbers of revolutions, one would need an extra phase space degree of
freedom. Furthermore, the group generated by this constraint is ${\rm
U}(1)$ which is compact. The number of revolutions around the ellipse,
therefore, has no physical meaning, in spite of the fact that the
gauge parameter may run over an infinite interval. We thus identify
states related by complete numbers of revolution.

\subsubsection{Evolving observables}

For the quantization of the model it turns out to be advantageous to use the following over-complete set of Dirac observables \cite{Rovmod}
\ba\label{angobs}
L_x=&\frac{1}{2}\left(p_1p_2+q_2q_1\right)\q,\q L_y=\frac{1}{2}\left(p_2q_1-p_1q_2\right)\q,\nn\\ &\text{and} \q\q L_z=\frac{1}{4}\left(p_1^2-p_2^2+q_1^2-q_2^2\right)\q,
\ea
which satisfy the constraint
\ba\label{angobscon}
L_x^2+L_y^2+L_z^2=\frac{M^2}{16}
\ea
and the usual angular momentum (Poisson) brackets. These variables may then be quantized via group quantization. The observable $L_y$ can be interpreted as the angular momentum of the system which also provides the orbits with an orientation.

In spite of the a priori timelessness of this model, one can give it
a (local) evolutionary interpretation. Given the timeless initial
data $\phi$ and $A$, the classical solution is completely specified
and prediction of relational information is possible. Choose a local
clock, say $q_1$, and evolve the other variables of interest, in
this case $q_2$ and $p_2$, with respect to $\tau$, where $\tau$ are
the possible values of $q_1$. The relational Dirac observables
corresponding to this evolution are, obviously, double valued, since
the orbit is closed and are given by \ba\label{corr}
q_2^\pm(\tau)&=&\sqrt{M/A-1}\left(\tau\cos\phi\pm\sqrt{A-\tau^2}\sin\phi\right)\q,\nn\\
p_2^\pm(\tau)&=&\sqrt{M/A-1}\left(-\tau\sin\phi\pm\sqrt{A-\tau^2}\cos\phi\right)\q.
\ea (where $\tau$ is now a parameter). The expressions with index $+$
refer to evolution forward in $q_1$-time, while the expressions with
index $-$ refer to backward evolution in $q_1$ (see Sec.~\ref{linloc} for additional discussion). The fact that these
correlations are double valued does not constitute a problem, since
the value of $\phi$ provides an orientation of the orbit. Starting
at a point of the ellipse at a given value of $q_1$, the direction
of relational evolution in $q_1$ is provided by the orientation and
one may evolve in this manner around the ellipse without having to
switch the clock at the classical level. Indeed, at the two turning
points of $q_1$ the relational momentum observable is non-vanishing
and, consequently, determines the direction of evolution. One can
simply switch, for instance, from $q_2^+$ to $q_2^-$ and change
the direction of $\tau$ since the system moves back in
$q_1$.\footnote{Continuation to larger absolute values of $\tau$
will produce meaningless complex correlations in Eq.~(\ref{corr}) which
simply indicates that the system will never reach such values of the
local clock.} This way a consistent relational evolution is obtained
along the trajectory which is entirely encoded within Dirac
observables and no use of any gauge parameter is made. For later
reference, it is useful to note that one could arrive at the same
predictions of correlations by providing --- instead of $\phi$ and
$A$ --- relational initial data, e.g., $q_2^+(\tau=\tau_0)$ and
$p_2^+(\tau=\tau_0)$, plus the orientation of the ellipse which is
encoded in the angular momentum $L_y$.
Notice that the
orientation must be specified since, given the values of $q_1,q_2,p_2$,
one can only solve for $p_1$ up to sign via Eq.~(\ref{cl-rov}). This is
due to the relativistic/quadratic nature of the constraint and
the reason why, in general, one needs to provide a time
direction in which to evolve (or equivalently a Hamiltonian) apart from the initial data
\cite{Hajlec}, in order to pose a well-defined initial value problem (IVP); purely relational information cannot
coordinatize the space of solutions of systems governed by
relativistic constraints.\footnote{In non-relativistic parametrized systems, where the
momentum conjugate to the time function appears linearly, the time
direction is automatically given.}

We will perform the precise analogue of this local relational evolution in the effective and quantum theory.

\subsubsection{Local relational evolution generated by physical Hamiltonians}\label{linloc}

If we interpret Eq.\ (\ref{corr}) as physical motion in $q_1$, we would
like to find a physical Hamiltonian which generates this motion in the
reduced phase space. Such a Hamiltonian is not the constraint, but
itself a Dirac observable which moves a given transversal surface
(time level) in phase space \cite{Bianca1,Bianca2,Haj1}. Given data on a
transversal surface, this data will be moved onto another transversal
surface in a direction determined by the Hamiltonian. More precisely,
the ``time direction'' is provided by its Hamiltonian vector
field. The trouble in the present model is, obviously, that these
transversal surfaces may be intersected twice or not at all by the
classical orbit. The two intersections of a trajectory with given
orientation also come with two different evolution directions because
the trajectory is closed. These two opposite directions can,
certainly, not both be generated by one and the same physical
Hamiltonian, since it moves the transversal surface in only one
direction in phase space. Thus,
unlike in systems with global clocks,
we are required to perform a change of
Hamiltonian at the turning points of the clock. In order to evolve
from the surface determined by the non-global clock $q_1$, we need two
Hamiltonians, one of which generates evolution for $q_2^+$ and $p_2^+$
in the positive $q_1$-direction until the turning point of $q_1$ and the
second of which then generates evolution for $q_2^-$ and $p_2^-$ in
the opposite direction, away from the turning point. Let us explore
this in more detail.

Choosing $q_1$ as local time, we may factorize Eq.\ (\ref{cl-rov}) classically into a pair of constraints linear in $p_1$,
\ba\label{lincon}
C=\left(p_1+H(\tau)\right)\left(p_1-H(\tau)\right)=\tilde{C}_+\tilde{C}_-\q,\nn\\
\text{where}\q\q H(\tau)=\sqrt{M-\tau^2-p_2^2-q_2^2}\q.
\ea
The dynamical equations now read $\{\,\cdot\,,C\}=\tilde{C}_+\{\,\cdot\,,\tilde{C}_-\}+\tilde{C}_-\{\,\cdot\,,\tilde{C}_+\}$. Away from the turning points in $q_1$-time we have $H(\tau)>0$ and, therefore, $C=0$ implies that one of the following two possibilities (but not both simultaneously) is true
\ba\label{cond}
&&\tilde{C}_+=0 \, \Leftrightarrow \, \tilde{C}_-=2p_1<0 \, \Rightarrow \, q_1'=\{q_1,C\}=2p_1<0 \nn\\ &&\text{and} \q\{\,\cdot\,,C\}\propto -\{\,\cdot\,,\tilde{C}_+\}\q,
\ea
or,
\ba
&&\tilde{C}_-=0 \, \Leftrightarrow \, \tilde{C}_+=2p_1>0 \, \Rightarrow \, q_1'=\{q_1,C\}=2p_1>0 \nn\\ &&\text{and} \q \{\,\cdot\,,C\}\propto +\{\,\cdot\,,\tilde{C}_-\}\q.
\ea
Hence, on the set defined by $\tilde{C}_\pm=0$ we may use $\tilde{C}_\pm$ as evolution generator, but notice that the flow generated by $\tilde{C}_+$ is directed opposite to the one generated by $C$. Furthermore, since $\{q_1,\tilde{C}_\pm\}=1$, $\tilde{C}_\pm$ and, thus, $\pm H(\tau)$ are evolution generators for $q_2$ and $p_2$ in $q_1$-time. In particular, on the part of the constraint surface, where $\tilde{C}_+$ vanishes and, thus, may be used as an evolution generator (whose Hamiltonian vector field points in opposite direction to the one determined by $C$), we have $q_1'=2p_1<0$ and, therefore, the system governed by $C$ moves back in $q_1$-time. As a consequence, while $-H(\tau)$ generates evolution for $q_2$ and $p_2$ forward in $q_1$-time, $+H(\tau)$ does precisely the opposite. Note, moreover, that the two Hamiltonians $\pm H(\tau)$ are themselves relational Dirac observables which generate the physical equations of motion
\ba\label{eom}
\dot{q}_2&=&\pm\{q_2,H(\tau)\}=\mp\frac{p_2}{H(\tau)} \q,\\ \dot{p}_2&=&\pm\{p_2, H(\tau)\}=\pm\frac{q_2}{H(\tau)}\q,
\ea
where $\,\dot{}\,$ denotes a time-derivative w.r.t.\ $\tau$. As can be easily checked by using Eq.\ (\ref{corr}), the solution to the equations of motion generated by $+H(\tau)$ will reproduce classically $q_2^-$ and $p_2^-$, while the solutions to the equations generated by $-H(\tau)$ will provide $q_2^+$ and $p_2^+$. Consequently, in the solutions $q_2^+$ and $p_2^+$ in (\ref{corr}) $\tau$ must run forward, while for $q_2^-$ and $p_2^-$ it must run backwards. Care must be taken at the turning point of $q_1$-time, where $p_1=H=0$. Here we have to perform the change from $-H(\tau)$ to $+H(\tau)$, or vice versa.

The situation here is quite different from the case of the free
relativistic particle for two reasons. Firstly, in the constraint for
the free relativistic particle the two momenta come with opposite
signs and $t'=\{t,C_{\rm particle}\}=\{t,-p_t^2+p^2\}=-2p_t$, which
entails that forward evolution in the clock $t$ is only possible where
$p_t<0$. Secondly, $p_t$ is a Dirac observable which implies that in
this model no change of Hamiltonian needs to be performed. Neither of
the two issues occurs in the non-relativistic case, where $p_t$
appears linearly and the time direction is automatically given.

\subsection{The quantum theory}\label{rovqt}

The constraint (\ref{cl-rov}), when promoted to a quantum operator in the Dirac procedure, reads
\ba\label{quant-rov}
\hat{C}=\hat{p}_1^2+\hat{p}_2^2+\hat{q}_1^2+\hat{q}_2^2-M\q.
\ea
The quantization of this model is straightforward, since zero lies in the discrete part of the spectrum of the constraint.
\footnote{We assume here that $M$ is chosen to the extent that there
exist $n_1$, $n_2$ such that $2\hbar(n_1+n_2+1)-M=0$ and zero actually
lies in the spectrum of $\hat{C}$.}
The physical Hilbert space is, therefore, a subspace of the kinematical Hilbert space $L^2(\mathbb{R}^2,
dq_1dq_2)$, where the physical inner product is identical to the kinematical inner product and simply given by
\ba\label{inprod}
\langle\psi,\phi\rangle_{\rm phys} =\int ^{+\infty}_{-\infty}dq_1dq_2\, \bar{\psi}(q_1,q_2)\phi(q_1,q_2)\q.
\ea
The general form of the physical states is
\ba\label{q-sol}
\psi_{\rm phys}(q_1,q_2)=\sum_{n=0}^{M/(2\hbar)-1}c_n\psi_n(q_1)\psi_{M/(2\hbar)-n-1}(q_2)\q,
\ea
($c_n={\rm const}$) and $\psi_n$ denotes the $n$-th eigenstate of the 1D harmonic oscillator. The Dirac observables in Eq.\ (\ref{angobs}) are also straightforwardly quantized, since there is no factor ordering ambiguity involved. For some aspects discussed here see also \cite{Rovbook,Rovmod}.

The inner product may easily be obtained from group averaging, where
$P=\int^{2\pi}_0ds\,e^{-i\hat{C}s/\hbar}$, in fact, is a true
projector. The integration range of $2\pi$ is due to the constraint
being a ${\rm U}(1)$ generator and compatible with the classical
identification of states on the orbit which are related by integer
numbers of revolution.

\subsubsection{Timelessness}

A priori, there should be no time evolution and no IVP since there is
no true time. Indeed, in the $(q_1,q_2)$-representation,
Eq.\ (\ref{quant-rov}) provides an elliptic PDE; thus, there is no
well-defined IVP for this quantum model, but rather a boundary value
problem. The ``initial data'' characterizing the quantum solution is
in a sense timeless. This is also highlighted by the inner product
(\ref{inprod}) which integrates out both configuration variables and,
therefore, cannot be captured by the standard inner products based on
constant time slicings. The latter are usually related to the
existence of a well posed IVP.

In spite of this a priori timelessness, we can give a local
dynamical interpretation to the quantum theory in analogous fashion
to the classical theory.
(The relational evolution to be discussed here is only an emergent
local evolutionary interpretation of a timeless model. Consequently,
the apparent non-unitarity in the non-global clock evolution and
possible decoherence effects related to this are an artefact of this
emergent interpretation. The model itself is neither non-unitary nor
decohering since there is no true time. For that reason, the issue of
``quantum illnesses'', raised, for instance, in \cite{Klauder}, is not
directly applicable here.)
The ensuing differences between the classical
and quantum theory are, as usual, merely due to the quantum
uncertainties; however, these have more severe implications in the
absence of a global clock.

Again, we can give a meaning to orientation in the quantum theory,
namely via $\hat{L}_y$, which --- being a Dirac observable --- is a
well defined operator on $\mathcal{H}_{\rm phys}$. Its positive and
negative eigenspaces distinguish the orientation which also provides a
direction of evolution. By superimposing the two, a superposition of
evolution in both directions is, in principle, possible.

However, owing to the quantum uncertainties, the relational concept of
evolution seems to be only of an essentially semiclassical and
certainly local nature when dealing with non-global clocks and even in
this regime, quantum effects have severe consequences. When asking for
the value of, say, $q_2$ when a certain value of $q_1$
is realized, one faces the problem that due to the spread, parts of the state may
already be ``beyond their turning point'' in $q_1$. Classically, this
results in a quite meaningless complex-valued correlation between the
two configuration variables (just extend $|\tau|$ beyond $A$ in
Eq.\ (\ref{corr})) which merely indicates that the system never reaches
this point. In the quantum theory, the correlation of the two
variables, thus, loses meaning earlier than in the classical theory;
the larger the quantum uncertainties, i.e., the larger the spread of
the state, the earlier the concept of the relational correlation
breaks down. At a given value of the clock $q_1$ part of the system is
lost and an apparent non-unitarity shows up. This, certainly, also
applies to semiclassical states and, therefore, one cannot fully reach
the classical turning point without changing the clock
beforehand. Here, one cannot simply switch between, e.g., $q_2^+$ and
$q_2^-$, as one could classically, and as a consequence relational
Dirac observables only have a local meaning.

By the same token, the peak of a coherent physical state may follow a
classical trajectory exactly while expectation values computed in an
internal time Schr\"odinger regime can only do so locally. Such a
Schr\"odinger regime results from a local deparametrization and is
aimed at locally approximating the timeless physical state and the
information contained in it by locally scanning through it, thereby
introducing a notion of quantum evolution. The Schr\"odinger regime
for this model, is explicitly discussed in Sec.~\ref{schrodreg}
below. For this regime we need an (emergent) inner product based on
constant internal time slicings (for only the part of a coherent
physical state which either corresponds to, e.g., $q_2^+$ or $q_2^-$)
and such a slicing becomes troublesome near the classical turning
point of the chosen clock due to the apparent non-unitarity, and
eventually breaks down. Since the breakdown occurs earlier the greater
the quantum uncertainties, it becomes apparent that the internal time
Schr\"odinger evolution is only meaningful here in a semiclassical
regime. And even then, an expectation value trajectory cannot
completely reproduce the corresponding classical trajectory near the
turning point, even though the peak of the coherent state may do so.

Thus, while the question for what value, say, $q_2$ takes when
$q_1$ reads such and such seems to be meaningless if the state is
extremely quantum, it is meaningful for a semiclassical
state, where at least locally the expectation value evaluated in some
``emergent'' inner product based on constant $q_1$-slicings follows a
classical trajectory until close to the $q_1$-turning point. For
highly quantum states in systems without globally valid clock
variables, however, the standard concept of (locally unitary)
relational evolution seems to disappear in conjunction with the
standard notion of relational time. For a more detailed general
discussion of this feature we refer the interested reader to
\cite{EffTime1}. The analysis of the present toy model supplies
several general statements in \cite{EffTime1} with concrete examples.

\begin{center}
\begin{figure}[htbp!]
\includegraphics[width=7.4cm]{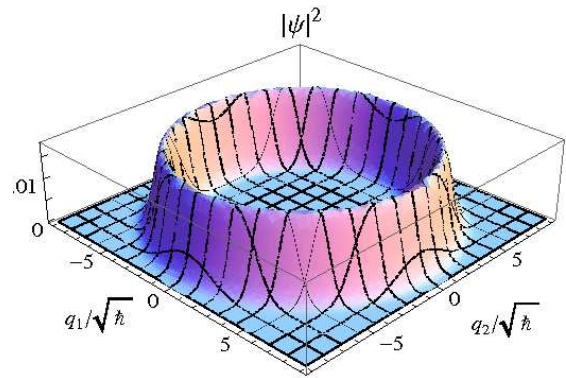}
\caption{\label{fig:cohstate} Square amplitude of a coherent
solution to the constraint~(\ref{quant-rov}), with $M=50\hbar$,
peaked about a circular configuration space trajectory.}
\end{figure}
\end{center}

Let us, therefore, investigate relational evolution via local
deparametrizations and how to reconstruct the information of the
physical state from it in the semiclassical regime. We refrain from
explicitly employing elliptic coherent physical states here, but in
order to visually facilitate the discussion we present an example of
such a state for this model in FIG.~\ref{fig:cohstate} (the interested
reader may find the recipe for the construction in this particular
model in \cite{ell}). In the semiclassical regime it is also
reasonable to consider only the solutions to Eq.\ (\ref{quant-rov})
which consist purely of positive or negative eigenstates of
$\hat{L}_y$ such that we avoid superposition of evolution in both
directions and are in a position to essentially repeat the same
procedure here as in the classical case.

We now have four methods for investigating the semiclassical regime:
the Dirac method, the reduction method,
\footnote{Since in the reduced phase space quantization the parameter
$\tau$ survives in the quantum theory, it is the only method in which
the timeless physical inner product (\ref{inprod}) may be used in
order to compute expectation values at a fixed value $\tau$ of $q_1$;
otherwise this physical inner product does not admit a sense of
evolution.}
evolution in an approximate
local Schr\"odinger regime or in the effective approach. This issue
has been partially analyzed in the reduction method (which in this
simple case turns out to be equivalent to the Dirac method) via group
quantization by Rovelli in \cite{Rovmod}, therefore, we will focus on
the local Schr\"odinger regime in Sec.\ \ref{schrodreg} and the
effective approach in Sec.\ \ref{effrovmod}, both truncated at order
$\hbar$, in this article. We will show that both yield equivalent
results.

\subsubsection{A local internal time Schr\"odinger regime}\label{schrodreg}

Since relational quantum evolution seems feasible for
semiclassical states, we would like to locally construct an internal time
Schr\"odinger regime which reproduces one branch of the timeless
physical state. This can be achieved by simply translating the local
relational
motion generated by the two Hamiltonians of Sec.\ \ref{linloc} into
the quantum theory and may, therefore, be understood as a local
deparametrization with a valid IVP.
To construct this Schr\"odinger regime, we require $q_1$ (or $q_2$)
--- in analogy to the parameter $\tau$ in (\ref{lincon}) ---
to appear as a parameter rather than as an operator, and the
corresponding states do not exist in the Hilbert space of the previous
subsection. We therefore need a new Hilbert space, with a new inner
product, in which we integrate only over $q_2$ at a fixed value of the
parameter $q_1$.  The Schr\"odinger regime using $q_2$ as an internal
clock naturally requires a further new Hilbert space, in which the
roles of $q_1$ and $q_2$ are reversed.  From the point of view of
standard Hilbert-space quantum theory, these Schr\"odinger regimes
thus constitute different quantizations of the classical theory: that
is, they are different and, in general, inequivalent quantum theories.
Even though solutions to the resulting Schr\"odinger equations violate
the quadratic quantum constraint with self-adjoint clock operator and
are not normalizable with (\ref{inprod}), they can be considered as
approximations to the original constrained problem by referring to the
analysis of~\cite{EffTime1} summarized in Sec.~\ref{sec:imtime}: the
WDW equation (\ref{quant-rov}) is, in fact, not violated if internal
time in this equation allows for an imaginary contribution.
Due to the apparent
non-unitarity alluded to above, the local Schr\"odinger regime will
break down on approach to the classical turning point of the clock,
and we can only hope to reconstruct/approximate the full physical
state by switching clocks and deparametrizations prior to the
breakdown of the respective clock. The results of this section will
become essential for understanding the effective approach, since the
local relational evolution of expectation values, i.e., of
fashionables, obtained in both approaches will prove to be
indistinguishable.

Choosing $\tilde{C}_+$ (and, thus, backward evolution in $q_1$) in Eq.\
(\ref{lincon}), standard quantization yields \ba\label{schrod}
i\hbar\frac{\partial}{\partial
q_1}\psi(q_1,q_2)&=&\hat{H}(\hat{q}_2,\hat{p}_2;
q_1)\psi(q_1,q_2)\nn\\
&=&\widehat{\sqrt{M-q_1^2-p_2^2-q_2^2}}\,\psi(q_1,q_2)\q,
\ea where $\hat{H}$ is defined via spectral decomposition. The
eigenfunctions of the latter are the harmonic oscillator eigenfunctions $\psi_n$ with
eigenvalues $H_n(q_1)=\sqrt{M-q_1^2-\hbar(2n+1)}$, and,
consequently, the operator is positive definite on the lower
energetic eigenstates, where the time dependent energy bound is
given by $M-q_1^2$.\footnote{This energy bound is related to the
upper limit of the sum in the physical state (\ref{q-sol}).} In
analogy with Eq.\ (\ref{lincon}) and in contrast to Eq.\ (\ref{quant-rov}),
$q_1$ has been reduced to a parameter here (see also
Sec.~\ref{sec:imtime} and~\cite{EffTime1} on this issue).

We solve Eq.\ (\ref{schrod}) in the standard way --- noting that
$[\hat{H}(\hat{q}_2,\hat{p}_2; q_1),\hat{H}(\hat{q}_2,\hat{p}_2;
q_1')]=0$ --- via \ba\label{linsol} \psi(q_2;q_1)&=&
e^{-\frac{i}{\hbar}\int_{{q_1}_0}^{q_1}dt\, \hat{H} (\hat{q}_2,
\hat{p}_2; t)} \psi_n(q_2;{q_1}_0) \nn\\
&=& e^{- \frac{i}{\hbar} E_n(q_1)}
\psi_n(q_2;{q_1}_0) \q, \ea where
\begin{widetext}
\ba\label{int}
\begin{split}
E_n(q_1) = &\int_{{q_1}_0}^{q_1}dt\,H_n(t) = \frac{1}{2} \left( q_1
\sqrt{M-q_1^2 - \hbar(2n+1)}- {q_1}_0 \sqrt{M-{q_1}_0^2 -
\hbar(2n+1)} \right.\\ &\left. + (M-\hbar(2n+1)) \left(\arctan\left(
\frac{q_1}{ \sqrt{M-q_1^2 - \hbar(2n+1)} } \right) - \arctan \left(
\frac{{q_1}_0}{\sqrt{M - {q_1}_0^2 - \hbar(2n+1)}} \right) \right)
\right) \q.
\end{split} \ea\end{widetext}

In order to better explore the semiclassical regime, let us attempt to construct coherent states. The
eigenstates of $\hat{H}$ are given by harmonic oscillator
eigenmodes; therefore,  it seems reasonable to make the following
standard ansatz for a coherent state\footnote{For convenience, we
shall henceforth employ bra and ket notation.} \ba\label{ansatz}
|z({q_1}_0)\rangle=e^{-|z|^2/2}e^{z\hat{a}^+}|0\rangle=e^{-|z|^2/2}\sum_{n\geq0}\frac{z^n}{\sqrt{n!}}|n\rangle\q,
\ea where $|n\rangle$ is the $n$-th eigenstate of the harmonic
oscillator, \ba \hat{a}=\frac{1}{2\hbar}(\hat{q}_2+i\hat{p}_2)\q \q \q
\hat{a}^+=\frac{1}{2\hbar}(\hat{q}_2-i\hat{p}_2) \ea are the usual
annihilation and creation operators of the harmonic oscillator, and
\ba z=\frac{{q_2}_0+i{p_2}_0}{\sqrt{2\hbar}}\q, \ea where ${q_2}_0$
and ${p_2}_0$ are the initial positions of the coherent state in
phase space.

The coherent state will be evolved with the (local) evolution
generator $\hat{H}$. Thus, \ba
|z(q_1)\rangle&=&e^{-\frac{i}{\hbar}\int_{{q_1}_0}^{q_1}dt\,\hat{H}(\hat{q}_2,\hat{p}_2;
t)}|z({q_1}_0)\rangle\nn\\&=&e^{-|z|^2/2}\sum_{n\geq0}\frac{z^n}{\sqrt{n!}}e^{-\frac{i}{\hbar}E_n(q_1)}|n\rangle\q.
\ea Furthermore, the states are normalized $\langle z(q_1)|
z(q_1)\rangle=1$ with respect to the standard inner product obtained by merely integrating out $q_2$.

The coherent states of the harmonic oscillator are dynamical
coherent states when evolved with the standard Hamiltonian. Here,
however, we are not evolving with the standard Hamiltonian and,
therefore, these states are only initially coherent states for our
local Schr\"odinger regime; the states are not eigenstates of $\hat{a}$
for all times, as can be seen from \ba
\hat{a}|z(q_1)\rangle=e^{-|z|^2/2}\sum_{n\geq0}\frac{z^{n+1}}{\sqrt{n!}}e^{-\frac{i}{\hbar}E_{n+1}(q_1)}|n\rangle\not\propto|z(q_1)\rangle\q, \ea and the form of Eq.\ (\ref{int}).

Expectation values as functions of $q_1$, i.e., {\it fashionables}, are now easily calculated
\begin{widetext}\ba\label{expec}
\begin{split}
\langle \hat{q}_2\rangle(q_1)&=\langle z(q_1)|\hat{q}_2|z(q_1)\rangle=\langle z(q_1)|\sqrt{\frac{\hbar}{2}}(\hat{a}+\hat{a}^+)|z(q_1)\rangle\\
&=e^{-|z|^2}\sum_{n\geq0}\frac{|z|^{2n}}{n!}\left({q_2}_0\cos\left(\frac{E_{n+1}(q_1)-E_n(q_1)}{\hbar}\right)+{p_2}_0\sin\left(\frac{E_{n+1}(q_1)-E_n(q_1)}{\hbar}\right)\right)\q,\\
\langle \hat{p}_2\rangle(q_1)&=\langle z(q_1)|\hat{p}_2|z(q_1)\rangle=\langle z(q_1)|\sqrt{\frac{\hbar}{2}}i(\hat{a}^+-\hat{a})|z(q_1)\rangle\\
&=e^{-|z|^2}\sum_{n\geq0}\frac{|z|^{2n}}{n!}\left({p_2}_0\cos\left(\frac{E_{n+1}(q_1)-E_n(q_1)}{\hbar}\right)-{q_2}_0\sin\left(\frac{E_{n+1}(q_1)-E_n(q_1)}{\hbar}\right)\right)\q.
\end{split}
\ea\end{widetext}
The explicit expressions for the fashionables of the moments $(\Delta q_2)^2, (\Delta p_2)^2$ and $\Delta(q_2p_2)$ as functions of $q_1$ are given in Appendix \ref{expmom}. The first two equations for $\langle \hat{q}_2\rangle$ and $\langle
\hat{p}_2\rangle$, certainly, reduce to the standard (classical) equations
of motion for the expectation values of the harmonic oscillator if
one replaces $E_n(q_1)$ with the usual eigenvalues of the harmonic
oscillator. Plots of these fashionables for a specific configuration
are provided in FIGs.\ \ref{comp} and \ref{momcomp} in Sec.\
\ref{effschrodcomp} below, combined with a comparison with the
effective results.

As an explicit example of the analysis summarized in
Sec.~\ref{sec:imtime}, let us discuss by how much we are
violating the WDW equation (\ref{quant-rov}) due to the fact that
$q_1$ is a real parameter here. To this end, we compute \ba
\langle z(q_1)|\hat{C}|z(q_1)\rangle&=&\langle z(q_1)|-\hbar^2\frac{\partial^2}{\partial q_1^2}-\hat{H}^2|z(q_1)\rangle\nn\\&=&\langle z(q_1)|i\hbar(\partial_{q_1}\hat{H})|z(q_1)\rangle\nn\\ &=&\langle z(q_1)|-i\hbar q_1(\hat{H})^{-1}|z(q_1)\rangle\nn\\
&=&-i\hbar
\,e^{-|z|^2}\sum_{n\geq0}\frac{|z|^{2n}}{n!}\frac{q_1}{\sqrt{M-q_1^2-\hbar(2n+1)}}\nn\\&=&i\hbar\frac{\partial}{\partial
q_1}\langle z(q_1)|\hat{H}|z(q_1)\rangle \q. \ea (The last line just
demonstrates the Ehrenfest theorem.) Linearizing in $\hbar$, one finds a violation of the quadratic constraint
\ba\label{viol}
 \langle z(q_1)|\hat{C}|z(q_1)\rangle=-\frac{i\hbar
q_1}{\sqrt{M-q_1^2}}+o(\hbar^2)\q. \ea To bridge this discrepancy,
we interpret $q_1$ as the operator (\ref{clockop}) with expectation
value having an imaginary contribution $-\frac{i\hbar}{2 \langle
\hat{p}_1 \rangle}$\ to order $\hbar$. Due to $(\Delta q_1)^2 = 0$,
one finds $\langle \hat{q}_1^2 \rangle = \langle \hat{q}_1 \rangle
^2 = q_1^2 - \frac{i \hbar q_1}{\langle \hat{p}_1 \rangle} +
O(\hbar^{\frac{3}{2}})$ and, with a little further calculation, it
turns out that the right hand side of Eq.\ (\ref{viol}) is precisely the
imaginary part of $\langle \hat{q}_1^2\rangle$. It may thus be
brought to the left hand side and interpreted as the imaginary
contribution to the expectation value of the clock $q_1$ in
Eq.\ (\ref{quant-rov}). Then, the quadratic constraint is satisfied to
this order and provides an explicit example for the general
derivation in~\cite{EffTime1}.

Similarly, to linear order in $\hbar$, Dirac observables of the quadratic constraint are, in general, constants of motion of the internal time Schr\"odinger regime only if the expectation value of the clock in the quadratic constraint is complex. For instance, the quantized Dirac observable $A$ of Eq.\ (\ref{aphi}) is given by $2\hat{A}=2(M-\hat{p}_2^2-\hat{q}_2^2)+\hat{C}$. The expectation value $ \langle z(q_1)|\hat{A}|z(q_1)\rangle$ is independent of $q_1$ only if the expectation value of $\hat{C}$ vanishes to semiclassical order since, employing Eq.\ (\ref{expec}) and the expressions in Appendix \ref{expmom}, one can easily convince oneself that the expectation value of $\hat{p}_2^2+\hat{q}_2^2$ is $q_1$-independent.

Finally, let us return to the issue of reconstructing the classical
trajectory or even the full physical state from the results in this
Schr\"odinger regime. The peak of a semiclassical state may follow a
classical trajectory almost precisely. However, the expectation
values can only follow the classical trajectory away from the
turning point. Due to the apparent non-unitarity of evolution in
$q_1$, the fashionables evaluated in the standard
Schr\"odinger type inner product with $q_1={\rm const}$ slicing must
become meaningless on approach to the turning point of $q_1$.
Heuristically, this may be understood by taking the expectation
value of the unit operator which may be interpreted as the
probability that the system is at some $q_2$ for a given value of
$q_1$. As long as the state is sufficiently semiclassical and the
peak is far enough away from the clock turning region, this
expectation value should always give 1. On approach to the turning
region, however, there will be parts of the state which are ``beyond
their turning point,'' precluding meaningful expectation values. Part
of the system is lost which implies that the expectation value of
the unit operator cannot give 1 anymore. Non-unitarity, therefore,
implies that the spread in $q_1$ cannot vanish close to the
classical turning point, since \ba
 (\Delta q_1)^2=\langle
q_1^2\rangle-\langle q_1\rangle^2=q_1^2\left(\langle
 \mathds{1}\rangle-\langle  \mathds{1}\rangle^2\right)\q,
\ea
which is non-vanishing when
the expectation value of the unit operator fails to be unity. This
provides an analogy in the internal time Schr\"odinger regime for why the
$q_1$-gauge, which among other conditions enforces $(\Delta
q_1)^2=0$, must break down on approach to the turning point of
$q_1$-time in the effective procedure.

As a consequence, in order to reproduce information from the full
physical state, we are forced to change from constant $q_1$- to
constant $q_2$-slicing, and thus from $q_1$- to $q_2$-time, prior to
the Schr\"odinger regime in $q_1$-time becoming invalid. Likewise, we
have to switch from $q_2$-time back to $q_1$-time again, prior to the
constant $q_2$-slicing subsequently becoming invalid and so on until
we have evolved once around the classical ellipse. In order for the
physical state to be reproduced, it then remains to be shown that the
expectation values of the quantum Dirac observables characterizing the
physical state, such as the three angular momentum operators
(\ref{angobs}), are invariant under the change of slicing.
Since the
two slicings used here are orthogonal to each other, one cannot
smoothly translate data from one slicing to the other. In fact, one
would expect jumps in the relational correlations when switching the
slicing. The necessary changes in slicing here are directly analogous to the
necessary changes between $q_1$- and $q_2$-gauge in the effective
approach in Sec.\ \ref{effrovmod} below and underline that
fashionables can only locally be made sense of.

\subsection{Effective procedure}\label{effrovmod}

To semiclassical order, the constraint (\ref{quant-rov}) translates into the following five constraints in the effective approach
\begin{widetext}\begin{align}\label{effrov}
C& = p_1^2+p_2^2+q_1^2+q_2^2+(\Delta p_1)^2+(\Delta p_2)^2+(\Delta q_1)^2+(\Delta q_2)^2-M= 0
\nonumber \\
C_{q_{1}}& = 2p_1\Delta(q_1p_1)+2p_2\Delta(q_1p_2)+2q_1(\Delta q_1)^2+2q_2\Delta(q_1q_2)+i\hbar p_1= 0 \nonumber \\
C_{p_1}& =
2p_1(\Delta p_1)^2+2p_2\Delta(p_1p_2)+2q_1\Delta(p_1q_1)+2q_2\Delta(p_1q_2)-i\hbar q_1= 0 \nonumber
\\
C_{q_2}& =2p_1\Delta(p_1q_2)+ 2p_2 \Delta(q_2p_2)+2q_1\Delta(q_1q_2)+2q_2(\Delta q_2)^2+i\hbar p_2 = 0 \nonumber \\
C_{p_2}& =2p_1\Delta(p_1p_2)+ 2p_2(\Delta p_2)^2+2q_1\Delta(q_1p_2)+2q_2\Delta(q_2p_2)-i\hbar q_2 = 0\q.
\end{align}\end{widetext}
Again, there are four linearly independent flows generated by these five constraints. The 14 dimensional Poisson manifold may, therefore, be reduced to five physical degrees of freedom. Dirac observables for this system are easily obtained by translating either Eqs.\ (\ref{aphi}) or (\ref{angobs}) into the quantum theory and taking their expectation values. For instance, the over-complete set (\ref{angobs}) now reads
\ba\label{effangobs}
L_x&=&\frac{1}{2}\left(p_1p_2+q_1q_2+\Delta(p_1p_2)+\Delta(q_1q_2)\right)\q,\nn\\
L_y&=&\frac{1}{2}\left(p_2q_1-p_1q_2+\Delta(q_1p_2)-\Delta(p_1q_2)\right)\q,\nn\\
L_z&=&\frac{1}{4}\left(p_1^2-p_2^2+q_1^2-q_2^2+(\Delta p_1)^2-(\Delta p_2)^2\right.\nn\\&&\left.+(\Delta q_1)^2-(\Delta q_2)^2\right)\q.
\ea
Owing to the definition of the effective Poisson bracket
(\ref{poisson}), also these effective observables satisfy the standard
angular momentum Poisson algebra. Moreover, due to Eq.\
(\ref{effobs}), the moments associated to these variables, $(\Delta
L_x)^2, (\Delta L_y)^2, (\Delta L_z)^2, \Delta(L_xL_y),
\Delta(L_xL_z)$ and $\Delta(L_yL_z)$, will provide the
$o(\hbar)$-observables. Since classically (\ref{angobs}) is an
over-complete set, also these nine observables here are, certainly,
over-complete. Indeed, to order $\hbar$, the constraint
(\ref{angobscon}) can easily be translated into four relations among
these effective observables, thus leaving us with the five physical
degrees of freedom to this order. The explicit expressions for the
moments, as well as the four relations among the full set of these
observables, are rather lengthy and not particularly illuminating. We,
therefore, abstain from showing them here. As regards relational
evolution, the angular momentum $L_y$ will provide an orientation to
the effective trajectories.

Due to the symmetry of the model in the indices $1$ and $2$, we will henceforth work with indices $i,j\in\{1,2\}$. In analogy to Eq.\ (\ref{ex2}), we impose the {\it $q_i$-gauge} (or the {\it Zeitgeist associated to} $q_i$)
\begin{align}\label{qi-gauge}
\phi_1&=(\Delta q_i)^2=0\nn\\
\phi_2&=\Delta(q_iq_j)=0\nn\\
\phi_3&=\Delta(q_ip_j)=0\q.
\end{align}

The remaining first class constraint with vanishing flow on the
variables $q_1$, $p_1$, $q_2$, $p_2$, $(\Delta q_j)^2$, $(\Delta
p_j)^2$, $\Delta (q_jp_j)$\ is directly proportional to $C_{q_i}$.
The solution of this constraint
\ba\label{Cqi} C_{q_i}\approx
2p_i\Delta(q_ip_i)+i\hbar p_i=0 \, \Rightarrow \,
\Delta(q_ip_i)=-\frac{i\hbar}{2}\q , \ea
again implies the saturation of the (generalized) uncertainty relation in $(q_i,p_i)$.

The Hamiltonian constraint reads
\ba\label{rovHam}
C_H=C+\alpha C_{p_i}+\beta C_{q_j}+\gamma C_{p_j}\q,
\ea
where on the gauge surface (\ref{qi-gauge})
\ba\label{rovcoeff}
\alpha=-\frac{1}{2p_i} \q,\q
\beta=\frac{q_j}{2p_i^2} \q\q\text{and}\q\q
  \gamma=\frac{p_j}{2p_i^2}\q.
  \ea
In addition to Eq.\ (\ref{Cqi}), we may solve $C_{p_i}$, $C_{q_j}$ and $C_{p_j}$ for the remaining non-physical moments
\ba\label{qimom}
(\Delta p_i)^2&=&\frac{p_j^2(\Delta p_j)^2+2q_jp_j\Delta(q_jp_j)+q_j^2(\Delta q_j)^2+i\hbar q_ip_i}{p_i^2}\q,\nn\\
\Delta(p_ip_j)&=&-\frac{2p_j(\Delta p_j)^2+2q_j\Delta(q_jp_j)-i\hbar q_j}{2p_i}\q,\nn\\
\Delta(q_jp_i)&=&-\frac{2q_j(\Delta q_j)^2+2p_j\Delta(q_jp_j)+i\hbar p_j}{2p_i}\q.
\ea
Making use of this, the relevant dynamical equations generated by $C_H$ simplify on the gauge surface (\ref{qi-gauge}) and are given by
\begin{widetext}\ba\label{simpeom1}
\dot{q}_i&=&\{q_i,C_H\}\approx2p_i-\frac{i\hbar q_i}{p_i^2}-2\,\frac{p_j^2(\Delta p_j)^2+2q_jp_j\Delta(q_jp_j)+q_j^2(\Delta q_j)^2}{p_i^3}\q,\nn\\
\dot{q}_j&=&\{q_j,C_H\}\approx2p_j+2\,\frac{q_j\Delta( q_jp_j)+p_j(\Delta p_j)^2}{p_i^2}\q,\nn\\
\dot{p}_i&=&\{p_i,C_H\}\approx-2q_i-\frac{i\hbar}{p_i}\q,\nn\\
\dot{p}_j&=&\{p_j,C_H\}\approx-2q_j-2\,\frac{q_j(\Delta q_j)^2+p_j\Delta(q_jp_j)}{p_i^2}\q,\nn\\
\dot{(\Delta q_j)^2}&=&\{(\Delta q_j)^2,C_H\}\approx4\,\frac{q_jp_j(\Delta q_j)^2+(p_i^2+p_j^2)\Delta(q_jp_j)}{p_i^2}\q,\nn\\
\dot{(\Delta p_j)^2}&=&\{(\Delta p_j)^2,C_H\}\approx-4\,\frac{q_jp_j(\Delta p_j)^2+(p_i^2+q_j^2)\Delta(q_jp_j)}{p_i^2}\q,\nn\\
\dot{\Delta(q_jp_j)}&=&\{\Delta(q_jp_j),C_H\}\approx2\,\frac{(p_i^2+p_j^2)(\Delta p_j)^2-(p_i^2+q_j^2)(\Delta q_j)^2}{p_i^2}\q.
\ea\end{widetext}
This set of coupled equations is rather complicated to solve analytically, but this is not necessary for our discussion here.

Although the dynamical equation for $p_i$ is not classical in nature, the $\hbar^0$-order part of $p_i$ must still vanish and $p_i\rightarrow o(\hbar)$ on approach to the turning point of $q_i$-time. In conjunction with Eq.\ (\ref{rovcoeff}), this implies that the {\it $q_i$-gauge} is inconsistent with the semiclassical truncation near the $q_i$ turning point as a result of the coefficients of the $o(\hbar)$-constraints becoming singular. In addition, we may note that due to the imaginary terms
\ba\label{cjbounce}
C_{q_j}\underset{\text{\tiny $p_i\rightarrow o(\hbar)$}}{\longrightarrow}2p_j\Delta(q_jp_j)+2q_j(\Delta q_j)^2+i\hbar p_j\approx 0\q,\nn\\
 C_{p_j}\underset{\text{\tiny $p_i\rightarrow o(\hbar)$}}{\longrightarrow}2p_j(\Delta p_j)^2+2q_j\Delta(q_jp_j)-i\hbar q_j\approx 0\q,
\ea
combined with the assumption of real valued $q_j$, $p_j$, $(\Delta q_j)^2$, $(\Delta p_j)^2$ and $\Delta(q_jp_j)$ implies a violation of $C_{q_j}$ and $C_{p_j}$ to semiclassical order at the turning point. But as previously discussed, this collapse of the $q_i$-gauge does not come unexpected, being related to a non-global clock.

In analogy to Eq.\ (\ref{pt}), combining $C_{p_i}$, $C_{q_j}$, $C_{p_j}$
and $C$ yields a further constraint proportional to $C_H$, which on
the constraint surface in the $q_i$-gauge reads \ba\label{pi4}
&p_i^4+\left(p_j^2+q_i^2+q_j^2-M+(\Delta p_j)^2+(\Delta
q_j)^2\right)p_i^2+i\hbar q_ip_i\nn\\&+p_j^2(\Delta
p_j)^2+2q_jp_j\Delta(q_jp_j)+q_j^2(\Delta q_j)^2=0\q. \ea We may use
this remaining constraint to discuss the imaginary contributions to
the variables we have chosen, as a result of the $i\hbar$-term in
Eq.\ (\ref{pi4}). For brevity, let us only state the (expected) result
here: in complete accordance with the general result of
Sec.~\ref{sec:imtime} and \cite{EffTime1}, it is inconsistent with the equations of
motion and the constraints in $q_i$-gauge to keep a real-valued
clock $q_i$ and to push the imaginary contributions to its conjugate
momentum $p_i$, while having real-valued variables associated to the
pair $(q_j,p_j)$. Instead, it is consistent to have both the
variables associated to the pair $(q_j,p_j)$ and $p_i$ real-valued,
as well as a complex clock with the standard imaginary contribution,
inherent to non-global clocks, \ba\label{imqi}
\Im[q_i]=-\frac{\hbar}{2{p_i}}\q. \ea A proof of this may be
found in Appendix \ref{rovimag}. Note, however, that it is also
possible that both $q_i$ and $p_i$ are complex simultaneously.

\subsubsection{Local evolution and comparison to the internal time Schr\"odinger regime}\label{effschrodcomp}

Since we are interested in a comparison of the effective approach
with the internal time Schr\"odinger regime, we solve the system of effective
equations (\ref{simpeom1}) numerically in the $q_1$-gauge and
compare the results with the ones obtained via Eq.~(\ref{expec}) and the
expressions in Appendix~\ref{expmom}. FIG.~\ref{comp} shows a
comparison of the classical, effective and Schr\"odinger regime
results for the configuration space ellipse for a specific
configuration, whose initial data is given in the caption of the
figure. These curves depict the relational Dirac observable
$q_2(q_1)$
in the classical case, the relationship $q_2(\Re[q_1])$ of expectation
values in the effective framework, and $\langle\hat{q}_2\rangle(q_1)$
from Eq.\ (\ref{expec}) in the Schr\"odinger regime where $q_1$ is a
real parameter.%
\footnote{Note that in the effective framework we evolve with respect
to the real part of $q_1$, in accordance with the discussion in
Sec.~\ref{genimagcomment} and the one concerning FIG.~\ref{reqiimqi}
below. For the effective curve, the axis label $q_1$, therefore,
actually refers to $\Re[q_1]$.}

The three curves are indistinguishable where
valid. Notice that the Schr\"odinger regime breaks down somewhat
earlier than the curve of effective expectation values, due to the
square roots in Eq.\ (\ref{int}) which become imaginary for larger values
of $q_1$ and states with higher $n$. The breakdown of the
correlations from the effective and Schr\"odinger regime emphasizes
the merely local nature of the fashionables. In
spite of this, the plot also demonstrates that, at least locally,
one can reconstruct a semiclassical orbit from the effective framework and the
Schr\"odinger regime.

For further --- non-trivial --- comparison of the Schr\"odinger regime
and the effective framework, we compare the relational evolution of
their respective moments, related to the pair $(q_2,p_2)$, in
$q_1$-time in FIG.\ \ref{momcomp} for the same initial data as
previously. The curves demonstrate that the relational evolution of
the moments of both approaches agrees perfectly to this order. Since
these relational moments are truly quantum in nature, this agreement
provides interesting non-trivial evidence for the equivalence of
these two different approaches to semiclassical order. It is also
found numerically, that the discrepancies between the results of the
two approaches are of $o(\hbar^2)$ or even smaller. Again, due to the
square roots in Eq.\ (\ref{int}), the Schr\"odinger regime in constant
$q_1$-slicing breaks down earlier than the $q_1$-Zeitgeist in the
effective framework. The eventual divergence of the effective moments
in FIG.\ \ref{momcomp} demonstrates the breakdown of the latter.

Finally, as regards the effective evolution in $q_1$, FIG.\
\ref{reqiimqi} shows the behavior of the real and imaginary parts of
$q_1$ with respect to the gauge parameter $s$ of (\ref{rovHam}) for
the same effective configuration. Away from the breakdown of the
$q_1$-Zeitgeist, signified by the divergence in both the real and
imaginary parts of $q_1$, the real part of $q_1$ is clearly
monotonic along the flow and may thus be used as a relational clock.
On the contrary, the imaginary contribution to $q_1$ does {\it not}
behave monotonically and, consequently, is not a useful clock here,
underlining the general argument for employing only the real part of
a clock for evolution, as advocated in Sec.~\ref{genimagcomment}.
Note that the real part of $q_1$ runs backwards in the flow
parameter, since we have chosen the initial data equivalently to the
Schr\"odinger regime, where for (\ref{schrod}) we had chosen the
quantization of $\tilde{C}_+$ in Eq.\ (\ref{lincon}), which generates
backwards evolution in $q_1$.
\begin{center}
\begin{figure*}[htbp!]
\includegraphics[scale=.45]{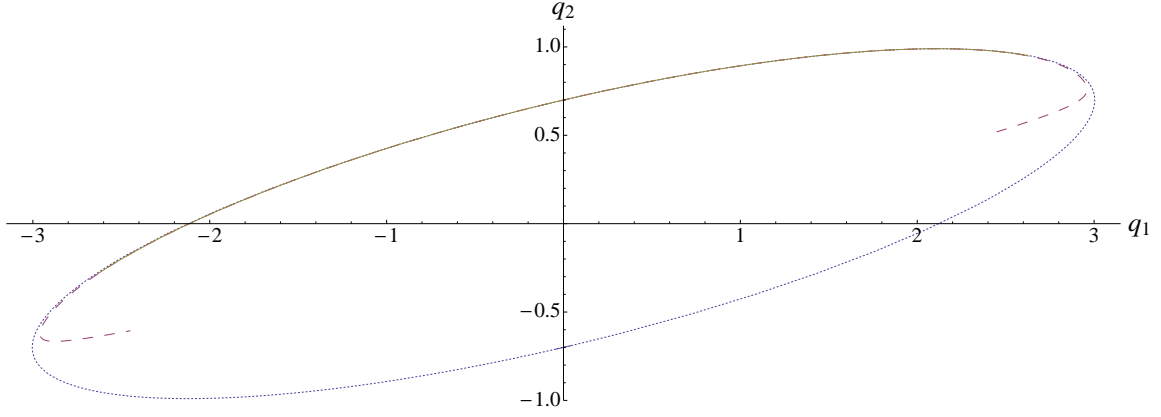}
    \caption{\label{comp}{\small Pictorial comparison of the classical
 relational Dirac observable $q_2(q_1)$ (full ellipse, blue curve) with the
 quantities $q_2(\Re[q_1])$ calculated in the effective theory using
 the $q_1$-gauge (violet dashed curve) and
 $\langle\hat{q}_2\rangle(q_1)$ in the Schr\"odinger regime
 (yellow solid curve).
Where valid, the three curves agree perfectly. The
 Schr\"odinger regime breaks down earlier than the $q_1$-gauge of the
 effective framework. The initial data match in all three
 cases: we chose ${q_2}_0=0.7$ and ${p_2}_0=-0.7$ for the
 Schr\"odinger regime, which via Eq.\ (\ref{eqexpmom}) yields $(\Delta
 q_2)^2(q_1=0)=(\Delta p_2)^2(q_1=0)=\frac{\hbar}{2}$ and
 $\Delta(q_2p_2)(q_1=0)=0$. We have set $M=10$ and, to amplify
 effects, $\hbar=0.03$. We take these values as initial data for the
 effective formalism as well, and, using Eq.\ (\ref{pi4}), we
 determine the initial value for ${p_1}_0=-2.998$ (the minus sign is
 necessary here, since in Eq.\ (\ref{schrod}) we quantized
 $\tilde{C}_+$ which evolves backwards in $q_1$). In the effective
 picture, due to the imaginary contribution to $q_1$ in the
 $q_1$-gauge, we have set the initial value of the clock to
 $q_1=-\frac{i\hbar}{2{p_1}_0}$, but employ $\Re[q_1]$ as relational
 clock (see also FIG.\ \ref{reqiimqi}). The initial data for the
 classical curve has been chosen accordingly. As regards the axis labels: for the effective
 framework both $q_1$ and $q_2$ refer to the expectation values of the
 corresponding operators (for $q_1$ the real part), while for the
 internal time Schr\"odinger regime $q_2$ refers to the expectation
 value from Eq.\ (\ref{expec}) and $q_1$ is the real evolution
 parameter.  }}
\end{figure*}
\end{center}
\noindent
\begin{center}
\begin{figure*}[htbp!]$
\begin{array}{ccc}
\includegraphics[scale=.36]{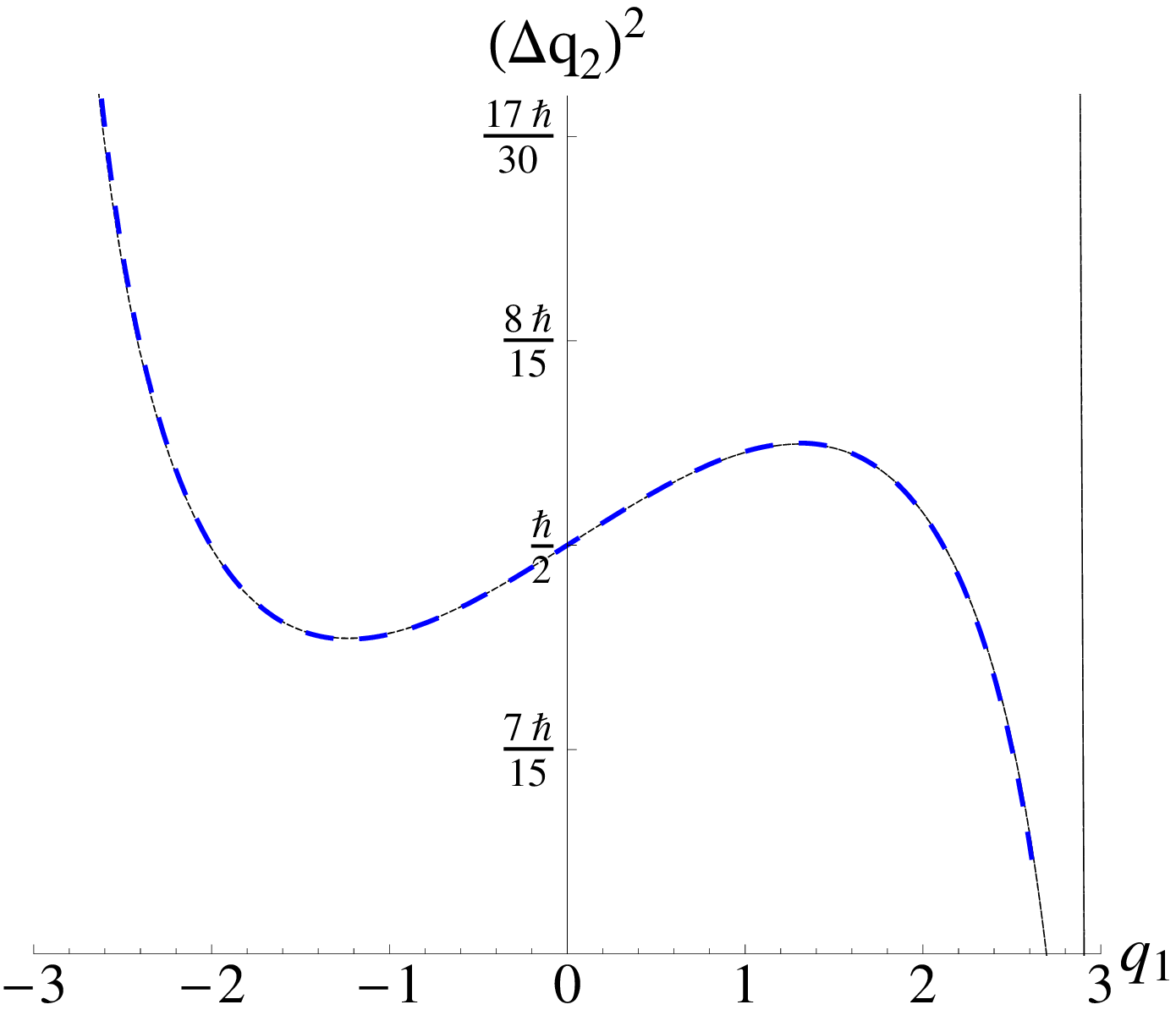} &\includegraphics[scale=.36]{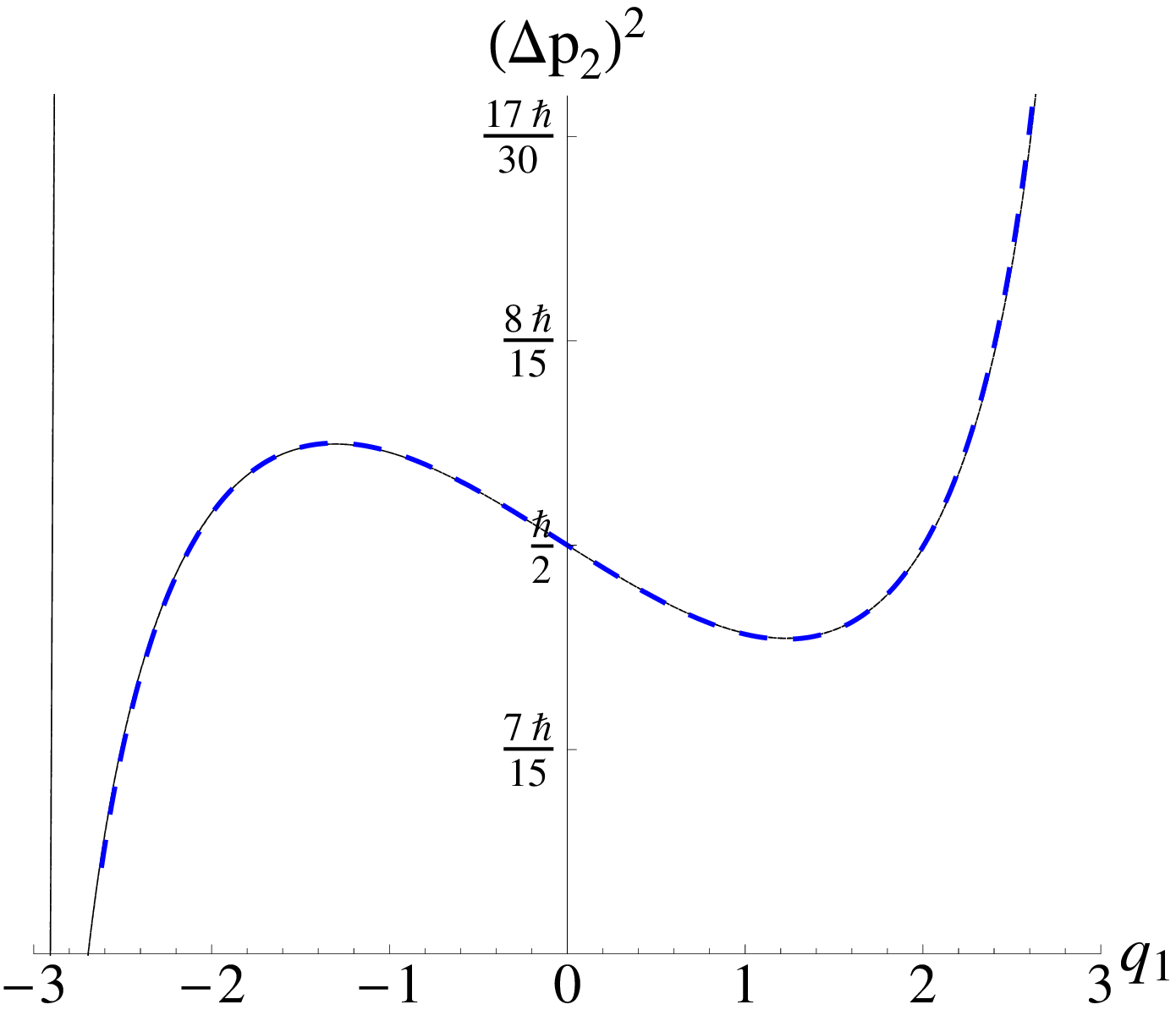} &\includegraphics[scale=.36]{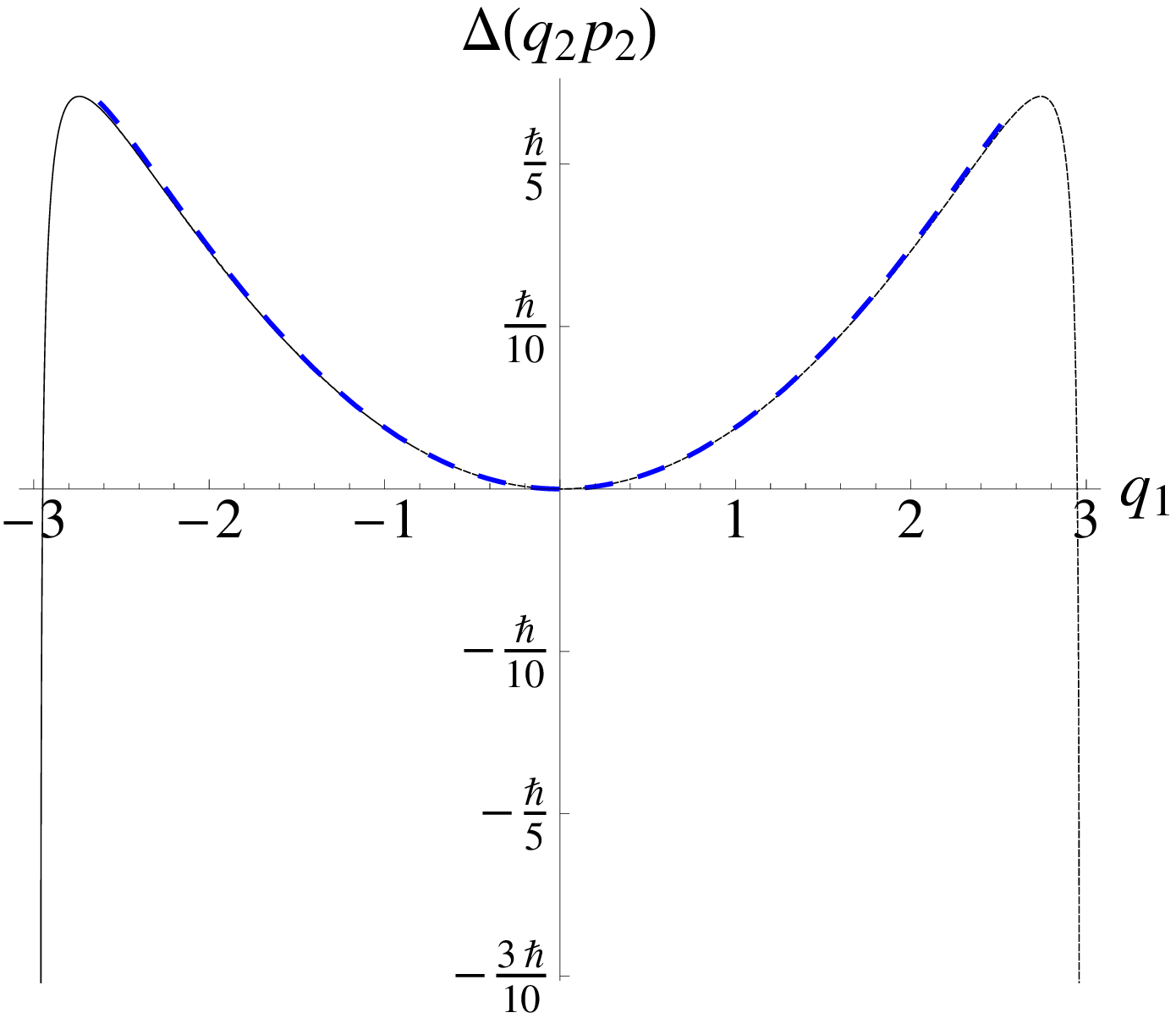}   \\
a)&b)&c)
\end{array}$
    \caption{\label{momcomp}{\small Comparison of the effective (black dotted curves) and internal time Schr\"odinger regime results (blue dashed curves) for the
fashionables in $q_1$-time
associated to moments: a) $(\Delta q_2)^2(q_1)$, b) $(\Delta p_2)^2(q_1)$ and c) $\Delta(q_2p_2)(q_1)$. The curves agree perfectly to order $\hbar$. As explained in the main text, the Schr\"odinger regime breaks down earlier than the $q_1$-gauge of the effective framework. The breakdown of the latter is clearly demonstrated by the divergence of the effective moments near $|q_1|= 3$. The initial data is identical to the one for FIG.\ \ref{comp}.}}
\end{figure*}
\end{center}
\noindent
\begin{center}
\begin{figure*}[htbp!]$
\begin{array}{cc}
\includegraphics[scale=.36]{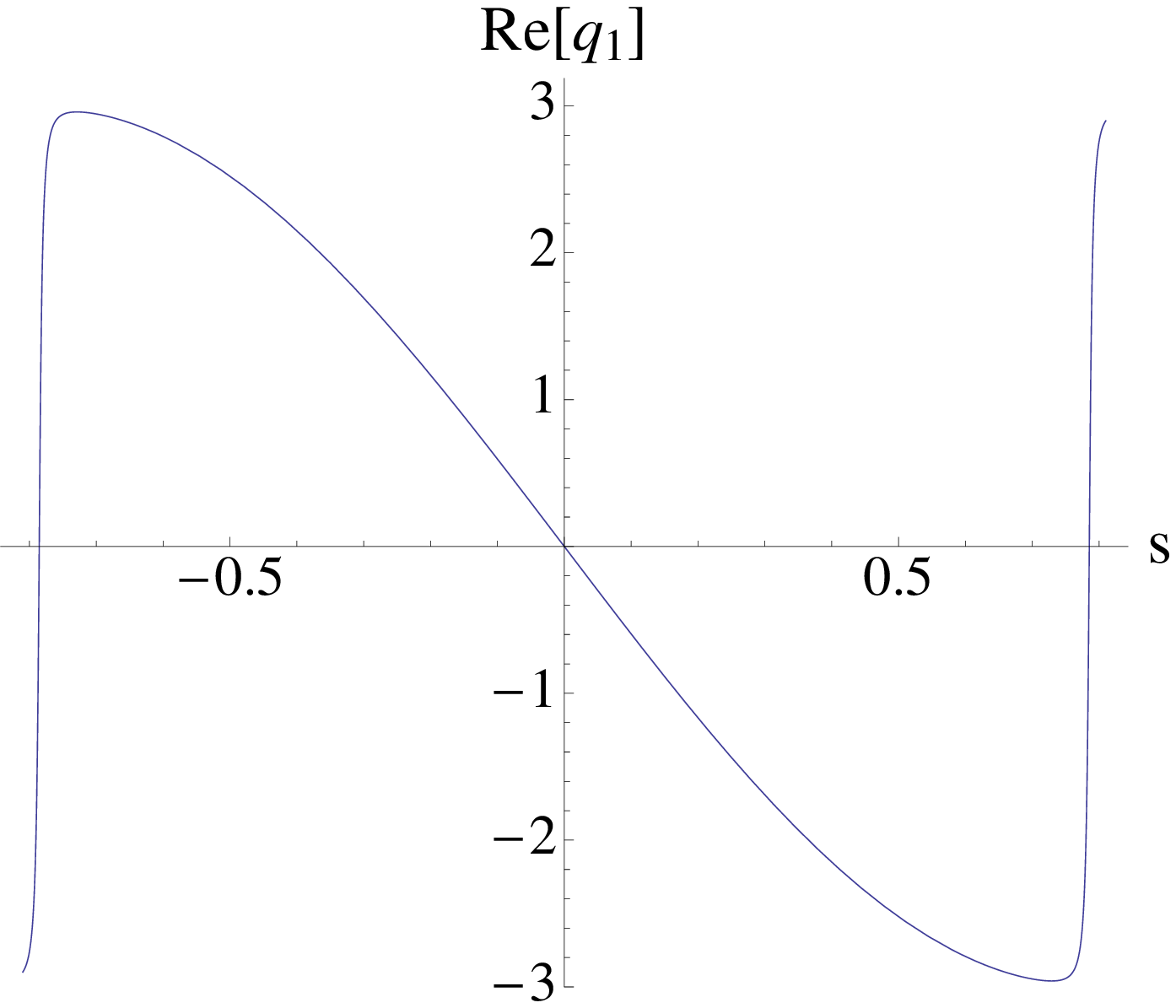}  &\includegraphics[scale=.36]{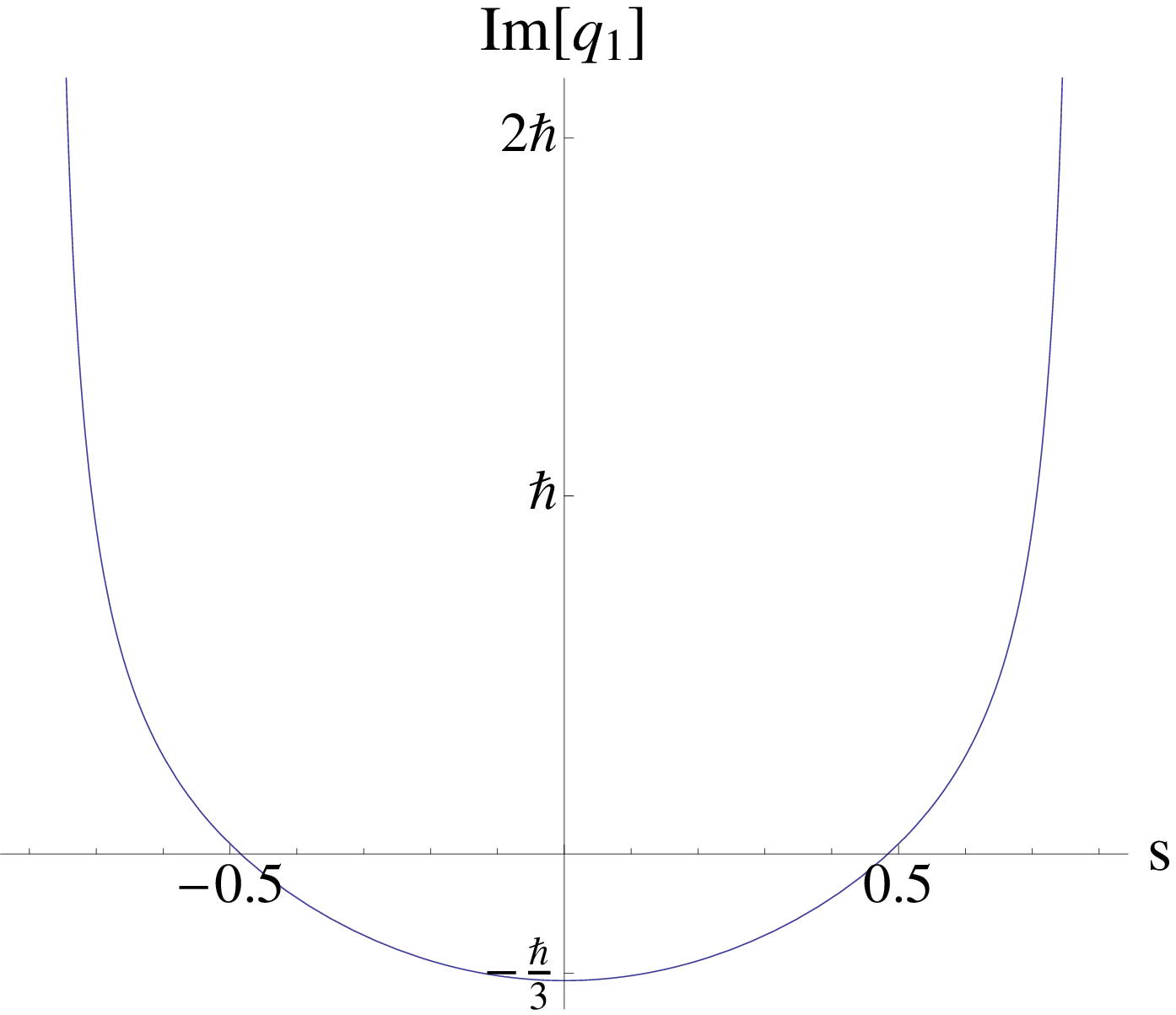} \\
a)&b)
\end{array}$
    \caption{\label{reqiimqi}{\small Behavior of a) the real and b) the imaginary part of the local clock $q_1$ with respect to the gauge parameter $s$ of $C_H$ for the effective configuration with initial data as given in the caption of FIG.\ \ref{comp}. Clearly, while $\Re[q_1]$ is monotonic along the flow of $C_H$ (as long as the $q_1$-gauge is valid) and, therefore, constitutes a useful local clock, $\Im[q_1]$ does not provide a suitable clock here. The divergence of both near $|s|= 0.79$ signifies the breakdown of the $q_1$-gauge. }}
\end{figure*}
\end{center}

\noindent

\subsubsection{Changing time and gauge transformations}\label{rovgt}

Just as in the model of Sec.~\ref{lt} we can use flows generated
by the constraint functions to perform  a gauge transformation from
$q_i$-gauge to $q_j$-gauge. In this way, we can evolve
the system through an entire closed orbit by switching the role of
time back and forth between the two configuration space variables.
In this section we calculate the corresponding gauge
transformations; evolution through the entire orbit is explored in
the following section.

Following the steps used in Sec.~\ref{sec:gaugec_lt} to construct the
gauge transformation between different Zeitgeister,
we find the effect of the flows on the other variables to be given by
\begin{eqnarray*}
X_{G_1}(q_i) = \frac{p_i q_i-2p_jq_j}{2p_ip_j^2} \quad&,&\quad X_{G_2}(q_i) = -\frac{1}{p_i}\\
X_{G_1}(p_i)=\frac{p_i}{2p_j^2} \quad&,&\quad X_{G_2}(p_i)=0
\end{eqnarray*}
\begin{eqnarray*}
X_{G_1}(q_j)=\frac{q_j}{2p_j^2} \quad&,&\quad X_{G_2}(q_j)=\frac{1}{p_j}\\
X_{G_1}(p_j)=-\frac{1}{2p_j} \quad&,&\quad X_{G_2}(p_j)=0\\
X_{G_1}\left((\Delta q_i)^2\right) =
-\frac{p_i^2}{p_j^2} \quad&,&\quad X_{G_2}\left((\Delta q_i)^2\right) = 0 \\
X_{G_1}\left( (\Delta p_i)^2 \right) =
\frac{q_i(2p_j q_j - p_i q_i)}{p_ip_j^2} \quad&,&\quad X_{G_2}\left( (\Delta p_i)^2
\right) = \frac{2q_i}{p_i}\\
X_{G_1}\left( \Delta(q_ip_i)
\right) =\frac{p_i q_i - p_j q_j}{p_j^2} \quad&,&\quad X_{G_2}\left( \Delta(q_ip_i) \right) =-1\q.
\end{eqnarray*}
This time the derivatives along the flow are not constant; however,
they depend only on expectation values. For the variables of interest, all of the derivatives in an
expansion of the flow actions of $\alpha_{G_1}$ and $\alpha_{G_2}$ via Eq.\ (\ref{gaugeaction}) are functions of expectation values only and are thus of
classical order $\hbar^0$. Second and higher derivative terms are
suppressed by second and higher powers of the flow parameter,
which is of order $\hbar$, since it goes from zero to $-(\Delta
q_j)^2_0$\ or $-\left( \Delta(q_jp_j)_0 +\frac{i\hbar}{2} \right)$.
Therefore, to order $\hbar$\ it is sufficient to take the terms up to first order in
derivatives in the flow expansion via Eq.\ (\ref{gaugeaction}) of $\alpha^s_G(f)(x_0):=\alpha_{G_2}^{- (\Delta(q_jp_j)_0+i\hbar/2)}\circ\alpha_{G_1}^{-(\Delta q_j)_0^2}(f)(x_0)$, i.e.\ we have $\alpha^s_G(f)(x_0)= f_0 - \left( X_{G_1}(f)
\right)_0 (\Delta q_j)_0^2-\left(X_{G_2}(f)\right)_0 \left(\Delta(q_jp_j)_0+i\hbar/2\right)+o(\hbar^2)$. The transformation to order $\hbar$\ thus obtained
has the form\footnote{In fact, the flows $\alpha_{G_1}$\ and $\alpha_{G_2}$\ have a relatively
simple form and can also be integrated analytically, yielding
identical results to order $\hbar$.} (dropping the $\alpha$'s for
brevity)
\begin{widetext}\begin{eqnarray}
\begin{split}
(\Delta q_i)^2 &= \frac{(p_i)_0^2 (\Delta q_j)^2_0}{(p_j)_0^2} \\
(\Delta p_i)^2 &= \frac{(p_j)_0^4 (\Delta p_j)^2_0 + \left(2
(p_j)_0 (q_j)_0 - 2 (p_i)_0 (q_i)_0 \right)
\Delta(q_jp_j)_0+(\Delta
q_j)^2_0 ((p_i)_0 (q_i)_0-(p_j)_0
(q_j)_0)^2}{(p_i)_0^2 (p_j)_0^2} \\
\Delta(q_ip_i) &= \frac{(\Delta q_j)^2_0 ((p_j)_0
(q_j)_0-(p_i)_0 (q_i)_0)}{(p_j)_0^2}+\Delta(q_jp_j)_0
 \\ q_i &= (q_i)_0 + \frac{i\hbar (p_j)_0^2+(\Delta
q_j)^2_0 (2 (p_j)_0 (q_j)_0-(p_i)_0 (q_i)_0)+2
(p_j)_0^2
\Delta(q_jp_j)_0}{2 (p_i)_0 (p_j)_0^2}  \\
p_i &= (p_i)_0 \left(1-\frac{(\Delta q_j)^2_0}{2
(p_j)_0^2}\right)  \\ q_j &= (q_j)_0-\frac{i\hbar
(p_j)_0+2 (p_j)_0 \Delta(q_jp_j)_0+(q_j)_0 (\Delta
q_j)^2_0}{2 (p_j)_0^2}\\ p_j &= (p_j)_0
\left(1+\frac{(\Delta q_j)^2_0}{2
(p_j)_0^2}\right)\q.\label{eq:rovelli_gtransf}
\end{split}
\end{eqnarray}\end{widetext}
These are the explicit expressions for the free variables of
$q_j$-gauge in terms of the free variables of the
$q_i$-gauge\footnote{Actually not all these variables are free, as $p_i$\ can
be eliminated in the $q_i$-gauge with the use of $C$. We display its
transformation for convenience, since we are using $(p_i)_0$\ and
$(p_j)_0$\ within the above expressions.}. We note that just as
in the model of Sec.~\ref{lt}, this transformation precisely
cancels out the imaginary part (\ref{imqi}) of the time variable
$q_i$, rendering it real in the $q_j$-gauge, while simultaneously
giving $q_j$ precisely the correct imaginary contribution expected
of a time variable, if its initial value $(q_j)_0$\ is real. See
Appendix~\ref{app:pos_rov} for the discussion of positivity of the
gauge transformed state.

\subsubsection{Evolution around the closed orbit}\label{fullorbit}

Finally, let us perform a sequence of gauge and clock changes until we
fully evolve around the configuration space ellipse. As a result of
the breakdown of the $q_i$-Zeitgeist near the $q_i$ turning point, the
changes between the gauges and $q_1$- and $q_2$-time are required. The
breakdown of the gauges and the necessity of gauge changes are
precisely the effective analog of the apparent non-unitarity in the internal time
Schr\"odinger regime in Sec.\ \ref{schrodreg} and the ensuing
breakdown of the constant $q_i$-slicing and the resulting obligation
to change the slicing and the clock. The jumps between the
correlations which one would obtain when changing slicing in the
Schr\"odinger regime translate into the jumps in correlations
encountered in the gauge changes in Sec.\ \ref{rovgt}. (As emphasized
in Sec.\ \ref{rovqt},
quantum relational
observables valid for all classically allowed values of the chosen
clock, therefore, do not exist.)

Apart from such quantum effects, the relational procedure works just
as in the classical case. Due to the relativistic nature of the
constraint, we are required to provide a time direction in which to
evolve, since imposing only the relational initial data
$q_j,p_j,(\Delta q_j)^2,(\Delta p_j)^2$ and $\Delta(q_jp_j)$ at a
fixed value of $q_i$ does not completely solve the IVP. As in the
classical model and the Dirac approach, providing $L_y$, being the
angular momentum, results in giving the required orientation to
evolution. Using Eq.\ (\ref{qimom}) and the expression for $C$ in Eq.\
(\ref{effrov}), $p_i$ is determined up to sign when providing the
relational initial data. The expression for $L_y$ in Eq.\
(\ref{effangobs}) then implies that additionally providing $L_y$ is
equivalent to imposing the sign of $p_i$. Note that, unlike in the
full quantum theory briefly described in Sec.\ \ref{rovqt} and in
complete accordance with semiclassicality, there cannot be a
superposition of evolution in the two opposite orientations in the
effective framework truncated at order $\hbar$.

Given this data, the system (\ref{simpeom1}) can be solved (at least
numerically) and we can relate the variables associated to $(q_j,p_j)$
to the clock $q_i$ and evolve forward in the $q_i$-Zeitgeist in the
given direction of evolution. Prior to the breakdown of this gauge, we
translate to $q_j$-gauge and, thus, to a different set of
fashionables. Then, just before the subsequent breakdown of the
$q_j$-Zeitgeist, we return to $q_i$-gauge and so forth, until fully
evolving around the ellipse. In this way, the initial data is
transported around the orbit independently of the gauge parameters,
although employing different gauges and even different sets of
fashionables in the different gauges
(see also Sec.~\ref{gaugec}).

It should be noted that, just as in Secs.\ \ref{linloc} and
\ref{schrodreg}, we could generate our physical evolution by a
physical Hamiltonian, which would be obtained by simply linearizing
Eq.\ (\ref{pi4}) in $p_i$. The resulting relational evolution would,
obviously, be identical to the one generated by $C_H$. Since the
system generated by $C_H$ is somewhat simpler to handle, we focus on
Eq.\ (\ref{simpeom1}) here. Notice also that the effective formalism
reintroduces a gauge parameter even in the quantum theory (the
parameter along the flow of $C_H$). Recall from the introduction that
this gauge parameter simplifies a patching solution to the global
problem of time in the classical case and that its absence in the
quantum theory is one of the reasons for the difficulties occurring
there. Nevertheless, the gauge parameter here is related to $C_H$
which depends on the $q_i$-Zeitgeist. When changing gauge, one
necessarily obtains a separate gauge parameter and since the gauges
break down prior to the classical turning points of the clocks, one
cannot use the effective gauge parameters in the classical way to
overcome the global problem of time.

As regards reconstructing the full coherent physical state from the
Schr\"odinger regime, it was noted in Sec.\ \ref{schrodreg} that one
would need to explore whether the quantum versions of the Dirac
observables (\ref{aphi}) or (\ref{angobs}), which characterize the
physical state, are constants of motion in a given constant
$q_i$-slicing and whether they are invariant under a change of
slicing. In the present effective case, the answer to this problem is
obvious: since the characterizing observables, for instance,
(\ref{effangobs}) and their moments are complete Dirac observables of
the effective system, they are invariant under the action of the
constraints (\ref{effrov}) and, therefore, also under the gauge
changes of Sec.\ \ref{rovgt}. Consequently, they are constant for the
given orbit which we are analyzing and, as a result, we are always
probing one and the same physical state. Since the internal time Schr\"odinger
regime corresponds to the effective framework to this order, we
conjecture that also in the Schr\"odinger regime, these observables
remain invariant, although this is more difficult to prove
explicitly.

\begin{center}
\begin{figure*}[htbp!]$
\begin{array}{cc}
\includegraphics[scale=.4]{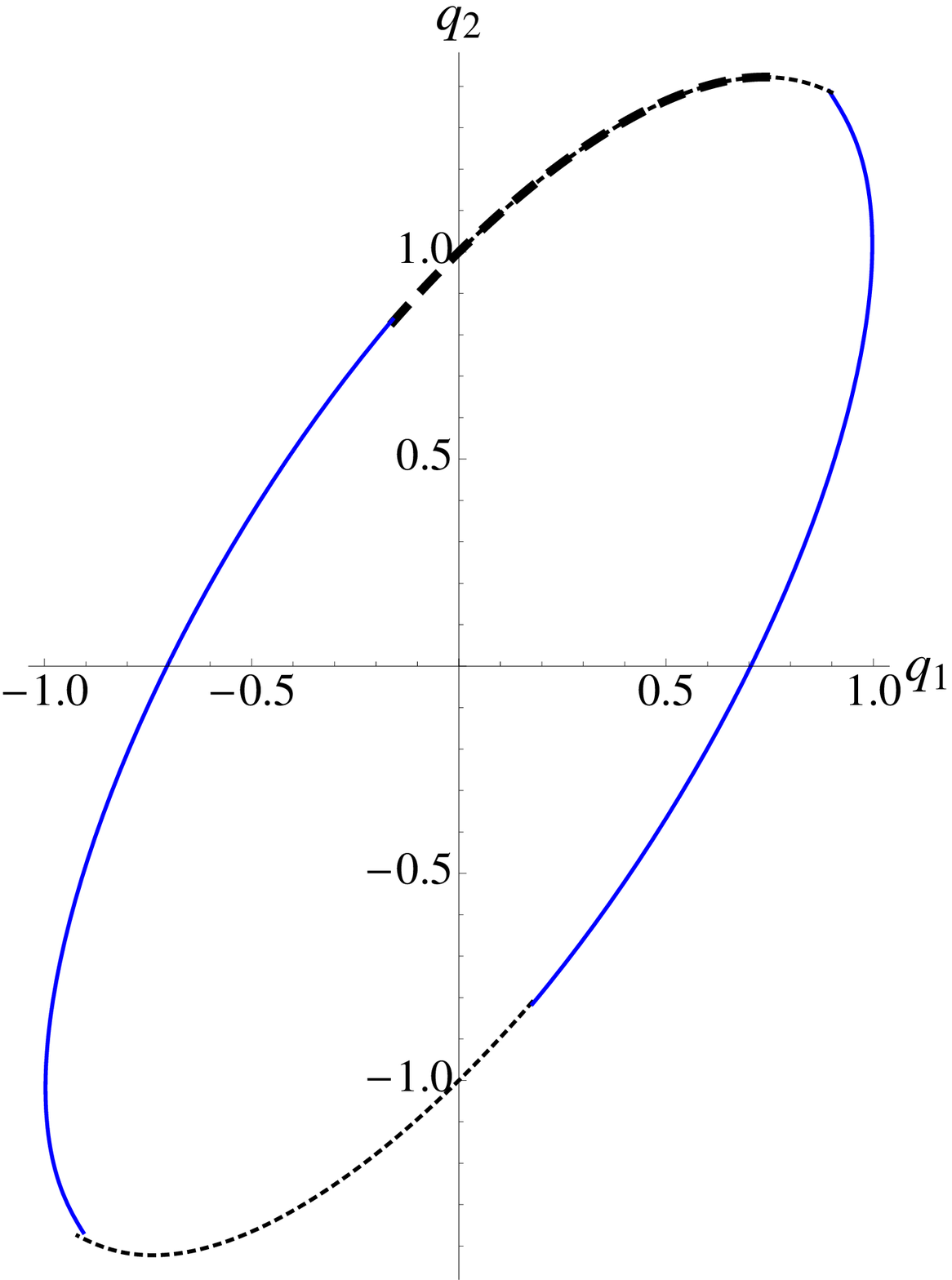} &\includegraphics[scale=.4]{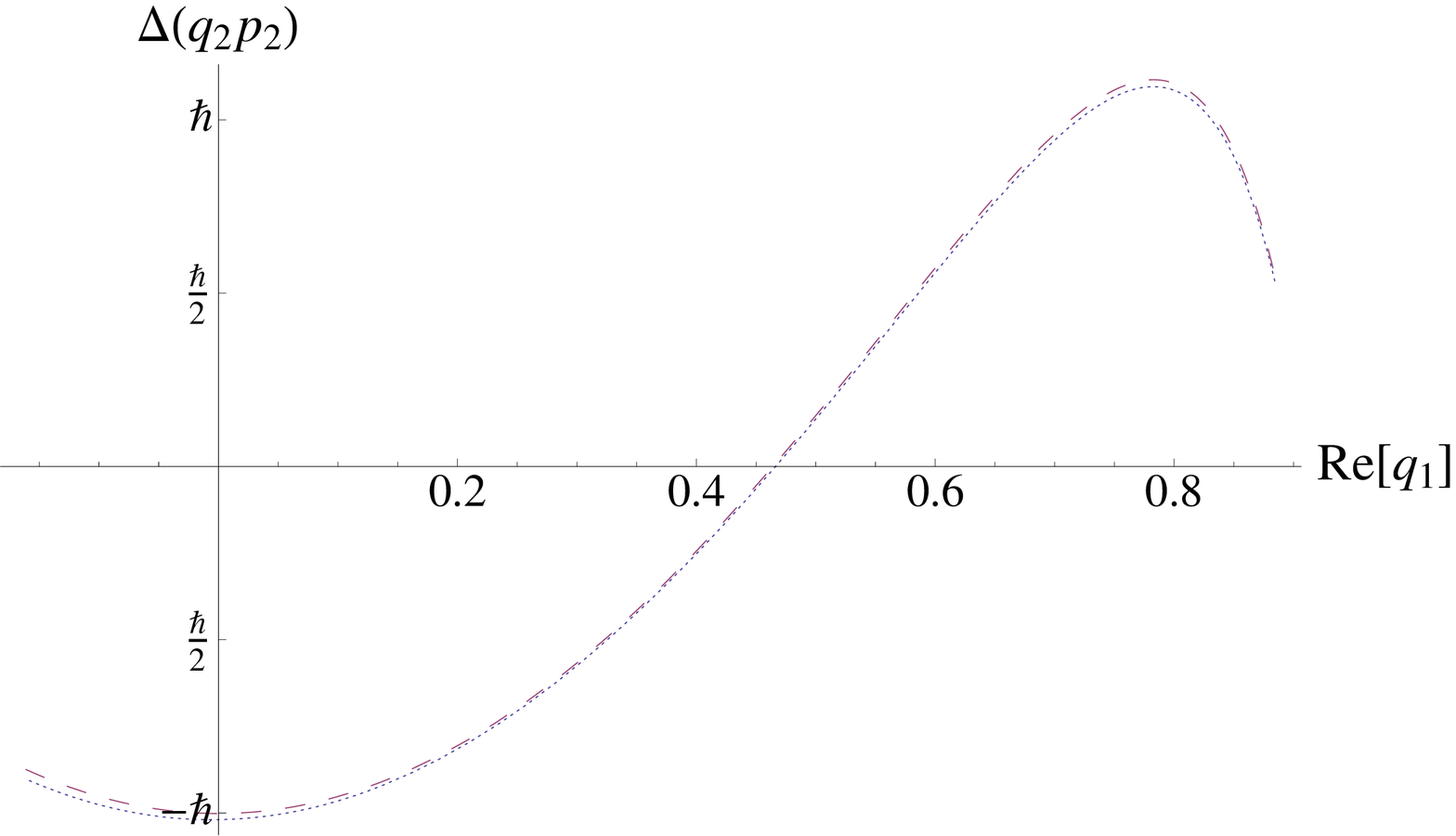}  \\
a) & b)
\end{array}$
    \caption{\label{fullorb}{\small a) Reconstruction of a
        semiclassical physical state via gauge switching in the
        effective framework. The jumps between the $q_1$-gauge (black
        dotted and dashed curves) and the $q_2$-gauge (blue solid
        curves) are a consequence of the $o(\hbar)$ jumps in the gauge
        transformations (\ref{eq:rovelli_gtransf}). The final
        evolution in $q_1$-Zeitgeist after the fourth clock change is
        given by the fat black dashed curve and coincides to
        $o(\hbar)$ with the initial evolution in $q_1$-gauge prior to
        the first clock change. For convenience we have labeled the
        axes by $q_1$ and $q_2$. It should be noted that for the
        curves in $q_i$-gauge, $q_i$ actually refers to $\Re[q_i]$. b)
        Comparison of $\Delta(q_2p_2)(\Re[q_1])$ in $q_1$-gauge before
        (dashed curve) and after (dotted curve) the complete
        revolution around the ellipse. The difference between the two
        curves is clearly of $o(\hbar^{2})$ or smaller. Initial data
        for both a) and b): ${q_1}_0=-\frac{i\hbar}{2},
        {p_1}_0={q_2}_0={p_2}_0=1, (\Delta q_2)_0^2=(\Delta
        p_2)_0^2=\frac{\hbar}{2}$. Furthermore, $M=3$ and, to enhance
        effects, we have set $\hbar=0.01$. The initial value for
        $\Delta(q_2p_2)$ follows from Eq.\ (\ref{pi4}).}}
\end{figure*}
\end{center}

As a specific example of an effective reconstruction of a
semiclassical physical state via gauge switching, we provide a plot
of the configuration space ellipse in FIG.\ \ref{fullorb}a for a
configuration whose initial data is provided in the caption of the
figure. We have started in the $q_1$-Zeitgeist and changed gauge four
times in the course of evolution, in order to reach the same gauge
after a complete revolution around the ellipse. Since revolution
numbers around the ellipse have no physical meaning in either the
classical or the quantum theory, we only evolve once around the
ellipse. In accordance with this, it is found that the discrepancy
between the variables in the $q_1$-gauge before and after one complete
revolution are of order $o(\hbar^{2})$ or smaller. For the particular
example of $\Delta(q_2p_2)(\Re[q_1])$ this is shown in FIG.\
\ref{fullorb} b); the two curves in the same gauge before and after
the complete revolution agree extremely well to order $\hbar$,
implying that they describe the same physical state. The jumps between
the curves in the two different gauges are a consequence of the
particular form of the gauge changes, as given in Sec.\ \ref{rovgt}.
In agreement with Sec.\ \ref{gtmoment}, it is also found numerically
that the end result does not depend on the precise instants of the
intermediate gauge changes, as long as the two gauges are valid before
and after the transformations. This shows consistency of the argument
in Sec.\ \ref{gtmoment} with the semiclassical approximation in this
particular example.

Validity of the semiclassical approximation and the new and old gauge
has to be checked when performing intermediate gauge changes. This is
not problematic as long as the ellipse is reasonably close to a
circle. For squeezed ellipses, however, when the turning points in
$q_1$- and $q_2$-time may lie very close to each other, one has to be
rather careful when precisely to carry out the gauge change, since in
spite of a semiclassical trajectory, the spread will play a more
restrictive role in this case. Nonetheless, this issue merely
constitutes a practical, but not a conceptual problem.

\section{Discussion and Conclusions} \label{sec:conclusions}

In this article we have described in two simple toy models the
effective approach of \cite{EffTime1} to coping with the general
problem of time in the semiclassical regime. A central additional
ingredient for the interpretation of this approach is the relational
concept of evolution. By employing an effective framework, one
benefits from the advantage of sidestepping many technical problems
associated to the general problem of time, thereby facilitating an
explicit investigation of various of its aspects, as well as their
repercussions for the usual Dirac quantization.

In particular, the effective approach avoids the {\it Hilbert space
  problem} altogether since no use of representations or physical
inner products has been made at any point of the algebraic
construction. The tedious problem of constructing physical states and
inner products, which is often even practically
impossible,\footnote{Ref.~\cite{PhysHilbert} notwithstanding, for
  the issue of defining physical evolution in the absence of global
  clocks has not been addressed in these approaches.} is replaced by
evaluating an (infinite) coupled set of quantum variables which can be
consistently truncated to a finite solvable system, for instance, at
semiclassical order; necessary physicality conditions for observables
are ultimately imposed just by reality conditions. At this stage, the
effective framework can be implemented numerically and its physical
properties can be studied in detail.

Although we can avoid practical problems in constructing physical
Hilbert spaces, we do not intend to suggest solutions of effective
constraints as full substitutes of physical states. Some questions,
such as the measurement problem, can only be addressed  with
Hilbert-space representations. Effective techniques at present do not
provide a complete description of quantum systems, but they can
capture representation-independent information which is sufficient for
many questions of interest.

The {\it multiple-choice problem}, furthermore, does not constitute a
problem at the effective level, since, from the point of view of the
Poisson manifold of the effective framework, all variables of a given
order are treated on an equal footing. Just as in the classical case,
we may choose whichever suitable (quantum) phase space clock function
we desire and deparametrize in this variable.  To simplify explicit
calculations and interpretations, it is helpful to further impose
gauge conditions on this effective constrained system, which are
closely related to the choice of the clock variable and which fix all
but one Hamiltonian gauge flow. Note that this gauge fixing happens
{\it after} quantization. At this level, choosing different clocks
means choosing different gauges and corresponding Zeitgeister in which
to evaluate the effective system. As explicitly demonstrated in two
examples, one can, moreover, translate between the different choices
for internal time by means of gauge transformations. In fact, in the
case of systems which admit the {\it global time problem} one is
forced to change the local clocks in the course of relational
evolution since gauges are, in general, not globally valid.
It should be emphasized that deparametrizations with respect to
different choices of internal time yield, in general, inequivalent
Hilbert-space representations, and thus different gauges at the
effective level generally correspond to different formulations of the
quantum theory.

The usual {\it operator-ordering problem} is not entirely
circumvented in this effective approach since we choose a particular
ordering for the constraint operator before treating it effectively.
This specific ordering, however, is not connected to the choice of a
(local) time variable which happens only {\it after} the effective
system has been constructed.

Of the technical problems briefly described in the introduction, it is
only the {\it global time problem} and the {\it problem of
  observables} which are not automatically sidestepped by the
effective approach. But by avoiding the other technical problems, the
effective approach greatly facilitates the construction of a
sufficient set of explicit fashionables since, although we face a
larger number of degrees of freedom, the problem can be addressed in
the usual classical manner which allowed for simple numerical
solutions in the toy models studied in this article. The effective
framework is, thus, amenable to techniques, usually aimed at a
solution to the classical {\it problem of observables}, such as
\cite{Bianca1,Haj1} and the perturbative expansions of \cite{Bianca2}.
Moreover, concrete evaluations of constrained systems are usually
done by employing gauge fixing, for which classical methods such as
those of \cite{GaugeFix} are useful.

Likewise, the effective approach enables us to perform a concrete
treatment of the {\it global time problem} and suggests a simple
patching solution. As discussed in Sec.\ \ref{rovmod}, the
relational concept is only of a local and semiclassical nature in the
absence of a global clock and, thus, the {\it problem of relational
observables} becomes a local one. Global relational observables valid
for all classical values of relational time do not exist in the
quantum theory. While in the absence of global clocks it is not at all
clear how to implement the relational concept and explicitly construct
relational Dirac observable operators in a Dirac quantization, some
simplification is offered by local deparametrization, resulting in a
local internal time Schr\"odinger regime. In contrast to this, it is clear how to
implement this scenario in a simple way within the effective
semiclassical analysis, which reproduces the results of the local
Schr\"odinger regime. An apparent non-unitarity leads to the breakdown
of a constant time slicing in this procedure and to the
failure of the gauge associated to the choice of local time in the
effective framework.  This is consistent with the related breakdown of
the relational observables in the reduction and in the Dirac method on
approach to a turning point \cite{Rovmod}. To achieve a consistent
evolution through turning points of a clock, we are forced to switch
to a different clock and a different set of variables to be evolved,
prior to reaching a turning point, which corresponds to switching to a
different local Schr\"odinger regime and a gauge change in the
effective approach. By switching to a good local clock, when another
time variable approaches a turning point, we can consistently
transport relational data along and thereby reconstruct the entire
information of a semiclassical physical state via local patches of
relational evolution. To our knowledge, there is no consistent method
for explicitly transferring data between different local
deparametrizations of one and the same model at a Hilbert space
level. Any such method is likely to be quite involved, to lead to
discontinuities in correlations and to be only applicable for states
that are sufficiently semiclassical.  On the other hand, the gauge
changes are easily implemented on the effective side, albeit
exhibiting jumps of order $\hbar$ in correlations, which underline the
merely local nature of relational observables. No sharp instant for the
change in time prior to a turning point has to be selected; the
transformation may be performed at any point, as long as the old and
new choice of time are valid before and after the clock change,
respectively.

As regards relational Hamiltonian evolution, in the second model we
have discussed the peculiarities associated to the IVP and the issue
of time direction in the absence of a global clock. While we may
classically keep one and the same relational time variable and only
have to switch the sign of the physical Hamiltonian at the turning
point of the clock, we are required to change the Hamiltonian operator
of the internal time Schr\"odinger regime to a new one adapted to a new local clock
{\it before} reaching the classical turning point. On the effective
side, we could proceed similarly by linearizing the Hamiltonian
constraint in the momentum conjugate to internal time in the gauge associated
to the chosen clock. Such an effective physical Hamiltonian,
obviously, changes together with the Hamiltonian constraint during
necessary gauge changes prior to turning points of non-global clocks.

A final striking consequence of the {\it global time problem} is the
inevitable appearance of a {\it complex internal time}. We have shown that
the particular form of the imaginary contribution to the time
variable is a quantum effect and a generic feature of the effective
approach. Similarly, we have collected strong evidence from an
expectation value calculation of the time operator in a Dirac
approach to the free relativistic particle and a comparison of the quadratic
Wheeler-DeWitt equation to an associated internal time Schr\"odinger equation that
this particular imaginary contribution is also a generic feature of
standard Hilbert-space quantizations. In particular, the
inequivalence between the Wheeler-DeWitt and Schr\"odinger equation
in the presence of a ``time potential'' is a result of the
assumption that time is real-valued in both equations. The two
equations can be locally reconciled if the expectation value of internal time
is allowed a particular imaginary contribution in the WDW case. By
the same token, as shown in the concrete example in Sec.\
\ref{schrodreg}, Dirac observables of the system governed by the
quadratic constraint are, in general, constants of motion of
the associated Schr\"odinger regime only if internal time is complex in the
Wheeler-DeWitt equation.

Despite the fact that the imaginary contribution to time also appears
for globally valid clocks, the imaginary contribution can be
disregarded altogether in this case, since it turns out to be a
constant of motion which is not necessary for the satisfaction of the
constraints. For non-global clocks, however, the imaginary
contribution turns out to be dynamical and cannot at all be ignored.
It is, therefore, rather a true non-global feature. When the local
clock eventually needs to be exchanged together with the corresponding
gauge at the effective level, the imaginary contribution is
consistently removed from the old clock which subsequently turns into
an evolving physical variable and pushed, accordingly, to the new
clock function.

Concerning relational evolution in the presence of a dynamical
imaginary contribution to internal time, we encounter the issue of a
``vector time'' with two separate degrees of freedom. In this article,
however, we argue, in agreement with common sense, to only employ the
real part of the internal clock as relational time, since the imaginary part
causes a number of additional problems, rendering it an even worse
clock than the already non-global real part.

In conclusion, the effective approach to the problem of time overcomes
a number of technical problems and substantially facilitates the
solution to various other problems, while simultaneously providing
further insight into standard Hilbert-space quantizations. In
particular, it is possible to master the {\it global time problem} at
the semiclassical level and to consistently evolve data through
turning points of non-global clocks. In this article and in
\cite{EffTime1}, we have, furthermore, argued that the standard notion
of relational time and the concept of relational evolution are, in
general, of merely local and semiclassical nature, which disappear
(together with complex relational time) for highly quantum states of systems
without global clock variables.

We emphasize that these results and conclusions are based on a
semiclassical analysis in simple toy models. It is, certainly,
dangerous to draw any general conclusions for full quantum gravity
from procedures which so far are only proven to work in simple
scenarios. Moreover, further technical problems, specifically related
to gravity, such as, e.g., the {\it spacetime reconstruction problem},
require significant advances in the effective formalism before they
may be tackled. Nevertheless, we believe that the present approach is
worth pursuing and promises some headway in evaluating quantum gravity
theories and models in a practical way. In this light, we expect
certain features, such as complex internal time, to be of a generic nature in
more general models, especially in quantum cosmology.

Owing to the advantage that the effective approach simultaneously
avoids many facets of the problem of time, it may be viewed as one
step in the quest to ``defeat the Ice Dragon'' of \cite{anderson},
symbolizing the conjunction of the apparently many faces of the
problem of time in quantum gravity.

\vspace{1cm}
\appendix

\section{Poisson algebra} \label{app:PB}

Expectation values satisfy the classical Poisson algebra and have
vanishing Poisson brackets with the moments of all orders.
Table~\ref{TabPoisson} lists the Poisson brackets between second order
moments generated by two canonical pairs of observables. The table has
originally appeared in the appendix of~\cite{EffConsRel} and is
reproduced here for convenience.

\begin{table*}[h]
\centering \caption{Poisson algebra of second order moments. First
terms in the bracket are labeled by rows, second terms are labeled
by columns. \label{TabPoisson}}
\begin{tabular}{|c|| c|c|c|c|c|c|c|c|c|c|} \hline & ${\scriptstyle(\Delta t)^2}$
& ${\scriptstyle\Delta(tp_t)}$ & ${\scriptstyle (\Delta p_t)^2}$ &
${\scriptstyle (\Delta q)^2}$ & ${\scriptstyle \Delta(qp)}$ &
${\scriptstyle (\Delta p)^2}$ & ${\scriptstyle \Delta(tq)}$ &
${\scriptstyle \Delta(p_tp)}$ & ${\scriptstyle \Delta(tp)}$ &
${\scriptstyle \Delta(p_tq)}$ \\ \hline\hline ${\scriptstyle (\Delta
t)^2}$ & ${\scriptstyle 0}$ & ${\scriptstyle 2(\Delta t)^2}$ &
${\scriptstyle 4 \Delta(tp_t)}$ & ${\scriptstyle 0}$ &
${\scriptstyle 0}$ & ${\scriptstyle 0}$ & ${\scriptstyle 0}$ &
${\scriptstyle 2 \Delta(tp)}$ & ${\scriptstyle 0}$ & ${\scriptstyle
2 \Delta(tq)}$
\\ \hline ${\scriptstyle \Delta(tp_t)}$ & ${\scriptstyle -2(\Delta
t)^2}$ & ${\scriptstyle 0}$ & ${\scriptstyle 2(\Delta p_t)^2}$ &
${\scriptstyle 0}$ & ${\scriptstyle 0}$ & ${\scriptstyle 0}$ &
${\scriptstyle -\Delta(tq)}$ & ${\scriptstyle \Delta(p_tp)}$ &
${\scriptstyle -\Delta(tp)}$ & ${\scriptstyle \Delta(p_tq)}$ \\
\hline ${\scriptstyle  (\Delta p_t)^2}$ & ${\scriptstyle
-4\Delta(tp_t)}$ & ${\scriptstyle -2(\Delta p_t)^2}$ &
${\scriptstyle 0}$ & ${\scriptstyle 0}$ & ${\scriptstyle  0}$ &
${\scriptstyle  0}$ & ${\scriptstyle  -2\Delta(p_tq)}$ &
${\scriptstyle  0}$ & ${\scriptstyle -2\Delta(p_tp)}$ &
${\scriptstyle 0}$ \\ \hline ${\scriptstyle  (\Delta q)^2}$ &
${\scriptstyle 0}$ & ${\scriptstyle 0}$ & ${\scriptstyle 0}$ &
${\scriptstyle 0}$ & ${\scriptstyle 2(\Delta q)^2}$ & ${\scriptstyle
4\Delta(qp)}$ & ${\scriptstyle 0}$ & ${\scriptstyle 2\Delta(p_tq)}$
& ${\scriptstyle 2 \Delta(tq)}$ & ${\scriptstyle 0}$ \\ \hline
${\scriptstyle \Delta(qp)}$ & ${\scriptstyle 0}$ & ${\scriptstyle
0}$ & ${\scriptstyle 0}$ & ${\scriptstyle -2(\Delta q)^2}$ &
${\scriptstyle 0}$ & ${\scriptstyle 2(\Delta p)^2}$ & ${\scriptstyle
-\Delta(tq)}$ & ${\scriptstyle \Delta(p_tp)}$ & ${\scriptstyle
\Delta(tp)}$ & ${\scriptstyle -\Delta(p_tq)}$ \\ \hline
${\scriptstyle (\Delta p)^2}$ & ${\scriptstyle 0}$ & ${\scriptstyle
0}$ & ${\scriptstyle 0}$ & ${\scriptstyle -4\Delta(qp)}$ &
${\scriptstyle -2(\Delta p)^2}$ & ${\scriptstyle 0}$ &
${\scriptstyle -2\Delta(tp)}$ & ${\scriptstyle 0}$ & ${\scriptstyle
0}$ & ${\scriptstyle -2\Delta(p_tp)}$ \\ \hline ${\scriptstyle
\Delta(tq)}$ & ${\scriptstyle 0}$ & ${\scriptstyle \Delta(tq)}$ &
${\scriptstyle 2\Delta(p_tq)}$ & ${\scriptstyle 0}$ & ${\scriptstyle
\Delta(tq)}$ & ${\scriptstyle 2\Delta(tp)}$ & ${\scriptstyle 0}$ &
${\scriptstyle \Delta(tp_t)}$ & ${\scriptstyle (\Delta t)^2}$ &
${\scriptstyle (\Delta q)^2}$ \\ & & & & & & & & ${\scriptstyle
+\Delta(qp)}$ & &
\\ \hline ${\scriptstyle \Delta(p_tp)}$ & ${\scriptstyle
-2\Delta(tp)}$ & ${\scriptstyle -\Delta(p_tp)}$ & ${\scriptstyle 0}$
& ${\scriptstyle -2\Delta(p_tq)}$ & ${\scriptstyle -\Delta(p_tp)}$ &
${\scriptstyle 0}$ & ${\scriptstyle -\Delta(tp_t)}$ & ${\scriptstyle
0}$ & ${\scriptstyle -(\Delta p)^2}$ & ${\scriptstyle -(\Delta
p_t)^2}$ \\ & & & & & & & ${\scriptstyle -\Delta(qp)}$ & & & \\
\hline ${\scriptstyle \Delta(tp)}$ & ${\scriptstyle 0}$ &
${\scriptstyle \Delta(tp)}$ & ${\scriptstyle 2\Delta(p_tp)}$ &
${\scriptstyle -2\Delta(tq)}$ & ${\scriptstyle -\Delta(tp)}$ &
${\scriptstyle 0}$ & ${\scriptstyle -(\Delta t)^2}$ & ${\scriptstyle
(\Delta p)^2}$ & ${\scriptstyle 0}$ & ${\scriptstyle \Delta(qp)}$ \\
& & & & & & & & & & ${\scriptstyle -\Delta(tp_t)}$ \\ \hline
${\scriptstyle \Delta(p_tq)}$ & ${\scriptstyle -2\Delta(tq)}$ &
${\scriptstyle -\Delta(p_tq)}$ & ${\scriptstyle 0}$ & ${\scriptstyle
0}$ & ${\scriptstyle \Delta(p_tq)}$ & ${\scriptstyle 2\Delta(p_tp)}$
& ${\scriptstyle -(\Delta q)^2}$ & ${\scriptstyle (\Delta p_t)^2}$ &
${\scriptstyle \Delta(tp_t)}$ & ${\scriptstyle 0}$ \\ & & & & & & &
& & ${\scriptstyle -\Delta(qp)}$ & \\ \hline
\end{tabular}
\label{tab:pb_moments}
\end{table*}

\section{Discussion of positivity}\label{positivity}

\subsection{Algebraic positivity}
Positivity is understood in the algebraic sense as the condition
$\langle \mathbf{AA}^{\ast} \rangle \geq 0, \ \ \forall \mathbf{A}
\in \mathcal{A}$, where $\mathcal{A}$ is some algebra. It relates
directly to the GNS construction of unitary representations for
$\ast$-algebras, and is also necessary for the measurement theory and
probabilistic interpretation of the state. In this appendix we focus
on the unital star algebra $\mathcal{A}$\ of all finite-order
polynomials generated by a single canonical pair $\hat{q}$\ and
$\hat{p}$\ subject to
\[
[ \hat{q}, \hat{p} ] = i\hbar \mathds{1} \quad {\rm and} \quad
\hat{q}^{\ast} = \hat{q}, \ \ \hat{p}^{\ast} = \hat{p}\q.
\]
We pose the following question:
\begin{itemize}
\item \emph{What are the necessary and sufficient conditions one needs to
place on a state on $\mathcal{A}$\ such that positivity holds to
order $\hbar$?}
\end{itemize}
By ``positivity holding to order $\hbar$'' we mean that
$|\Im[\langle \mathbf{AA}^{\ast} \rangle]| \propto
\hbar^{\frac{3}{2}}$\ and $\Re[ \langle \mathbf{AA}^{\ast} \rangle]
\geq -\hbar^{\frac{3}{2}}$. The answer turns out to be simple, in
addition to normalization $\langle \mathds{1} \rangle = 1$, we need
to impose
\begin{eqnarray}
&& q, p, (\Delta q)^2, (\Delta p)^2, \Delta(qp) \in \mathbb{R} \nonumber \\
&& (\Delta p)^2, (\Delta q)^2 \geq  0 \nonumber \\ && (\Delta q)^2
(\Delta p)^2 - \left(\Delta(qp)\right)^2 \geq \frac{1}{4} \hbar^2\q.
\label{eq:pos_conditions}
\end{eqnarray}
We only outline the demonstration of \emph{necessity}, as these are
standard results in ordinary quantum mechanics:
\begin{itemize}
\item We recall that positivity can be used to derive $\langle
\mathbf{A}^{\ast} \rangle  = \overline{\langle \mathbf{A} \rangle}$,
where bar denotes the complex conjugate. This immediately implies
$q, p, (\Delta q)^2, (\Delta p)^2, \Delta(qp) \in \mathbb{R}$.
\item $\langle \left(\hat{q} -
\langle \hat{q} \rangle \mathds{1} \right)  \left(\hat{q} - \langle
\hat{q} \rangle \mathds{1} \right)^{\ast} \rangle \geq 0$\
immediately gives $(\Delta q)^2 \geq 0$, we similarly get $(\Delta
p)^2 \geq 0$.
\item The uncertainty relation can be obtained by first deriving the
Schwarz-type inequality $|\langle \mathbf{AB}^{\ast} \rangle|^2
\leq \langle \mathbf{AA}^{\ast} \rangle \langle \mathbf{BB}^{\ast}
\rangle$, and substituting $\mathbf{A} = \hat{q} - q\mathds{1}$\ and
$\mathbf{B} = \hat{p} - p\mathds{1}$.
\end{itemize}

Before we demonstrate \emph{sufficiency}, we derive an inequality implied
by~(\ref{eq:pos_conditions}), which we will use on several occasions
in this section and the following ones:
\begin{equation}\label{eq:inequality}
\alpha^2 (\Delta q)^2 + \beta^2 (\Delta p)^2 + 2\alpha \beta
\Delta(qp) \geq 0 \q , \q \forall \ \alpha, \ \beta \in \mathbb{R}
\q.
\end{equation}
This follows as
\begin{widetext}
\begin{eqnarray*}
\alpha^2 (\Delta q)^2 + \beta^2 (\Delta p)^2 + 2\alpha \beta
\Delta(qp) &\geq& \alpha^2 (\Delta q)^2 + \beta^2 (\Delta p)^2 -
2|\alpha| |\beta| |\Delta(qp)| \\ &\geq& |\alpha|^2 (\Delta q)^2 +
|\beta|^2 (\Delta p)^2 - 2|\alpha| |\beta| \sqrt{(\Delta q)^2
(\Delta p)^2} \geq \left( |\alpha| \sqrt{(\Delta q)^2} - |\beta|
\sqrt{(\Delta p)^2} \right)^2 \geq 0 \ .
\end{eqnarray*}
\end{widetext}

To demonstrate \emph{sufficiency} to order $\hbar$, we adopt a rather
direct approach.  Any finite order polynomial in $\hat{q}$\ and
$\hat{p}$\ can be expanded using the symmetrized products $\left(
\hat{q}^m \hat{p}^{n} \right)_{\rm Weyl}$
\[
\hat{f} = \sum_{m,n \geq 0} \alpha_{mn} \left( \hat{q}^m \hat{p}^{n}
\right)_{\rm Weyl} =: f( \hat{q}, \hat{p})\q.
\]
Here, $f(\hat{q}, \hat{p})$\ is understood as a map from the algebra
to itself; in particular, it keeps track of the ordering, which we
chose to be completely symmetric in this case. In general,
$\alpha_{mn} \in \mathbb{C}$, for self-adjoint elements $\alpha_{mn}
\in \mathbb{R}$. We now expand the polynomial in terms of a
different set of elements $\widehat{\Delta q} := \hat{q} - q$\ and
$\widehat{\Delta p}:= \hat{p} - p$. Evidently
\begin{widetext}
\begin{eqnarray*}
\hat{f} &=& f (\hat{q}, \hat{p}) = f(q + \widehat{\Delta q}, p +
\widehat{\Delta p}) \\ &=& f(q, p) + \frac{\partial f}{\partial q}
(q, p) \widehat{\Delta q} + \frac{\partial f}{\partial p} (q, p)
\widehat{\Delta p} + \frac{1}{2} \frac{\partial^2 f}{\partial q^2}
(q, p) (\widehat{\Delta q})^2 + \frac{1}{2} \frac{\partial^2
f}{\partial p^2} (q, p) (\widehat{\Delta p})^2 \\ & & +
\frac{\partial^2 f}{\partial q \partial p} (q, p) (\widehat{\Delta
q} \widehat{\Delta p})_{\rm Weyl} + \left({\rm higher \ powers \ of
\ }\widehat{\Delta q}, \ \widehat{\Delta p} \right)\q.
\end{eqnarray*}
$q$\ and $p$\ can be any real numbers, below we set them to the
expectation values $\langle \hat{q} \rangle$\ and $\langle \hat{p}
\rangle$, which enables us to utilize semiclassical truncation.
Keeping terms of order $\hbar$\ we find the expectation value of
$\hat{f}$
\[
\langle \hat{f} \rangle = f(q, p) + \frac{1}{2} \frac{\partial^2
f}{\partial q^2} (q, p) (\Delta q)^2 + \frac{1}{2} \frac{\partial^2
f}{\partial p^2} (q, p) (\Delta p)^2 + \frac{\partial^2 f}{\partial
q \partial p} (q, p) \Delta(qp) + O(\hbar^{3/2})\q,
\]
so that, again to order $\hbar$, we have
\begin{eqnarray*}
|\langle \hat{f} \rangle|^2 &=& |f|^2 + \frac{1}{2} \left[ f \left(
\overline{ \frac{ \partial^2 f}{\partial q^2} }\right) + \bar{f}
\left( \frac{
\partial^2 f}{\partial q^2} \right) \right] (\Delta q)^2 +
\frac{1}{2} \left[ f \left( \overline{ \frac{
\partial^2 f}{\partial p^2} }\right) + \bar{f} \left( \frac{
\partial^2 f}{\partial p^2} \right) \right] (\Delta p)^2 \\ && + \left[ f
\left( \overline{ \frac{\partial^2 f}{\partial q \partial p} }
\right) + \bar{f} \left( \frac{\partial^2 f}{\partial q \partial p}
\right) \right] \Delta(qp) + O(\hbar^{\frac{3}{2}})\q.
\end{eqnarray*}
We note that since $|\langle \hat{f} \rangle|^2 \geq 0$, the
truncated expression for $|\langle \hat{f} \rangle|^2$, satisfies
the inequality to order $\hbar$\ in the sense discussed earlier. Now
consider positivity of the state evaluated on $\hat{f}$:
\begin{eqnarray*}
\langle \hat{f} \hat{f}^{\ast} \rangle &=& \left\langle \left( f +
\frac{\partial f}{\partial q} \widehat{\Delta q} + \frac{\partial
f}{\partial p} \widehat{\Delta p} + \frac{1}{2} \frac{\partial^2
f}{\partial q^2} (\widehat{\Delta q})^2 + \frac{1}{2}
\frac{\partial^2 f}{\partial p^2} (\widehat{\Delta p})^2 +
\frac{\partial^2 f}{\partial q \partial p} (\widehat{\Delta q}
\widehat{\Delta p})_{\rm Weyl} \right) \right. \\ && \ \ \left.
\left( \bar{f} + \overline{\frac{\partial f}{\partial q}}
\widehat{\Delta q} + \overline{\frac{\partial f}{\partial p}}
\widehat{\Delta p} + \frac{1}{2} \overline{ \frac{\partial^2
f}{\partial q^2}} (\widehat{\Delta q})^2 + \frac{1}{2} \overline{
\frac{\partial^2 f}{\partial p^2}} (\widehat{\Delta p})^2 +
\overline{ \frac{\partial^2 f}{\partial q \partial p}}
(\widehat{\Delta q} \widehat{\Delta p})_{\rm Weyl} \right) \right\rangle + O(\hbar^{3/2}) \\
&=& |f|^2 + \frac{1}{2} \left[ f \left( \overline{ \frac{ \partial^2
f}{\partial q^2} }\right) + \bar{f} \left( \frac{
\partial^2 f}{\partial q^2} \right) \right] (\Delta q)^2 +
\frac{1}{2} \left[ f \left( \overline{ \frac{
\partial^2 f}{\partial p^2} }\right) + \bar{f} \left( \frac{
\partial^2 f}{\partial p^2} \right) \right] (\Delta p)^2 \\ && + \left[ f
\left( \overline{ \frac{\partial^2 f}{\partial q \partial p} }
\right) + \bar{f} \left( \frac{\partial^2 f}{\partial q \partial p}
\right) \right] \Delta(qp) + \left| \frac{\partial f}{\partial q}
\right| (\Delta q)^2 + \left| \frac{\partial f}{\partial p} \right|
(\Delta p)^2 + 2\Re \left[\frac{\partial f}{\partial q} \overline{
\frac{\partial f}{\partial p}} \right] \Delta(qp) + O(\hbar^{{3/2}})
\\ &=& |\langle \hat{f} \rangle |^2 + \left| \frac{\partial f}{\partial q}
\right| (\Delta q)^2 + \left| \frac{\partial f}{\partial p} \right|
(\Delta p)^2 + 2\Re \left[\frac{\partial f}{\partial q} \overline{
\frac{\partial f}{\partial p}} \right] \Delta(qp) +
O(\hbar^{{3/2}})\q.
\end{eqnarray*}
Now $|\langle \hat{f} \rangle |^2 \geq 0$, and the next three terms are
positive by inequality~(\ref{eq:inequality})
\begin{eqnarray*}
\left| \frac{\partial f}{\partial q} \right| (\Delta q)^2 + \left|
\frac{\partial f}{\partial p} \right| (\Delta p)^2 + 2\Re
\left[\frac{\partial f}{\partial q} \overline{ \frac{\partial
f}{\partial p}} \right] \Delta(qp) &\geq& \left| \frac{\partial
f}{\partial q} \right| (\Delta q)^2 + \left| \frac{\partial
f}{\partial p} \right| (\Delta p)^2 - 2 \left| \frac{\partial
f}{\partial q} \right| \left| \frac{\partial f}{\partial p} \right|
|\Delta(qp)| \geq 0\q.
\end{eqnarray*}
So that, as claimed earlier, $\langle \hat{f} \hat{f}^{\ast} \rangle
\geq 0$\ to order $\hbar$.
\end{widetext}

\subsection{Positivity in the model of Section~\ref{lt}}

Here we use the explicit form of gauge invariant functions to prove
the following statements to order $\hbar$\ for the relativistic
particle in a $\lambda t$\ potential:
\begin{itemize}
\item the positivity of a state is preserved by the dynamics in $t$-gauge,
\item it is also preserved by gauge transformation between $q$-gauge
and $t$-gauge,
\item finally it is preserved by the dynamics in $q$-gauge.
\end{itemize}
The constraint in this model is
\[
\hat{C} = \hat{p}_t^2 - \hat{p}^2 - m^2 \mathds{1} +
\lambda\hat{t}\,.
\]
A complete set of Dirac observables may be constructed from the
canonical pair:
\[
\hat{\mathcal{Q}} := \hat{q} - \frac{2}{\lambda} \hat{p}\hat{p}_t
\quad {\rm and} \quad \hat{\mathcal{P}}:=\hat{p}, \quad {\rm
satisfying \ \ } [ \hat{\mathcal{Q}}, \hat{\mathcal{P}} ] = i\hbar
\mathds{1}\q,
\]
which commute with the constraint $[\hat{\mathcal{Q} },\hat{C} ] = 0
= [ \hat{\mathcal{P}}, \hat{C} ]$. Below we provide the expectation
values and second order moments of these observables:
\begin{widetext}
\begin{eqnarray*}
\mathcal{Q} &=& q - \frac{2}{\lambda}\left(pp_t + \Delta (p_t p)
\right), \quad \mathcal{P} = p, (\Delta \mathcal{P} )^2 = (\Delta
p)^2 \ \ , \ \ \Delta(\mathcal{QP}) = \Delta(qp) - \frac{2}{\lambda}
\left( \Delta(p_tpp) + p_t(\Delta p)^2 + p\Delta(p_tp) \right) \\
(\Delta \mathcal{Q})^2 &=& (\Delta q)^2 - \frac{4}{\lambda} \left(
\Delta(p_t q p) + p_t \Delta(qp) + p \Delta(p_tq) \right) +
\frac{4}{\lambda^2} \left[ \Delta(p_tp_tpp) + 2p_t\Delta(p_tpp) + 2p
\Delta(p_tp_tp) + p_t^2 (\Delta p)^2 \right. \\ && \left. + p^2
(\Delta p_t)^2 + \left( 2p_tp-\Delta(p_tp) \right)\Delta(p_tp)
\right] \q.
\end{eqnarray*}
\end{widetext}
Poisson brackets of these functions with constraint functions must
vanish to the given order, since the operators that generate them
commute with the constraint operator (see Eq.\ (\ref{effobs})). Additionally, we note that
$p=\mathcal{P}$\ is a constant of motion, while $p_t$\ evolves as
$p_t(s) = -\lambda s + p_{t0}$\ and is preserved by the
transformation between the gauges, therefore, the condition $p_t, p
\in \mathbb{R}$\ is preserved in all situations considered here.

\subsubsection{Dynamics in the $t$-gauge}

Below are the expressions for the same invariants truncated at order
$\hbar$, evaluated in the $t$-gauge, with the moments generated by
$\hat{p}_t$\ eliminated through constraint functions:
\begin{eqnarray*}
\mathcal{Q} &=& q - \frac{2}{\lambda}\left(pp_t +
\frac{p}{p_t}(\Delta p)^2 \right)\ \ , \quad \mathcal{P} = p\ \ ,
\\ (\Delta \mathcal{Q})^2 &=& (\Delta q)^2 - 2\theta \Delta(qp) + \theta^2(\Delta p)^2\ ,
\ \
(\Delta \mathcal{P} )^2 = (\Delta p)^2 \\
\Delta(\mathcal{QP}) &=& \Delta(qp) - \theta(\Delta p)^2\q, \\ {\rm
where} && \ \ \theta = \frac{2(p_t^2+p^2)}{\lambda p_t}\q.
\end{eqnarray*}
We now re-express the gauge dependent moments in  terms of these
invariants:
\begin{eqnarray*}
(\Delta q)^2 &=& (\Delta \mathcal{Q})^2 + \theta^2 (\Delta
\mathcal{P} )^2 + 2 \theta \Delta(\mathcal{QP}) \\ (\Delta p)^2 &=&
(\Delta \mathcal{P} )^2 \\ \Delta(qp) &=& \Delta(\mathcal{QP}) +
\theta (\Delta\mathcal{P})^2\q.
\end{eqnarray*}
Assuming that $\theta$\ is real (which holds provided $p_t$\ and $p$
are real), one can see that:
\begin{itemize}
\item reality of invariant moments implies reality of evolving moments,
\item trivially $(\Delta \mathcal{P} )^2 > 0 \Longrightarrow (\Delta p)^2>0$,
\item $(\Delta q)^2 > 0$ follows directly from the inequality~(\ref{eq:inequality}),
\item finally one finds
\[
(\Delta q)^2(\Delta p)^2 - \left( \Delta(qp) \right)^2 =(\Delta
\mathcal{Q} )^2(\Delta \mathcal{P} )^2 - \left( \Delta(\mathcal{QP})
\right)^2 \geq \frac{\hbar^2}{4}.
\]
\end{itemize}
In short, positivity of the observables implies positivity of
$t$-gauge variables, provided $\theta$\ is real. The converse is
also true: positivity of $t$-gauge observables (together with $p_t
\in \mathbb{R}$) implies positivity of the invariants. The Dirac
observables are invariant under gauge transformations and, in
particular, under the $t$-gauge dynamics, which must then preserve
positivity of the invariant moments and, therefore, also of the
evolving moments.

\subsubsection{Dynamics in the $q$-gauge}

We now verify the equivalent statement in the $q$-gauge. In this
gauge, the invariant moments to order $\hbar$\ are given by:
\begin{eqnarray*}
(\Delta \mathcal{Q} )^2 &=& \frac{1}{\theta\nu - 1}\left((\Delta
t)^2 + \theta^2 (\Delta p_t)^2 + 2\theta \Delta(tp_t) \right)
\\ (\Delta \mathcal{P})^2 &=& \frac{1}{\theta\nu - 1} \left( (\Delta p_t)^2
+ 2\nu \Delta(tp_t) + \nu^2(\Delta t)^2 \right) \\
\Delta(\mathcal{QP}) &=& \frac{-1}{\theta\nu - 1} \left( ( \theta\nu
+ 1) \Delta(tp_t) + \theta (\Delta p_t)^2 + \nu (\Delta t)^2
\right)\ ,
\end{eqnarray*}
where $\theta = \frac{2(p_t^2+p^2)}{\lambda p_t}$\ and $\nu =
\frac{\lambda}{2p_t}$, so that $\frac{1}{\theta \nu -1} =
\frac{p_t^2}{p^2}$. These relations are tricky to invert by hand,
but the final result is exactly symmetrical, it just so happens that
the above transformation is its own inverse:
\begin{eqnarray}
(\Delta t )^2 &=& \frac{1}{\theta\nu - 1}\left((\Delta
\mathcal{Q})^2 + \theta^2 (\Delta \mathcal{P})^2 + 2\theta \Delta(
\mathcal{QP}) \right) \nn \\ (\Delta p_t)^2 &=& \frac{1}{\theta\nu -
1} \left( (\Delta \mathcal{P})^2 + 2\nu \Delta( \mathcal{QP} ) +
\nu^2(\Delta \mathcal{Q})^2 \right) \\\Delta( tp_t) &=&
\frac{-1}{\theta\nu - 1} \left( ( \theta\nu + 1) \Delta(
\mathcal{QP}) + \theta (\Delta \mathcal{P})^2 + \nu (\Delta
\mathcal{Q})^2 \right) .\nn \label{eq:qgauge_moments}
\end{eqnarray}
If $p_t$\ and $p$\ are real and if $p \neq 0$, then
$\frac{1}{\theta\nu - 1} \geq 0$, with equality only when $p_t = 0$.
We can use the same arguments as before to show that positivity of
the invariants implies positivity of the $q$-gauge moments (for $p_t
= 0$\ case we substitute the expressions for $\theta$\ and $\nu$\ in
terms of $p_t$\ and $p$ first). In particular,
\[
(\Delta t)^2(\Delta p_t)^2 - \left( \Delta(tp_t) \right)^2 =(\Delta
\mathcal{Q} )^2(\Delta \mathcal{P} )^2 - \left( \Delta(\mathcal{QP})
\right)^2 \geq \frac{\hbar^2}{4}.
\]
We note that, once we enforce $p_t, p \in \mathbb{R}$, the reality
of $t$\ in this gauge follows directly from setting $\langle \hat{C}
\rangle = 0$\ and the reality of the moments of $\hat{t}$\ and
$\hat{p}_t$. Eliminating $( \Delta p)^2$\ through other constraints
and imposing the $q$-gauge conditions, $\langle \hat{C} \rangle =
0$\ gives
\begin{eqnarray*}
t = \frac{1}{\lambda} \biggl[ p^2 + m^2 - p_t^2 &+& \frac{p_t^2 -
p^2}{p^2} (\Delta p_t)^2 \\ &+& \frac{\lambda p_t}{p^2} \Delta(tp_t)
+ \frac{\lambda^2}{4p^2} (\Delta t)^2 \biggr]\q.
\end{eqnarray*}
Reality of $\mathcal{Q}$\ then provides a condition on the imaginary
part of $q$, since in this gauge
\[
\mathcal{Q} = q - \frac{2}{\lambda}pp_t - \frac{2p_t}{\lambda p}
(\Delta p_t)^2 - \frac{1}{p} \Delta(t p_t) + \frac{i\hbar}{2p}\q,
\]
so that $\mathcal{Q} \in \mathbb{R}$\ implies $\Im[q] =
-\frac{i\hbar}{2p}$, which is compatible with the transformation
between the two gauges derived in Sec. \ref{lt}.

We have demonstrated that the positivity of the invariant
observables together with $p_t \in \mathbb{R}$\ results in the
positivity of the evolving $q$-gauge observables and yields the
imaginary part of $q$. The converse can also be demonstrated,
namely, starting with the positivity of the $q$-gauge observables
and $\Im[q] = -\frac{i\hbar}{2p}$, one discovers that the invariants
are positive (to demonstrate that $p \in \mathbb{R}$\ one needs to
select the solution to the constraint functions compatible with the
semiclassical approximation). This shows that positivity is
preserved by the dynamics in $q$-gauge.

\subsubsection{Gauge transformation}

The gauge transformation of the second order moments from $t$-gauge
to $q$-gauge can be written as
\begin{eqnarray*}
(\Delta t)^2 &=& (\Delta q)^2_0\frac{ p_t^2}{p^2} \\
(\Delta p_t)^2 &=& \frac{p^2}{p_t^2} \left( (\Delta p)^2_0 +
\mu^2(\Delta q)^2_0 - 2 \mu
\Delta(qp)_0 \right)  \\
\Delta(tp_t) &=& \Delta(qp)_0 - \mu(\Delta q)^2_0\q, \\
{\rm where} & & \ \ \mu = \frac{\lambda p_t}{2 p^2}\q.
\end{eqnarray*}
Assuming $p_t>0$, and that $p$\ and $\lambda$\ are real (which also
means that $\mu$\ is real), it follows in a similar way that
\begin{itemize}
\item $(\Delta q)^2_0>0 \Longrightarrow (\Delta t)^2>0$,
\item once again, $(\Delta p_t)^2>0$\ follows from the
inequality~(\ref{eq:inequality}),
\item one also finds
\[
(\Delta t)^2(\Delta p_t)^2 - \left(\Delta(tp_t) \right)^2 = (\Delta
q)^2(\Delta p)^2 - \left( \Delta(qp) \right)^2 \geq
\frac{\hbar^2}{4}\ .
\]
\end{itemize}
So that a positive state in $t$-gauge transforms to a positive state
in $q$-gauge. The reverse gauge transformation can be analyzed
identically.

\subsection{Positivity in the timeless model of
Sec.~\ref{rovmod}} \label{app:pos_rov}

We will not establish the positivity-preserving properties of
effective dynamics within this model, instead, we point out its
close relation with a local internal time Schr\"odinger evolution, which by
construction preserves positivity so long as it remains valid.

We briefly show that the gauge
transformation~(\ref{eq:rovelli_gtransf}) of Sec.~\ref{rovgt}
consistently transfers positivity between the two sets of physical
variables to order $\hbar$. Firstly, we note that the only initial
parameter that has an imaginary part is $(q_i)_0$. The imaginary
contribution~(\ref{imqi}) is of order $\hbar$\ and leads to the
imaginary contributions to the final values of $q_i$, $p_i$,
$(\Delta q_i)^2$, $(\Delta p_i)^2$, $\Delta(q_ip_i)$\ only at order
$\hbar^2$. Hence, to order $\hbar$\ these variables are real in the
$q_j$-gauge. In addition:
\begin{itemize}
\item $(\Delta q_j)^2_0 \geq 0$\ implies $(\Delta q_j)^2 \geq 0$,
\item $(\Delta p_i)^2 \geq 0$\ follows once again from the
inequality~(\ref{eq:inequality}),
\item The uncertainty relation follows after some straightforward algebraic
manipulations.
\end{itemize}

\section{Explicit moments for the Schr\"odinger regime of Sec.\ \ref{schrodreg}}\label{expmom}

In Eq.\ (\ref{expec}), we provided the explicit form of the expectation values for $\hat{q}_2$ and $\hat{p}_2$ as functions of $q_1$, i.e., as fashionables, in the internal time Schr\"odinger regime. Below we also provide the explicit form of the moments associated to these two operators.
\begin{widetext}\ba\label{eqexpmom}
\begin{split}
(\Delta q_2)^2(q_1)&=\langle \hat{q}_2^2\rangle (q_1)-\langle \hat{q}_2\rangle^2(q_1)=\frac{\hbar}{2}\langle z(q_1)|\hat{a}^2+\hat{a}^+{}^2+2\hat{a}\hat{a}^++\hat{\mathbf{1}}|z(q_1)\rangle-\langle \hat{q}_2\rangle^2(q_1)\\
&=e^{-|z|^2}\sum_{n\geq0}\frac{|z|^{2n}}{n!}\left(\frac{{q_2}_0^2-{p_2}_0^2}{2}\cos\left(\frac{E_n(q_1)-E_{n+2}(q_1)}{\hbar}\right)-{q_2}_0{p_2}_0\sin\left(\frac{E_n(q_1)-E_{n+2}(q_1)}{\hbar}\right)\right)\\
&+\frac{{q_2}_0^2+{p_2}_0^2}{2}+\frac{\hbar}{2}-\langle \hat{q}_2\rangle^2(q_1)\q,\\
(\Delta p_2)^2(q_1)&=\langle \hat{p}_2^2\rangle (q_1)-\langle \hat{p}_2\rangle^2(q_1)=\frac{\hbar}{2}\langle z(q_1)|-\hat{a}^2-\hat{a}^+{}^2+2\hat{a}\hat{a}^++\hat{\mathbf{1}}|z(q_1)\rangle-\langle \hat{p}_2\rangle^2(q_1)\\
&=-e^{-|z|^2}\sum_{n\geq0}\frac{|z|^{2n}}{n!}\left(\frac{{q_2}_0^2-{p_2}_0^2}{2}\cos\left(\frac{E_n(q_1)-E_{n+2}(q_1)}{\hbar}\right)-{q_2}_0{p_2}_0\sin\left(\frac{E_n(q_1)-E_{n+2}(q_1)}{\hbar}\right)\right)\\
&+\frac{{q_2}_0^2+{p_2}_0^2}{2}+\frac{\hbar}{2}-\langle \hat{p}_2\rangle^2(q_1)\q,\\
\Delta(q_2p_2)(q_1)&=\frac{1}{2}\langle (\hat{q}_2-\langle \hat{q}_2\rangle)(\hat{p}_2-\langle \hat{p}_2\rangle)+(\hat{p}_2-\langle \hat{p}_2\rangle)(\hat{q}_2-\langle \hat{q}_2\rangle)\rangle=\langle(\hat{q}_2-\langle \hat{q}_2\rangle)(\hat{p}_2-\langle \hat{p}_2\rangle)\rangle-\frac{i\hbar}{2}\\
&=\langle\sqrt{\frac{\hbar}{2}}(-\langle \hat{p}_2\rangle+i\langle \hat{q}_2\rangle)\hat{a}-\sqrt{\frac{\hbar}{2}}(\langle \hat{p}_2\rangle+i\langle \hat{q}_2\rangle)\hat{a}^++\langle \hat{q}_2\rangle\langle \hat{p}_2\rangle+\frac{i\hbar}{2}(\hat{a}^+{}^2-\hat{a}^2)\rangle\\
&=e^{-|z|^2}\sum_{n\geq0}\frac{|z|^{2n}}{n!}\left(\left(\langle \hat{q}_2\rangle(q_1) {q_2}_0-\langle \hat{p}_2\rangle(q_1){p_2}_0\right)\sin\left(\frac{E_{n+1}(q_1)-E_n(q_1)}{\hbar}\right)\right.\\
&\left.-\left(\langle \hat{p}_2\rangle(q_1) {q_2}_0+\langle \hat{q}_2\rangle(q_1){p_2}_0\right)\cos\left(\frac{E_{n+1}(q_1)-E_n(q_1)}{\hbar}\right)+\frac{{q_2}_0^2-{p_2}_0^2}{2}\sin\left(\frac{E_n(q_1)-E_{n+2}(q_1)}{\hbar}\right)\right.\\
&\left.+{q_2}_0{p_2}_0\cos\left(\frac{E_n(q_1)-E_{n+2}(q_1)}{\hbar}\right)\right)+\langle \hat{q}_2\rangle(q_1)\langle \hat{p}_2\rangle(q_1)\q.
\end{split}
\ea\end{widetext}

\section{Imaginary contributions in the $q_i$-gauge of Sec.\ \ref{effrovmod}}\label{rovimag}

Here we want to summarize the analysis, which leads to the standard imaginary contribution (\ref{imqi}) to the clock $q_i$ in $q_i$-Zeitgeist.

Linearizing $q_i={q_i}_{cl}+\hbar\,{}^{(1)}q_i$ and $p_i={p_i}_{cl}+\hbar\,{}^{(1)}p_i$ and similarly for $q_j$ and $p_j$ yields to first order
\begin{widetext}\ba\label{pi1}
\begin{split}
\hbar\,{}^{(1)}p_i=-\left(\frac{(\Delta q_j)^2+(\Delta p_j)^2}{2{p_i}_{cl}}+\hbar \frac{2{p_i}_{cl}({p_j}_{cl}\,{}^{(1)}p_j+{q_i}_{cl}\,{}^{(1)}q_i+{q_j}_{cl}{}^{(1)}q_j)}{2{p_i}_{cl}^2}+\frac{i\hbar{q_i}_{cl}}{2{p_i}_{cl}^2}\right.\\
\left.+\frac{{p_j}_{cl}^2(\Delta p_j)^2+{q_j}_{cl}^2(\Delta q_j)^2+2{q_j}_{cl}{p_j}_{cl}\Delta(q_jp_j)}{2{p_i}_{cl}^3}\right)\q.
\end{split}
\ea\end{widetext}
Since the coefficients (\ref{rovcoeff}) are of zeroth order, it is consistent to replace all $q_i$, $q_j$, $p_i$ and $p_j$ appearing in terms of order $\hbar$ in (\ref{simpeom1}) by their zero-order (or classical) parts which in (\ref{pi1}) we have denoted by a subscript $cl$, and whose solutions are given in (\ref{cl-sol}). To order $\hbar$ this does not modify the equations and helps for their solutions. Furthermore, remembering that all zero-order variables are kept real-valued, (\ref{simpeom1}) and (\ref{pi1}) imply that either ${}^{(1)}p_i$ or ${}^{(1)}q_i$ or both must contain imaginary contributions while all variables associated to the canonical pair $(q_j,p_j)$ are consistently real-valued as a result of real-valued equations of motion.

Requiring $p_i$ to be real, it is obvious that \ba\label{imgqeom}
\frac{d\,\Im[q_i]}{ds}=-\frac{\hbar {q_i}_{cl}}{{p_i}_{cl}^2}\q. \ea
Using Eq.\ (\ref{cl-sol}) and integrating this equation, precisely
yields the standard imaginary contribution (\ref{imqi}) which is
also consistent with the constraint (\ref{pi1}) and cancels the
imaginary term in the equation of motion for $p_i$ in Eq.\
(\ref{simpeom1}). Requiring $q_i$ to be real-valued, however, and
repeating the same analysis shows that the solution for $\Im[p_i]$
would {\it not} reproduce the imaginary term
$-i\hbar{q_i}_{cl}/(2{p_i}_{cl}^2)$ in Eq.\ (\ref{pi1}). It is,
hence, inconsistent to keep $q_i$ real-valued and push the imaginary
contribution to $p_i$. In accordance with the analysis in
Sec.~\ref{sec:imtime} and \cite{EffTime1}, we, thus, find the
generic $o(\hbar)$ imaginary contribution inherent to all non-global
clocks in the effective framework.

\begin{acknowledgments}
We would like to thank Bianca Dittrich for useful comments, Igor
Khavkine and Renate Loll for interesting discussions and Emilia
Kubalova for reading a version of the manuscript.
Moreover, it is a special pleasure to thank Karel Kucha\v{r} for
preparing many careful and valuable handwritten comments on this
approach.
This work was
supported in part by NSF grant PHY0748336 and a grant from the
Foundational Questions Institute (FQXi). PAH is grateful for the
support of the German Academic Exchange Service (DAAD) through a
doctoral research grant and acknowledges a travel grant of
Universiteit Utrecht.  Furthermore, he would like to express his
gratitude to the Albert Einstein Institute in Potsdam for hospitality
during the final stages of this work.
Finally, we would like to thank an anonymous referee for constructive
criticism.
\end{acknowledgments}


\begin{thebibliography}{99}
\parskip -2pt


\bibitem{EffTime1} M.~Bojowald, P.~A.~H\"ohn and A.~Tsobanjan, Class.\
  Quantum Grav.\ {\bf 28} 035006 (2011), (arXiv:1009.5953[gr-qc])

\bibitem{Kuc1} K.~V.~Kucha\v{r}, in \textit{Proc.\ 4th Canadian Conference on General Relativity and Relativistic Astrophysics}, edited by G.~Kunstatter, D.~Vincent and J.~Williams (World Scientific, Singapore, 1992)
\bibitem{Ish} C.~J.~Isham, in \textit{Integrable Systems, Quantum Groups, and Quantum Field Theories} (Kluwer Academic Publishers, London, 1993) (arXiv:gr-qc/9210011),
C.~J.~Isham, in {\it Canonical Gravity: From Classical to Quantum}, edited by J.~Ehlers and H.~Friedrich, Lect.\ Notes Phys.\ {\bf 434} (Springer Verlag Berlin, 1994) 150
\bibitem{anderson} E.~Anderson, arXiv:1009.2157[gr-qc]
\bibitem{Rovbook} C.~Rovelli, \textit{ Quantum Gravity} (CUP, Cambridge, 2004)
\bibitem{Bar} J.~Barbour and B.~Z.~Foster, arXiv:0808.1223[gr-qc]
\bibitem{Kuc2} K.~V.~Kucha\v{r}, in \textit{Proc.\ 13th Intern.\ Conf.\ on General Relativity and Gravitation}, edited by R.~J.~Gleiser, C.~N.~Kozameh and O.~M.~Moreschi(Bristol: Institute of Physics, 1992) pp 119 (arXiv:gr-qc/9304012)
\bibitem{Rovmod} C.~Rovelli, Phys.\ Rev.\ {\bf D 42} 2638 (1990), Phys.\ Rev.\ {\bf D 43} 442 (1991), P.~H\'aj\'{\i}\v{c}ek, Phys.\ Rev.\ {\bf D 44} 1337 (1991), C.~Rovelli, Phys.\ Rev.\ {\bf D 44} 1339 (1991), C.~Rovelli, in \textit{Conceptual Problems of Quantum Gravity, proc.\ of the Osgood Hill Conf.\ Boston}, edited by A.~Ashtekar and J.~Stachel(Birkhauser, Boston,1991)



\bibitem{Bianca1} B.~Dittrich, Gen.\ Rel.\ Grav.\ {\bf 39} 1891 (2007) (arXiv:gr-qc/0411013), Class.\ Quant.\ Grav.\ {\bf 23} 6155 (2006) (arXiv: gr-qc/0507106)

\bibitem{Bianca2} B.~Dittrich and J.~Tambornino, Class.\ Quant.\ Grav.\ {\bf 24} 4543 (2007) (arXiv:gr-qc/0702093), Class.\ Quant.\ Grav.\ {\bf 24} 757 (2007) (arXiv:gr-qc/0610060)

\bibitem{Haj1} P.~H\'aj\'{\i}\v{c}ek, J.\ Math.\ Phys.\ {\bf 36} 4612 (1995) (arXiv:gr-qc/9412047), Class.\ Quant.\ Grav.\ {\bf 13} 1353 (1996) (arXiv:gr-qc/9512026), Nucl.\ Phys.\ B (Proc.\ Suppl.) {\bf 57} 115 (1997) (arXiv:gr-qc/9612051)

\bibitem{Hartle} J.~B.~Hartle, Class.\ Quant.\ Grav.\ {\bf 13} 361 (1996) (arXiv:gr-qc/9509037)

\bibitem{gampul} R.~Gambini, R.~A.~Porto and J.~Pullin, Gen.\ Rel.\ Grav.\ {\bf 39} 1143 (2007) (arXiv:gr-qc/0603090), R.~Gambini and J.~Pullin, Found.\ Phys.\ {\bf 37} 1074 (2007) (arXiv:quant-ph/0608243), R.~Gambini, R.~A.~Porto, J.~Pullin and S.~Torterolo, Phys.\ Rev.\ {\bf D 79} 041501(R) (2009) (arXiv:0809.4235[gr-qc]), R.~Gambini, L.~P.~Garc\'{\i}a-Pintos and J.~Pullin, arXiv:1002.4209[quant-ph]

\bibitem{Hajlec} P.~H\'aj\'{\i}\v{c}ek, in {\it Canonical Gravity: From Classical to Quantum}, edited by J.~Ehlers and H.~Friedrich, Lect.\ Notes Phys.\ {\bf 434} (Springer Verlag Berlin, 1994) 113

\bibitem{Haj2} P.~H\'aj\'{\i}\v{c}ek, J.\ Math.\ Phys.\ {\bf 30} 2488 (1989), M.~Sch\"on and P.~H\'aj\'{\i}\v{c}ek, Class.\ Quant.\ Grav.\ {\bf 7} 861 (1990), P.~H\'aj\'{\i}\v{c}ek, Class.\ Quant.\ Grav.\ {\bf 7} 871 (1990)

\bibitem{bojsisk} M.~Bojowald, P.~Singh and A.~Skirzewski, Phys.\ Rev.\ {\bf D 70} 124022 (2004) (arXiv:gr-qc/0408094)

\bibitem{Klauder} C.~Zhu and J.~R.~Klauder, Am.\ J.\ Phys.\ {\bf 61} 605 (1993)

\bibitem{EffCons} M.~Bojowald, B.~Sandh\"ofer, A.~Skirzewski and A.~Tsobanjan, Rev.\ Math.\ Phys.\ {\bf 21} 111 (2009) (arXiv:0804.3365[math-ph])

\bibitem{EffConsRel} M.~Bojowald and A.~Tsobanjan, Phys.\ Rev.\ {\bf D 80} 125008 (2009) (arXiv:0906.1772[math-ph])


\bibitem{PhysHilbert} D.~Marolf, arXiv:gr-qc/9508015, R.~M.~Wald, Phys.\ Rev.\ D {\bf 48} 2377 (R) (1993) (arXiv:gr-qc/9305024), T.~Thiemann, Class.\ Quant.\ Grav.\ {\bf 23} 2211 (2006) (arXiv:gr-qc/0305080),
B.~Dittrich and T.~Thiemann, Class.\ Quant.\ Grav.\
{\bf 23} 1025 (2006) (arXiv:gr-qc/0411138)

\bibitem{EffAc} M.~Bojowald and A.~Skirzewski, Rev.\ Math.\ Phys.\ {\bf 18}
713--745 (2006) (math-ph/0511043)

\bibitem{EffConsComp} M.~Bojowald and A.~Tsobanjan,
Class.\ Quant.\ Grav.\ {\bf 27} 145004 (2010) (arXiv:0911.4950[gr-qc])



\bibitem{HenTeit} M.~Henneaux and C.~Teitelboim, \textit{Quantization of Gauge Systems} (Princeton University Press, Princeton,1992)

\bibitem{ell} J.~Pollet, O.~M\'eplan and C.~Gignoux, J.\ Phys.\ A: Math.\ Gen.\ {\bf 28} 7287 (1995),
see especially p.\ 7288


\bibitem{GaugeFix} J.~M.~Pons, D.~C.~Salisbury and K.~A.~Sundermeyer, Phys.\ Rev.\  D {\bf 80}  084015 (2009) (arXiv:0905.4564[gr-qc])



\end{thebibliography}
\end{document}